\documentclass[longauth]{aa} 

\usepackage{graphicx}
\usepackage{array,multirow,makecell}
\usepackage{txfonts}
\usepackage{natbib}
\usepackage{amsmath}
\usepackage{ulem} 
\usepackage{booktabs}
\usepackage{arydshln}
\usepackage{textcomp}
\usepackage{gensymb}
\usepackage{color}
\usepackage{afterpage}
\usepackage[caption = false]{subfig}
\usepackage{float}
\bibpunct{(}{)}{;}{a}{}{,} 

\newcommand{\Msol}{\ensuremath{M_{\odot}}}

\newcommand{\hii}{H\mbox{\sc ~ii} }

\newcommand*{\rom}[1]{\expandafter\@slowromancap\romannumeral #1@}

\usepackage[colorlinks=True, filecolor=blue, linkcolor=blue]{hyperref}
\usepackage[capitalise]{cleveref}
\hypersetup{colorlinks,linkcolor={blue},citecolor={blue},urlcolor={cyan}}

\usepackage[caption=false]{subfig}

\usepackage[normal]{threeparttable}

\usepackage{amsmath} 
\DeclareFontEncoding{LS1}{}{}
\DeclareFontSubstitution{LS1}{stix}{m}{n}
\DeclareMathAlphabet\mathscr{LS1}{stixscr}{m}{n}

\usepackage{enumitem} 
\usepackage{longtable,lscape} 
\usepackage{stfloats} 
\usepackage[dvipsnames]{xcolor} 

\newcommand{\bsens}{\texttt{bsens}\xspace}
\newcommand{\cleanest}{\texttt{cleanest}\xspace}
\newcommand{\denoised}{\texttt{denoised}\xspace}
\newcommand{\original}{\texttt{original}\xspace}

\makeatletter
\renewcommand*\aa@pageof{, page \thepage{} of \pageref*{LastPage}}
\makeatother

\begin{document}

\title{ALMA-IMF III - Investigating the origin of stellar masses: Top-heavy core mass function in the W43-MM2\&MM3 mini-starburst\thanks{
    Full Tables E.1 and E.2 are only available at the CDS via anonymous ftp to \url{cdsarc.u-strasbg.fr} (130.79.128.5) or via \url{http://cdsweb.u-strasbg.fr/cgi-bin/qcat?J/A+A/}}
}
\titlerunning{ALMA-IMF III: Top-heavy core mass function in the W43-MM2\&MM3 mini-starburst}

\author{Y. Pouteau\inst{1}
    \and F. Motte\inst{1}
    \and T. Nony\inst{2}
    \and R. Galv\'an-Madrid\inst{2}
    \and A. Men'shchikov\inst{3}
    \and S. Bontemps\inst{4}
    \and J.-F. Robitaille\inst{1}
    \and F. Louvet\inst{5,1}
    \and A. Ginsburg\inst{6}
    \and F. Herpin\inst{4}
    \and A. L\'opez-Sepulcre \inst{1,7}
    \and P. Dell'Ova\inst{8,9}
    \and A. Gusdorf\inst{8,9}
    \and P. Sanhueza\inst{10,11}
    \and A. M.\ Stutz\inst{12,13}
    \and N. Brouillet\inst{4}
    \and B. Thomasson\inst{1}
    \and M. Armante\inst{8,9}
    \and T. Baug\inst{14}
    \and M. Bonfand\inst{4}
    \and G. Busquet\inst{15, 1}
    \and T. Csengeri\inst{4}
    \and N. Cunningham\inst{1}
    \and M. Fern\'andez-L\'opez\inst{16}
    \and H.-L. Liu\inst{12, 17}
    \and F. Olguin\inst{18}  
    \and A. P.\ M.\ Towner\inst{6}
    \and J. Bally\inst{19}
    \and J. Braine\inst{4}
    \and L. Bronfman\inst{5}
    \and I. Joncour\inst{1}
    \and M. Gonz\'alez\inst{1, 20}
    \and P. Hennebelle\inst{3}
    \and X. Lu\inst{21}
    \and K. M.\ Menten\inst{22}
    \and E. Moraux\inst{1}
    \and K. Tatematsu\inst{23,11}
    \and D. Walker\inst{24}
    \and A. P.\ Whitworth\inst{25}
    }

\institute{Univ. Grenoble Alpes, CNRS, IPAG, 38000 Grenoble, France  
    \and Instituto de Radioastronom\'ia y Astrof\'isica, Universidad Nacional Aut\'onoma de M\'exico, Morelia, Michoac\'an 58089, M\'exico 
    \and AIM, IRFU, CEA, CNRS, Universit{\'e} Paris-Saclay, Universit{\'e} Paris Diderot, Sorbonne Paris Cit{\'e}, F-91191 Gif-sur-Yvette, France 
    \and Laboratoire d'astrophysique de Bordeaux, Univ. Bordeaux, CNRS, B18N, all\'ee Geoffroy Saint-Hilaire, 33615 Pessac, France  
    \and Departamento de Astronom\'{i}a, Universidad de Chile, Casilla 36-D, Santiago, Chile   
    \and Department of Astronomy, University of Florida, PO Box 112055, USA 
    \and Institut de RadioAstronomie Millim\'etrique (IRAM), Grenoble, France    
    \and Laboratoire de Physique de l'{\'E}cole Normale Sup{\'e}rieure, ENS, Universit{\'e} PSL, CNRS, Sorbonne Universit{\'e}, Universit{\'e} de Paris, 75005, Paris, France 
    \and Observatoire de Paris, PSL University, Sorbonne Universit{\'e}, LERMA, 75014, Paris, France 
    \and National Astronomical Observatory of Japan, National Institutes of Natural Sciences, 2-21-1 Osawa, Mitaka, Tokyo 181-8588, Japan 
    \and Department of Astronomical Science, SOKENDAI (The Graduate University for Advanced Studies), 2-21-1 Osawa, Mitaka, Tokyo 181-8588, Japan   
    \and Departamento de Astronom\'{i}a, Universidad de Concepci\'{o}n, Casilla 160-C, 4030000 Concepci\'{o}n, Chile 
    \and Max-Planck-Institute for Astronomy, K\"{o}nigstuhl 17, 69117 Heidelberg, Germany   
    \and S. N. Bose National Centre for Basic Sciences, Block JD, Sector III, Salt Lake, Kolkata 700106, India    
    \and Departament de F\'isica Qu\`antica i Astrof\'isica, Institut de Ci\`encies del Cosmos, Universitat de Barcelona (IEEC-UB), c/ Mart\'i i Franqu\`es 1, E-08028, Barcelona, Catalonia, Spain 
    \and Instituto Argentino de Radioastronom\'\i a (CCT-La Plata, CONICET; CICPBA), C.C. No. 5, 1894, Villa Elisa, Buenos Aires, Argentina   
    \and Department of Astronomy, Yunnan University, Kunming, 650091, PR China    
    \and Institute of Astronomy, National Tsing Hua University, Hsinchu 30013, Taiwan   
    \and Department of Astrophysical and Planetary Sciences, University of Colorado, Boulder, Colorado 80389, USA    
    \and Universidad Internacional de Valencia (VIU), C/Pintor Sorolla 21, E-46002 Valencia, Spain 
    \and Shanghai Astronomical Observatory, Chinese Academy of Sciences, 80 Nandan Road, Shanghai 200030, People’s Republic of China  
    \and Max Planck Institute for Radio Astronomy, Auf dem H\"{u}gel 69, 53121 Bonn,  Germany  
    \and Nobeyama Radio Observatory, National Astronomical Observatory of Japan, National Institutes of Natural Sciences, Nobeyama, Minamimaki, Minamisaku, Nagano 384-1305, Japan 
    \and University of Connecticut, Department of Physics, 196A Auditorium Road, Unit 3046, Storrs, CT 06269 USA 
    \and School of Physics and Astronomy, Cardiff University, Cardiff, UK 
    }

\date{Received December 19, 2021; accepted \today}

\abstract
{}
%
{The processes that determine the stellar initial mass function (IMF) and its origin are critical unsolved problems, with profound implications for many areas of astrophysics. The W43-MM2\&MM3 mini-starburst ridge hosts a rich young protocluster, from which it is possible to test the current paradigm on the IMF origin. }
%
{The ALMA-IMF Large Program observed the W43-MM2\&MM3 ridge, whose 1.3~mm and 3~mm ALMA 12~m array continuum images reach a $\sim$2\,500~au spatial resolution. We used both the best-sensitivity and the line-free ALMA-IMF images, reduced the noise with the multi-resolution segmentation technique \textsl{MnGSeg}, and derived the most complete and most robust core catalog possible. 
Using two different extraction software packages, \textsl{getsf} and \textsl{GExt2D}, we identified $\sim$200 compact sources, whose $\sim$100 common sources have, on average, fluxes consistent to within 30\%. We filtered sources with non-negligible free-free contamination and corrected fluxes from line contamination, resulting in a W43-MM2\&MM3 catalog of 205 \textsl{getsf} cores. With a median deconvolved FWHM size of 3\,400~au, core masses range from $\sim$0.1$~\Msol$ to $\sim$70$~\Msol$ and the \textsl{getsf} catalog is 90\% complete down to $0.8~\Msol$.}
%
{The high-mass end of the core mass function (CMF) of W43-MM2\&MM3 is top-heavy compared to the canonical IMF. Fitting the cumulative CMF with a single power-law of the form $N(>\log M)\propto M^{\rm \alpha}$, we measured $\alpha=-0.95 \pm 0.04$, compared to the canonical $\alpha = -1.35$ Salpeter IMF slope. The slope of the CMF is robust with respect to map processing, extraction software packages, and reasonable variations in the assumptions taken to estimate core masses. We explore several assumptions on how cores transfer their mass to stars (assuming a mass conversion efficiency) and subfragment (defining a core fragment mass function) to predict the IMF resulting from the W43-MM2\&MM3 CMF. While core mass growth should flatten the high-mass end of the resulting IMF, core fragmentation could steepen it.}
%
{In stark contrast to the commonly accepted paradigm, our result argues against the universality of the CMF shape. More robust functions of the star formation efficiency and core subfragmentation are required to better predict the resulting IMF, here suggested to remain top-heavy at the end of the star formation phase. If confirmed, the IMFs emerging from starburst events could inherit their top-heavy shape from their parental CMFs, challenging the IMF universality.}


\maketitle


\section{Introduction} \label{sect:intro}

The stellar initial mass function (IMF), which characterizes the mass distribution of stars between $0.01~\Msol$ and $>100~\Msol$, has long been considered universal \citep[see, e.g., reviews by][]{bastian2010, kroupa2013}. The IMF, which is therefore qualified as canonical, is often represented by a lognormal function peaking at stellar masses around $0.2-0.3~\Msol$, connected to a power-law tail, $\frac{{\rm d}N}{{\rm d}\log M} \propto M^{-1.35}$, that dominates for masses larger than $1~\Msol$ \citep{chabrier2005}. 
Following the functional description of the IMF by \cite{salpeter1955} and \cite{scalo1986}, \cite{kroupa1993} proposed another representation based on a series of three broken power-laws. In this representation, which was later refined by \cite{kroupa2002}, the form of the IMF would follow $\frac{{\rm d}N}{{\rm d}\log M} \propto M^{0.7}$ in the range $0.01-0.08~\Msol$, $\frac{{\rm d}N}{{\rm d}\log M} \propto M^{-0.3}$ in the range $0.08-0.5~\Msol$, and $\frac{{\rm d}N}{{\rm d}\log M} \propto M^{-1.3}$ for $M>0.5~\Msol$. The power-laws at the high-mass end of these two representations correspond, within the limits of observational uncertainties, to the description of \cite{salpeter1955}, $\frac{{\rm d}N}{{\rm d}\log M} \propto M^{-1.35}$, which becomes $N(>\log M)\propto M^{-1.35}$ in its complementary cumulative distribution form. 
The IMF universality, which has been postulated on the basis of studies of field stars and young stellar clusters in the solar vicinity (up to a few hundred of parsecs), has recently been challenged in more extreme environments. Observations of young massive clusters in the Milky Way \citep{lu2013, maia2016, hosek2019}, in nearby galaxies \citep{schneider2018}, and of high-redshift galaxies \citep{smith2014, zhang2018} measured top-heavy IMFs with a large proportion of high-mass stars compared to low-mass stars (see review by \citealt{hopkins2018}). Conversely, bottom-heavy IMFs have been measured for metal-rich populations, indicating that the IMF may vary with metallicity \citep[e.g.,][]{marks2012, martin-navarro2015}. 

The physical processes at the origin of the IMF and the questions of whether and how the IMF is linked to its environment are still a matter of debate (see reviews by \citealt{offner2014, krumholz2015, ballesteros2020, lee2020}). 
Over the past two decades a plethora of studies of the core populations in nearby star-forming regions revealed that their mass distribution, called the core mass function (CMF), has a shape that resembles that of the IMF. This result has been consistently found through (sub)millimeter continuum observations with ground-based single-dish telescopes \citep[e.g.,][]{motte1998, motte2001, stanke2006, enoch2008} and interferometers \citep[e.g.,][]{testi-sargent1998}. It has been confirmed with deep, far-infrared to submillimeter images obtained by the \textit{Herschel} space observatory \citep[e.g.,][]{konyves2015, benedettini2018, massi2019, ladjelate2020} and a handful of near-infrared extinction maps and molecular line integrated images \citep{alves2007, onishi2001, takemura2021}. 
The astonishing similarity between the IMF and the observed CMFs, all of which are consistent with each other, suggests that the IMF may inherit its shape from the CMF \citep[e.g.,][]{motte1998, andre2014}. 

The IMF would arise from a global shift of the CMF by introducing, for individual cores, a conversion efficiency of core mass into star mass, also called star formation efficiency ($\epsilon_{\rm core}$). 
CMF studies in low-mass star-forming regions suggest a broad range of mass conversion efficiencies, from $\epsilon_{\rm core}\sim 15\%$ \citep{onishi2001} to $\epsilon_{\rm core}\sim 30-40\%$ \citep{alves2007, konyves2015, pezzuto2021} or even $\epsilon_{\rm core}\sim 100\%$ \citep{motte1998, benedettini2018}. These differences could simply be related to the spatial resolution of the observations, which defines cores as peaked cloud structures with full width at half maximum (FWHM) sizes $1-3$ times the resolution element \citep{reid2010, louvet2021simu, tatematsu2021}. 
Cores identified in low-mass star-forming regions generally have sizes of $1\,000-20\,000$~au ($0.005-0.1$~pc) and masses of $0.01-10~\Msol$. We here adapt the terminology of \cite{motte2018a} to gas structures in massive protoclusters and assume that clumps have sizes of $\sim$0.1~pc (or 20\,000~au), cores of $\sim$0.01~pc (or 2\,000~au), and fragments of $\sim$500~au.

In contrast with the vast majority of published CMF studies, \cite{motte2018b} and \cite{kong2019} revealed that the CMF of two high-mass star-forming clouds, W43-MM1 and G28.37+0.07, presented an excess of high-mass cores, challenging the classical interpretation of the IMF origin. Combined CMFs, each built from a dozen to several dozen massive clumps, are also top-heavy \citep{csengeri2017b, liu2018, sanhueza2019, lu2020, sadaghiani2020, oneill2021}. However, these CMF measurements are most likely biased by mass segregation because clumps, which were observed with single pointings \citep[except for][]{sanhueza2019}, are overpopulated with massive cores that cluster at their centers \citep{kirk2016, plunkett2018, dibHenning2019, nony2021}. Systematic studies of massive protoclusters imaged at submillimeter wavelengths over their full extent, possibly a few square parsecs,
are necessary to determine whether they generally display a canonical or top-heavy CMF.

Although it is obvious that the star mass originates from the gas mass in molecular clouds, the gas reservoir used to form a star is difficult to define from observations. Most CMF studies are based on the concept of cores in the framework of the core-collapse model \citep{shu1987, andre2014}. Cores would be the quasi-static mass reservoirs for the self-similar collapse of protostars that will form a single star or, at most, a small stellar system originating from disk fragmentation. From recent studies \citep[e.g.,][]{csengeri2011, olguin2021, sanhueza2021}, it has become obvious, however, that cores are dynamical entities that are not isolated from their surroundings. In the framework of competitive accretion, hierarchical global collapse, or coalescence-collapse scenarios, cores generally acquire most of their mass during the protostellar collapse \citep[e.g.,][]{bonnell2006, leeHennebelle2018a, vazquez2019, pelkonen2021}. 
Despite the ill-defined concept of a core, constraining the CMF shape is crucial to show its universality or lack thereof. In particular, the CMFs of high-mass star-forming regions need to be constrained to investigate whether they follow the shape found in nearby, low-mass star-forming clouds \citep[e.g.,][]{konyves2015, ladjelate2020, pezzuto2021} or whether they are, at least in some cases, top-heavy. We here take the CMF as a metric, useful for comparing the distribution of small-scale structures, the cores of different clouds, and discuss the potential consequences of its shape on that of the IMF.

Predicting the IMF from an observed CMF requires, among other things, a precise knowledge of the turbulent core subfragmentation, also called core multiplicity. 
The fragmentation of cores of size $\sim$2\,000~au into fragments of a few hundred astronomical units, however, remains a very young area of research. This is even more the case for the disk fragmentation process, which is expected to take over at scales smaller than $\sim$100~au.
As a consequence, only a handful of studies investigated the effect of core multiplicity on the IMF, and they were only based on stellar multiplicity prescriptions \citep{swift2008, hatchellFuller2008, alcockParker2019, clarkWhitworth2021}. The authors used a wide range of core mass distributions between subfragments, also called mass partitions, varying from equipartition to a strong imbalance.

The history of star formation can also significantly complicate the potentially direct relationship between the CMF and the IMF. The CMF represents a $\sim$10$^5$~yr snapshot, only valid for the cores involved in one star formation event, which lasts for one to two clump free-fall times \citep{motte2018a}. In contrast, the IMF results from the sum, over $\sim$10$^6$~yr in young star clusters to $10^9-10^{11}$~yr in galaxies \citep{heiderman2010, krumholz2015}, of the stars formed by many, $10-10^6$, star formation events.

The ALMA-IMF\footnote{
    ALMA project \#2017.1.01355.L, see \url{http://www.almaimf.com}.}
Large Program (PIs: Motte, Ginsburg, Louvet, Sanhueza) is a survey of 15 nearby Galactic protoclusters that aims to obtain statistically meaningful results on the origin of the IMF \citep[see companion papers, Paper~I and Paper~II,][]{motte2021, ginsburg2021}. 
The W43-MM2 cloud is the second most massive young protocluster of ALMA-IMF \citep[$\sim$$1.2\times10^4~\Msol$ over 6~pc$^2$,][]{motte2021}. With its less massive neighbor, W43-MM3, also imaged by ALMA-IMF, W43-MM2 constitutes the W43-MM2\&MM3 ridge, which has a total mass of $\sim$3.5$\,\times10^4~\Msol$ \citep{nguyen2013} over a $\sim$14~pc$^2$ area. Located at $5.5$~kpc from the Sun \citep{zhangB2014}, the W43-MM2\&MM3 ridge is part of the exceptional W43 molecular cloud, which is at the junction of the Scutum-Centaurus spiral arm and the Galactic bar \citep{nguyen2011a, motte2014}. As expected from the high-density filamentary parsec-size structures that we call ridges \citep[see][]{hill2011, hennemann2012, motte2018a}, W43-MM2\&MM3 hosts a rich protocluster efficiently forming high-mass stars, thus qualifying as a mini-starburst \citep{nguyen2011b,motte2021}. In the W43-MM1 ridge, which is located 10~pc north of W43-MM2\&MM3, a mini-starburst protocluster has also been observed \citep{louvet2014, motte2018b, nony2020}. The W43-MM1 and W43-MM2\&MM3 clouds could therefore be the equivalent progenitors of the Wolf-Rayet and OB-star cluster \citep{blum1999,bik2005}
located between these two ridges and powering a giant \hii region. Despite the presence of gas heated by this giant \hii region, the W43-MM2\&MM3 ridge is mainly constituted of cold gas (21-28~K, see Fig.~2 of \citealt{nguyen2013}). 
In Paper~I \citep{motte2021} W43-MM1 and W43-MM2 are qualified as young protoclusters, while the W43-MM3 cloud represents a more evolved evolutionary stage, quoted as intermediate.

From the ALMA observations presented in Sect.~\ref{sect:obs and DR}, we set up a new extraction strategy that results in a census of 205 cores in the W43-MM2\&MM3 ridge (see Sect.~\ref{sect:extraction of compact sources}). The thermal dust emission of cores is carefully assessed and their masses are estimated (see Sect.~\ref{sect:core nature mass estim}). In Sect.~\ref{sect:cmf results}, we present the top-heavy CMF found for the W43-MM2\&MM3 protocluster and discuss its robustness. In Sect.~\ref{sect:discussion on the origin of stellar masses}, we then predict the core fragmentation mass function and IMF resulting from various mass conversion efficiencies and core fragmentation scenarios. We summarize the paper and present our conclusions in Sect.~\ref{sect:conclusions}.

{\renewcommand{\arraystretch}{1.5}%
\begin{table*}[ht]
\centering
\begin{threeparttable}[c]
\caption{Observational data summary for the W43-MM2 and W43-MM3 12~m array images and their combination.}
\label{tab:observation table}
\begin{tabular}{cccccccc}
    \hline\hline
    \multirow{2}{*}{ALMA band} & \multirow{2}{*}{Field} & \multirow{2}{*}{Mosaic size} & \multirow{2}{*}{$\Theta_{\rm maj}\times\Theta_{\rm min}$} & \multirow{2}{*}{BPA} & Continuum & Original & Denoised \\
     & & & & & bandwidth & RMS & RMS \\
     & & [$\arcsec\times\arcsec$] & [$\arcsec\times\arcsec$] & [$\degree$] & [GHz] & [mJy$\,$beam$^{-1}$] &[mJy$\,$beam$^{-1}$] \\
    (1) & (2) & (3) & (4) & (5) & (6) & (7) & (8) \\
    \hline
     & \multirow{2}{*}{W43-MM2} & \multirow{2}{*}{$92\times97$} & \multirow{2}{*}{$0.52\times0.41$} & \multirow{2}{*}{106} & 1.655 (\cleanest) & 0.175 & -- \\
     & & & & & 3.448 (\bsens) & 0.132 & -- \\
    \cline{2-8}
    $1.3~$mm & \multirow{2}{*}{W43-MM3} & \multirow{2}{*}{$92\times97$} & \multirow{2}{*}{$0.51\times0.43$} & \multirow{2}{*}{
    89} & 3.172 (\cleanest) & 0.101 & -- \\
    $228.4~$GHz & & & & & 3.448 (\bsens) & 0.093 & -- \\
    \cline{2-8}
     & \multirow{2}{*}{W43-MM2\&MM3} & \multirow{2}{*}{$158\times120$} & \multirow{2}{*}{$0.51\times0.42$} & \multirow{2}{*}{98} & $-$ (\cleanest) & $\sim$0.15 & -- \\
     & & & & & 3.448 (\bsens) & $\sim$0.11 & $\sim$ 0.08 \\
    \hline\hline
     & \multirow{2}{*}{W43-MM2} & \multirow{2}{*}{$202\times180$} & \multirow{2}{*}{$0.30\times0.24$} & \multirow{2}{*}{107} & 1.569 (\cleanest) & 0.041 & -- \\
     & & & & & 2.906 (\bsens) & 0.026 & -- \\
    \cline{2-8}
    $3.0~$mm & \multirow{2}{*}{W43-MM3} & \multirow{2}{*}{$202\times180$} & \multirow{2}{*}{$0.42\times0.28$} & \multirow{2}{*}{94} & 2.528 (\cleanest) & 0.045 & -- \\
    $99.66~$GHz & & & & & 2.906 (\bsens) & 0.031 & -- \\
    \cline{2-8}
     & \multirow{2}{*}{W43-MM2\&MM3} & \multirow{2}{*}{$275\times202$} & \multirow{2}{*}{$0.46\times0.46$} & \multirow{2}{*}{101} & $-$ (\cleanest) & $\sim$0.048 & -- \\
     & & & & & 2.906 (\bsens) & $\sim$ 0.028 &  $\sim$ 0.021 \\
    \hline
\end{tabular}
\begin{tablenotes}[flushleft]
\item (4) Major and minor sizes of the beam at half maximum. $\Theta_{\rm beam}$ is the geometrical average of these two quantities.
\item (5) Position angle of the beam, measured counterclockwise from north to east.
\item (6) Spectral bandwidth used to estimate the continuum emission level, with the name of the associated image in parentheses (see their definition in Sect.~\ref{sect:obs and DR}).
\item (7) Noise level measured in the original map unities and thus with different beam sizes (see Col.~4).
\item (8) Noise level measured in the \textsl{MnGSeg} \denoised images (see Sect.~\ref{sect:extraction of compact sources} and \citealt{robitaille2019}).
\end{tablenotes}
\end{threeparttable}
\end{table*}}

\section{Observations and data reduction} \label{sect:obs and DR}

Observations were carried out between December 2017 and December 2018 as part of the ALMA Large Program named ALMA-IMF (project \#2017.1.01355.L, see \citealt{motte2021}). The 12~m and 7~m ALMA arrays were used at both 1.3~mm and 3~mm (central frequencies $\nu_{\rm c} \simeq 228.4$~GHz in band~6 and $\simeq 99.66$~GHz in band~3, see \cref{tab:observation table}). The W43-MM2 and W43-MM3 fields have the same extent and were imaged by the ALMA 12~m and 7~m arrays with mosaics composed of 27 (respectively 11) pointings at 1.3~mm and 11 (respectively 3) pointings at 3~mm. For the 12~m array images, the maximum recoverable scales are $\sim$5.6~$\arcsec$ at 1.3~mm and $\sim$8.1~$\arcsec$ at 3~mm \citep{motte2021}, corresponding to $0.15-0.2$~pc at 5.5~kpc.
At 1.3~mm and 3~mm, eight (respectively four) spectral windows were selected for the ALMA-IMF setup; they sum up to bandwidths of 3.7~GHz and 2.9~GHz, respectively. \cref{tab:observation table} summarizes the basic information of 12~m array observations for each field and each continuum waveband. A more complete description of the W43-MM2 and W43-MM3 data sets can be found in Paper~I \citep{motte2021} and Paper~II \citep{ginsburg2021}.

The present W43-MM2 and W43-MM3 data sets were downloaded from the ALMA archive before they were corrected for system temperature and spectral data normalisation\footnote{
    ALMA ticket: \url{https://help.almascience.org/kb/articles/607}, \url{https://almascience.nao.ac.jp/news/amplitude-calibration-issue-affecting-some-alma-data}}.
This, however, has no significant impact on the continuum data as shown in Section~2 of Paper~II \citep{ginsburg2021}.
The data were first calibrated using the default calibration pipelines of the CASA\footnote{
    ALMA Pipeline Team, 2017, ALMA Science Pipeline User’s Guide, ALMA Doc 6.13.
    See \url{https://almascience.nrao.edu/processing/science-pipeline}.} software.
We then used an automatic CASA~5.4 pipeline script\footnote{
    \url{https://github.com/ALMA-IMF/reduction}}
developed by the ALMA-IMF consortium and fully described in Paper~II \citep{ginsburg2021} to produce self-calibrated images. In short, this pipeline performs several iterations of phase self-calibration using custom masks in order to better define the self-calibration model and clean more deeply using the TCLEAN task and refined parameters after each pass. This process results in quantitatively reducing interferometric artifacts and leads to a noise level reduced by 12-20\% at 1.3~mm and 8-12\% at 3~mm for the 12~m array images for W43-MM2 and W43-MM3, respectively. 
The data we used for this analysis are different from those presented in Paper~I and Paper~II \citep{motte2021,ginsburg2021}, which are from an updated version of the pipeline using, among other things, CASA~5.7 instead of CASA~5.4 and an updated version of ALMA data products. 
We compared the images presented here to those in Paper~I and Paper~II \citep{motte2021,ginsburg2021} and found that the flux differed by $<$5\% for all continuum peaks. The difference is largely accounted for by small differences ($<$5\%) in beam area, which arise from changes in the baseline weighting during the processing that corrected for system temperature and spectral data normalisation. Greater differences were observed in the extended emission, but this has no impact on our analysis since, as described in Sect.~\ref{sect:extraction of compact sources}, the extended emission is filtered out when source identification is performed. 
We used the \texttt{multiscale} option of the TCLEAN task to minimize interferometric artifacts associated with missing short spacings. With the \texttt{multiscale} parameters of 0, 3, 9, 27 pixels (up to 81 at 3~mm) and with $4-5$ pixels per beam, it independently cleaned structures with characteristic sizes from the geometrical average of the beam size, $\Theta_{\rm beam}\simeq$0.46$\arcsec$, to $6$ and $17$ times this value, which means $\sim$2.7$\arcsec$ at 1.3~mm and up to $\sim$8$\arcsec$ at 3~mm, respectively. The combined 12~m$\,+\,$7~m images have a noise level higher by a factor of $\sim$3.4\footnote{
    The higher noise level of the combined ALMA 12~m $+$ ACA 7~m images is due to a) the higher noise level of the 7~m data, b) the structural noise resulting from larger-scale emission, and c) the lower efficiency of the self-calibration process when applied to 7~m data.}
and will thus not be used in this work.

\begin{figure*}[hbtp!]
    \centering
    \vskip -0.5cm
    \begin{minipage}{1.\textwidth}
      \centering
      \includegraphics[width=.94\textwidth]{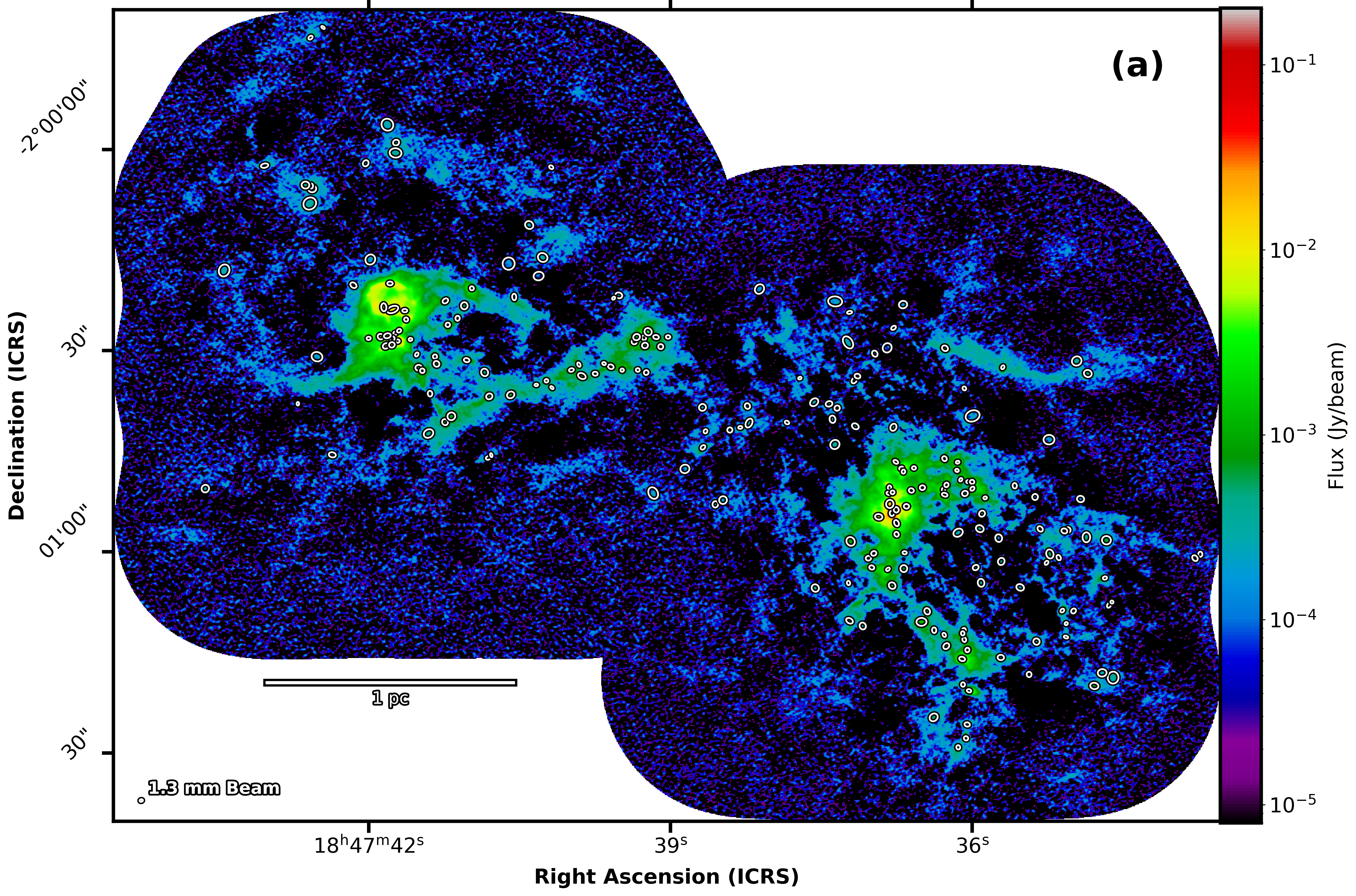}
    \end{minipage}
    \begin{minipage}{1.\textwidth}
      \hspace{14pt} 
      \includegraphics[width=.85\textwidth]{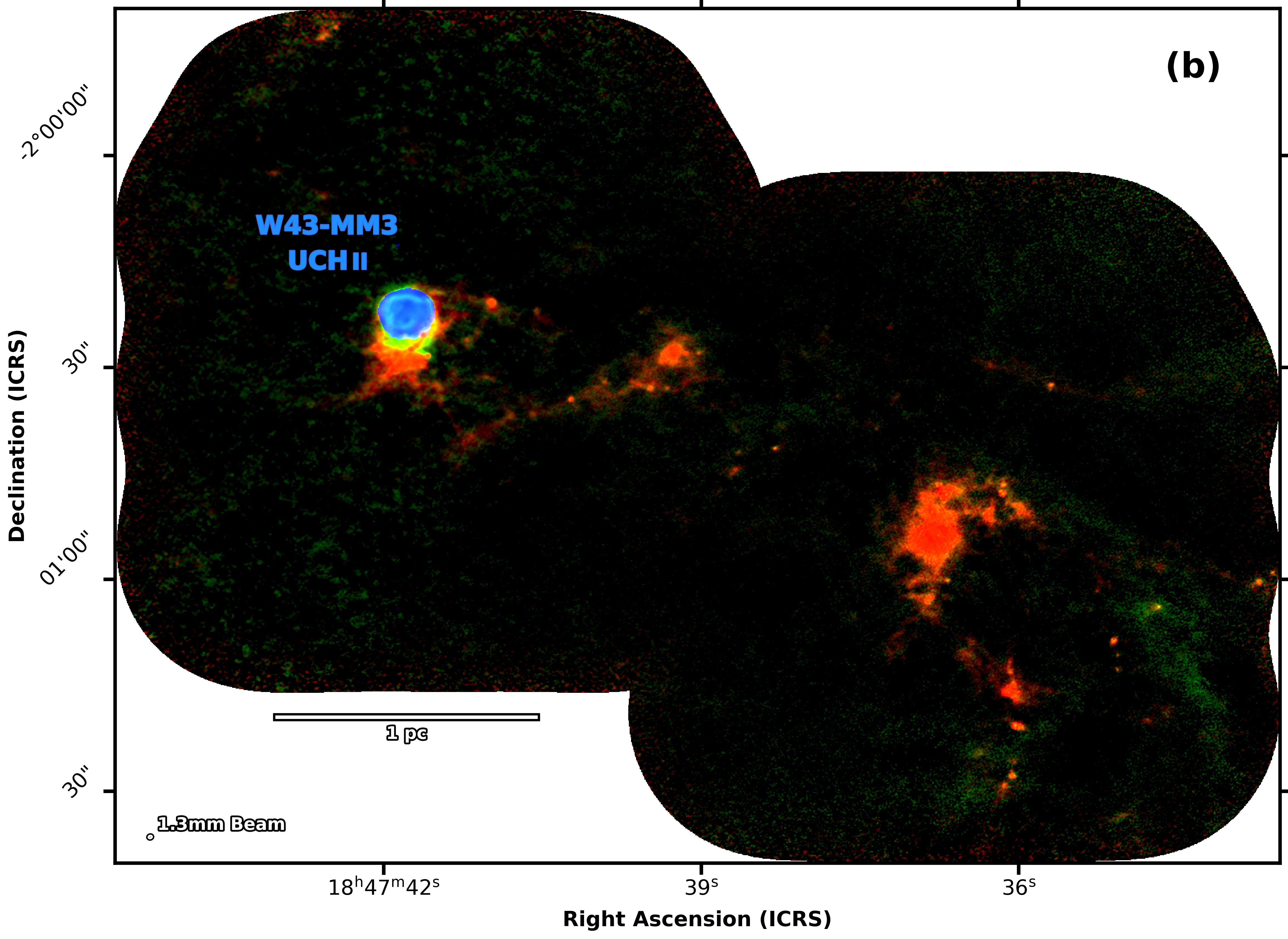}
    \end{minipage}%
    \caption{W43-MM2\&MM3 protocluster cloud. Panel \textsl{(a)}: 1.3~mm image obtain by the ALMA 12~m array (best-sensitivity image, prior to primary-beam correction). W43-MM2 is to the west and W43-MM3 is to the east. White ellipses outline the FWHM size of compact cores extracted by \textsl{getsf}. Panel \textsl{(b)}: Three-color ALMA image. Red and green display the \bsens continuum images at 1.3~mm and 3~mm, respectively, scaled by the theoretical ratio of thermal dust emission (see Eq.~\ref{eq:theo thermal ratio}). Blue corresponds to the free-free continuum emission image at the frequency of the H41$\alpha$ recombination line (Galv\'an-Madrid et al. in prep.). Filaments and cores appear in orange (red $+$ green), tracing thermal dust emission; the UC\hii region appears in blue or cyan (blue $+$ green), indicating free-free emission. Ellipses in the lower left corners represent the angular resolution of the \bsens 1.3~mm image and scale bars indicate the size in physical units.}
    \label{fig:1.3mm and trichrone}
\end{figure*}

The ALMA-IMF pipeline produces two different estimates of the continuum images \citep[see][]{ginsburg2021}. The first, called the \cleanest image, was produced using the \texttt{findContinuum} routine of CASA which excludes, before the TCLEAN task, the channels associated with lines to estimate the continuum level. The \cleanest image is thus a continuum image free of line contamination. 
In the case of the ALMA-IMF data of W43-MM2 and W43-MM3, the bandwidths of the \cleanest images are, respectively, a fraction of $\sim$50\% and $\sim$90\% of the total bandwidths at 1.3~mm and 3~mm (see \cref{tab:observation table} and Fig.~3 of \citealt{ginsburg2021}). The second continuum image produced by the ALMA-IMF pipeline uses all channels of all the spectral bands to estimate the continuum at 1.3~mm and 3~mm. With a $\sim$30\% decrease in the rms noise level, it corresponds to the best-sensitivity image and is thus called the \bsens image (see \cref{tab:observation table}). 

The W43-MM2 and W43-MM3 ALMA fields share a common area in both bands: $\sim$10$\arcsec \times 90\arcsec$ at 1.3~mm and $\sim$100$\arcsec \times 180\arcsec$ at 3~mm within their respective primary-beam responses down to 15\%. 
We combined the individually cleaned images in the image plane because CASA~5.4 cannot clean two fields with two different phase centers using the \texttt{multiscale} option. Although we requested the same angular resolution for both 1.3~mm and 3~mm mosaics, the latter were observed at a much higher resolution (see \cref{tab:observation table}). We thus smoothed the W43-MM2 and W43-MM3 \cleanest and \bsens images at 3~mm to the angular resolution of the 1.3~mm images, $\sim$0.46$\arcsec$, or 2\,500~au at the 5.5~kpc distance of W43. Because the beam orientations are similar (see \cref{tab:observation table}), we assumed that the median of the W43-MM2 and W43-MM3 parallactic angles are good approximations for the beams of the combined images. We then used the primary-beam shape of each individual mosaic to weight\footnote{
    The combined primary-beam corrected image, $I_{\rm MM2+MM3}^{\rm PBcor}$, is the sum of individual primary-beam corrected images, $I_{\rm MM2}^{\rm PBcor}$ and $I_{\rm MM3}^{\rm PBcor}$ weighted by their combined primary-beam maps, PB$_{\rm MM2}$ and PB$_{\rm MM3}$, following the equation
    \begin{equation*}
    I_{\rm MM2+MM3}^{\rm PBcor} = \frac{ I_{MM2}^{\rm PBcor} \times ({\rm PB_{\rm MM2}})^2 + I_{\rm MM3}^{\rm PBcor} \times ({\rm PB_{\rm MM3}})^2 } {({\rm PB_{\rm MM2}})^2 + ({\rm PB_{\rm MM3}})^2}.
    \end{equation*}} 
the flux of pixels in the common area and define the combined primary-beam corrected image. This approach is valid because the noise level, 
when measured in the common area of maps with the same beam and uncorrected by the primary beam, is similar to within 20\% between maps, which is smaller than the 35\% difference measured on the whole map (see \cref{tab:observation table}). 

Figures~\ref{fig:1.3mm and trichrone}a and \ref{appendixfig:3mm image with cores} present the W43-MM2\&MM3 ridge, covered by the combined image of the W43-MM2 and W43-MM3 protoclusters observed by ALMA-IMF. They display the 12~m array \bsens image at 1.3~mm and 3~mm, respectively. Figure~\ref{fig:1.3mm and trichrone}b presents a three-color image, which separates the thermal dust emission of star-forming filaments from the free-free emission associated with \hii regions, as done in Paper~I \citep{motte2021}. It uses ALMA-IMF images of the 1.3~mm and 3~mm continuum and of the H41$\alpha$ recombination line, tracing the free-free continuum emission of ionized gas (see Sect.~\ref{sect:obs and DR} and \citealt{motte2021}).
Several filaments cross the image and the W43-MM2 cloud displays a centrally concentrated structure reminiscent of hubs \citep[e.g.,][]{myers2009, peretto2013, didelon2015}. 
In single-dish studies, W43-MM2 has a $2.4\times 10^4~L_\odot$ bolometric luminosity, integrated over 0.23~pc, and coincides with a 6.67~GHz methanol maser \citep{walsh1998, motte2003}. The W43-MM3 clump, itself characterized by \cite{elia2021}, has a 0.24~pc size and $5.7\times 10^4~L_\odot$ bolometric luminosity.
In \cref{fig:1.3mm and trichrone}b, it harbors an ultra-compact \hii (UCH{\sc ~ii}) region, whose bubble forms a ring-like structure. Its $\sim$0.12~pc diameter, or $\sim$4.8$\arcsec$ at 5.5~kpc, is in good agreement with its size estimated from single-dish millimeter continuum \citep{motte2003}. 
Many compact sources are found along the dust emission of filaments of the W43-MM2\&MM3 ridge, suggesting that they could be dense cloud fragments such as cores.

\section{Extraction of compact sources} \label{sect:extraction of compact sources}
Since our goal is to extract cores from their surrounding cloud, we need to use software packages that identify and characterize cores as emission peaks, whose size is limited by their structured background and neighboring cores. Many source extraction algorithms have been used in star formation studies \citep[see][]{joncour2020,men2021getsf}. Here we use two completely independent methods, \textsl{getsf} and \textsl{GExt2D}.

The \textsl{getsf}\footnote{
    \url{https://irfu.cea.fr/Pisp/alexander.menshchikov/}} 
method \citep{men2021getsf} employs a spatial decomposition of the observed images to better isolate various spatial scales and separate the structural components of relatively round sources and elongated filaments from each other and from the background. The new method has many common features with its predecessors \textsl{getsources}, \textsl{getfilaments}, and \textsl{getimages} \citep{men2012multi, men2013getfilament, men2017getimages}. It has a single free parameter, the maximum size of the sources to be extracted. The detection provides a first-order estimate of the source footprints, sizes, and fluxes. As a second step, robust measurements of the sizes and fluxes of sources are done on background-subtracted images computed at each wavelength and, possibly, on other auxiliary images. The resulting catalog contains the size and fluxes of each source for each image.

\textsl{GExt2D} (Bontemps et al. in prep.), like the \textsl{CuTeX} algorithm \citep{molinari2017}, uses second derivatives to identify the local maxima of the spatial curvature, which are then interpreted as the central positions of compact sources. The outskirts of each source are then determined, at each wavelength independently, from the inflexion points that are observed as the emission decreases away from the source peak. For each wavelength, the background under each source is evaluated by interpolating the emission along the source outskirts. Then, for all identified compact sources, their sizes and fluxes are measured by fitting Gaussians to their positions in the emission maps from which the associated background has been subtracted.

Both algorithms allow multiple input images and separate the source detection step (see Sect.~\ref{sect:source detection}) from the step that characterize the sources in terms of size and flux measurements (see Sect.~\ref{sect:source characterization}).

\subsection{Source detection}\label{sect:source detection}
With the objective to build the most complete and most robust core catalog in the W43-MM2\&MM3 protocluster cloud, the core positions and footprints should be defined in the detection image that provides the optimum image sensitivity. This corresponds to the \bsens image at 1.3~mm (see Sect.~\ref{sect:obs and DR}). 
To further improve the sensitivity of the image chosen to detect cores, we removed the noise associated with cloud structures, which are incoherent from one scale to another. To do this we used the Multi-resolution non-Gaussian Segmentation software (\textsl{MnGSeg}) that separates the incoherent structures, referred to as Gaussian, of a cloud from the coherent structures associated with star formation \citep[][see also \cref{appendixsect:mngseg}]{robitaille2019}. The removed Gaussian component corresponds to structural noise associated with the small-scale structures of cirrus that lie along the line of sight to the W43-MM2 and W43-MM3 protoclusters. In detail, the \denoised image chosen for source extraction no longer contains incoherent components at scales larger than the beam size; it therefore consists of the sum of all the coherent cloud structures associated with star formation plus the white instrumental noise, which is a flux component needed to quantify the signal-to-noise ratio of extracted cores. 
We hereafter call \denoised \& \bsens and \denoised \& \cleanest the images passed through \textsl{MnGSeg} since their noise level decreases. As shown in \cref{appendixsect:mngseg}, images denoised by \textsl{MnGSeg} are indeed more sensitive and do not introduce spurious sources, meaning sources that are not part of the synthetic core population. In the case of the combined ALMA images of W43-MM2 and W43-MM3 the noise level decreased by about $\sim$30\% at both 1.3~mm and 3~mm wavelengths (see \cref{tab:observation table}), and thus allows the $5\,\sigma$ detection of point-like cores with masses of $\sim$0.20~$\Msol$ (see Eq.~\ref{eq:optically thin mass} and adopted assumptions). 

{\renewcommand{\arraystretch}{1.5}%
\begin{table*}[ht]
\centering
\begin{threeparttable}[c]
\caption{Number of sources extracted by \textsl{getsf} in the W43-MM2\&MM3 protocluster, using different detection images (all 12~m array 1.3~mm uncorrected by the primary beam) and various measurement images (all 12~m array 1.3~mm and 3~mm primary-beam-corrected).}
\label{tab:sensivity stat}
\begin{tabular}{l|c|cc|cc}
    \hline\hline
    Detection image & \cleanest & \multicolumn{2}{c|}{\bsens} & \multicolumn{2}{c}{\denoised \& \bsens}\\
    Measurement images  & \cleanest & \cleanest & \bsens  & \denoised \& \cleanest & \denoised \& \bsens\\
    \hline
    Number of sources,                           &     &      &      &      &       \\
    with robust 1.3~mm measurements\tnote{*}     & 75  & 100  & 120  & 158  & 208   \\
    with measurable 3~mm fluxes\tnote{$\dagger$} & 46  & 63   & 93   & 86   & 121   \\
    \hline
\end{tabular}
\begin{tablenotes}
\item[*] They are 1.3~mm sources that pass the recommended filtering of \textsl{getsf}: monochromatic goodness and significance above 1 in the detection image, small ellipticity, $a_{\rm 1.3mm}/b_{\rm 1.3mm}\leq 2$, and robust flux measurements at 1.3~mm, $S^{\rm peak}_{\rm 1.3mm} \geq 2 \sigma^{\rm peak}_{\rm 1.3mm}$, and $S^{\rm int}_{\rm 1.3mm} \geq 2 \sigma^{\rm int}_{\rm 1.3mm}$ in the measurement image. We also imposed a small average diameter, $\sqrt{a_{\rm 1.3mm} \times b_{\rm 1.3mm}}\leq 4\times \Theta_{\rm beam}$.
\item[$\dagger$] The 3~mm fluxes of sources robustly detected at 1.3~mm are considered measurable when they correspond to small and low-ellipticity sources, $\sqrt{a_{\rm 3mm} \times b_{\rm 3mm}}\leq 4\times \Theta_{\rm beam}$ and $a_{\rm 3mm}/b_{\rm 3mm}\leq 2$, detected above $1\,\sigma_{\rm 3mm}$, $S^{\rm peak}_{\rm 3mm} > \sigma^{\rm peak}_{\rm 3mm}$, and $S^{\rm int}_{\rm 3mm} > \sigma^{\rm int}_{\rm 3mm}$.
\end{tablenotes}
\end{threeparttable}
\end{table*}}

Hereafter the master source catalogs will be those from the extraction performed with \textsl{getsf} (v210414), using the listed input images for the following:
\begin{itemize}
    \item detection: 1.3~mm \denoised \& \bsens 12~m array image, not corrected by the primary beam;
    \item 1.3~mm measurements: \denoised \& \bsens and \denoised \& \cleanest 12~m array images, corrected by the primary beam;
    \item 3~mm measurements: \denoised \& \bsens and \denoised \& \cleanest 12~m array images, corrected by the primary beam;
\end{itemize}
To facilitate core extraction, the noise level of the detection image is flattened, using images that are uncorrected by the primary beam.
\cref{appendixtab:core detection table} lists the sources detected by \textsl{getsf} at 1.3~mm and identified by their peak coordinates, RA and Dec, along with their characteristics measured at 1.3~mm and at 3~mm in the \denoised \& \bsens images: non-deconvolved major and minor diameters at half maximum, $a_{\rm 1.3mm} \times b_{\rm 1.3mm}$ and $a_{\rm 3mm} \times b_{\rm 3mm}$; position angles, PA$_{\rm 1.3mm}$ and PA$_{\rm 3mm}$; peak and integrated fluxes, $S^{\rm peak}_{\rm 1.3mm}$, $S^{\rm peak}_{\rm 3mm}$, $S^{\rm int}_{\rm 1.3mm}$ and $S^{\rm int}_{\rm 3mm}$; two tags to identify cores also extracted by \textsl{GExt2D} and cores identified as suffering from line contamination (see Sect.~\ref{sect:line contamination}).
The \textsl{getsf} package extracted 208 cores that passed the basic recommended filtering\footnote{
    The monochromatic goodness and significance of \textsl{getsf} sources, defined in \cite{men2021getsf}, should be larger than 1. For robust flux measurements, \cite{men2021getsf} recommends $S^{\rm peak}\geq 2 \sigma^{\rm peak}$ and $S^{\rm int} \geq 2 \sigma^{\rm int}$. Lastly, sources that have high ellipticity are filtered imposing $a/b\geq2$. These internal parameters of \textsl{getsf} are used to assess the quality of the detection of a source and the measurements of its size and fluxes.} 
\citep{men2021getsf}. \cref{tab:sensivity stat} gives the number of sources extracted by \textsl{getsf} when using different detection and measurement images, from the \cleanest to the \bsens and finally \denoised \& \bsens images, at 1.3~mm and 3~mm.
The 208 sources of \cref{appendixtab:core detection table} are $\sim$1.6 times more numerous than the sources detected in the \original \& \bsens image and $\sim$2.8 times more numerous that those detected in the \original \& \cleanest image. 
In order to check the robustness of the \textsl{getsf} catalog of \cref{appendixtab:core detection table}, \textsl{GExt2D} (v210208) is used.
Applied to the \bsens 12~m array 1.3~mm image, not corrected by the primary beam, and after the recommended post-filtering\footnote{
    To guarantee a reliable catalog, it is recommended to only keep \textsl{GExt2D} sources, whose signal-to-noise ratio measured in an annulus around each source is greater than 4 (see Bontemps et al. in prep.). The flux quality 
    that quantifies the ratio of the second derivative isotropic part to its elliptical part, should also be higher than 1.85. It is used to exclude small flux variations along filaments. Lastly, sources that have high ellipticity are filtered imposing $a/b\geq1.5$.} 
(Bontemps et al. in prep.), \textsl{GExt2D} provides a catalog of 152 cores.

\subsection{Source characterization}\label{sect:source characterization}
The \textsl{getsf} and \textsl{GExt2D} measurements of source characteristics, that is to say their sizes and fluxes, were made in the 12~m array 1.3~mm and 3~mm images, which are primary-beam corrected. According to the good results of the \textsl{MnGSeg} denoising procedure applied on simulations of \textsl{getsf} extractions (see \cref{appendixsect:mngseg}), we kept the \textsl{getsf} measurements made in the \denoised images. 
Since we need to estimate, and later on correct, the line contamination of fluxes of the sources extracted in the \bsens image (see Sect.~\ref{sect:line contamination}), extraction was performed in the \denoised \& \cleanest images in addition to that performed in the \denoised \& \bsens images. 
Using the maximum size free parameter of \textsl{getsf}, we excluded five sources with FWHM larger than four times the beam, $\sqrt{a_{\rm 1.3mm}\times b_{\rm 1.3mm}} > 4\times\Theta_{\rm beam}$. 
They correspond to $\sim$10\,000~au at $d=5.5$~kpc, which thus would be much larger than the typical core size expected to be a few 1\,000~au in the dense W43 protoclusters \citep[e.g.,][]{bontemps2010, palau2013, motte2018b}. They have low 1.3~mm fluxes, with a median mass of $\sim$2~$\Msol$ (see Eq.~\ref{eq:optically thin mass}), and are located at the outskirts of the protocluster cloud. 

In summary, the \textsl{getsf} catalog of \cref{appendixtab:core detection table} contains 208 sources, which are detected at 1.3~mm with robust flux measurements. Given the lower sensitivity of our 3~mm continuum images, 121 have 3~mm fluxes that are qualified as ``measurable'' because they are above $1\,\sigma$ (see \cref{tab:sensivity stat}). 
Of the 208 \textsl{getsf} sources, 100 are qualified as ``robust'' because they are also identified by \textsl{GExt2D} and $\sim$90\% of these common sources have no significant differences in their integrated fluxes, that is, their fluxes are at worst a factor of two larger or smaller than each other. 
The sources that have 1.3~mm fluxes consistent to within 30\% are considered even more robust, as indicated in \cref{appendixtab:core detection table}.

\begin{figure}[ht]
    \centering
    \includegraphics[width=1.\linewidth]{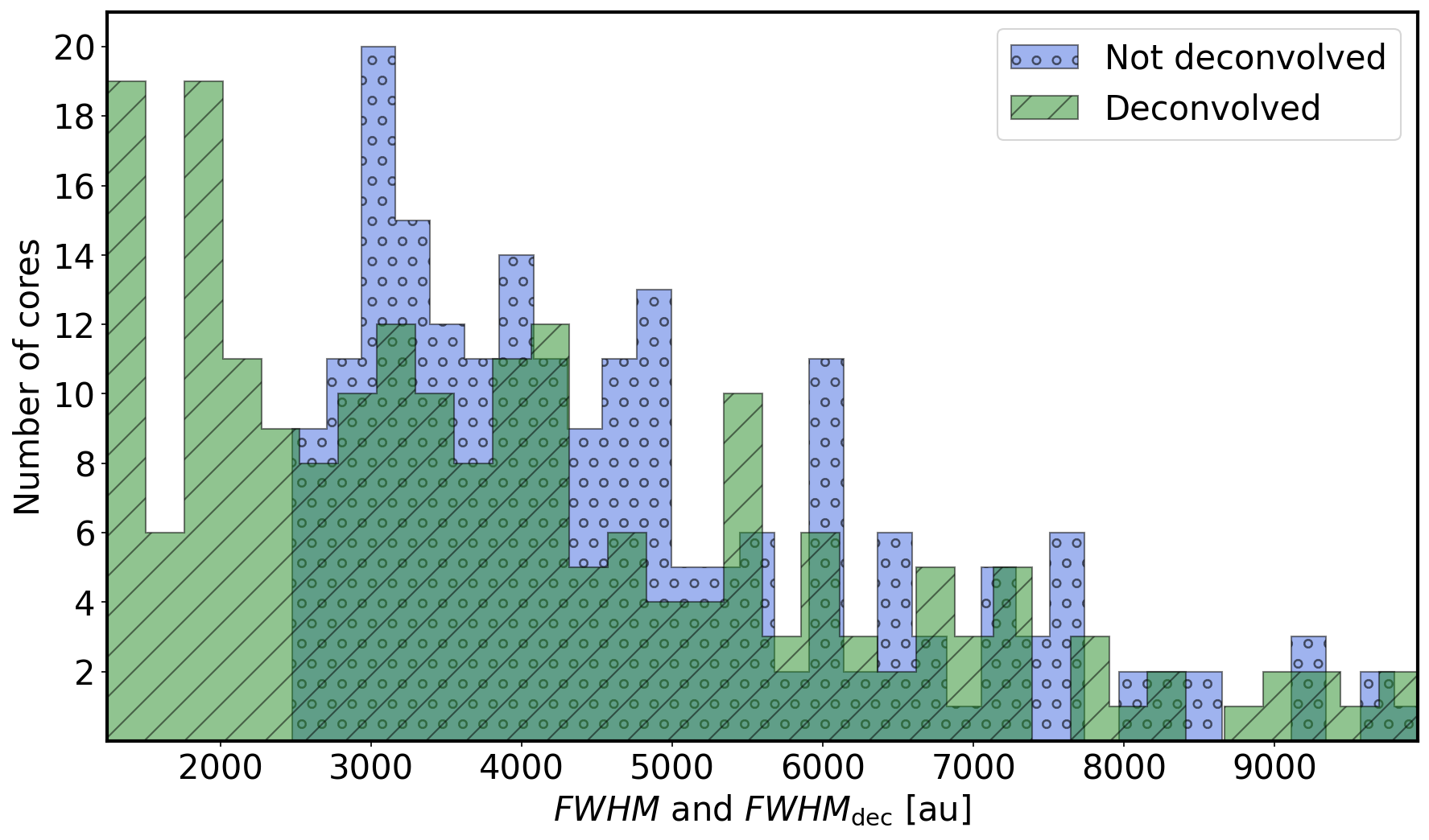}
    \caption{Distribution of the FWHM and FWHM$^{\rm dec}$ of the \textsl{getsf} sources as measured at 1.3~mm. A minimum size of $1\,300$~au is assumed for FWHM$^{\rm dec}$. The median value of the core deconvolved sizes is about $0.75\arcsec\simeq 1.6\times\Theta_{\rm beam}$ with $\Theta_{\rm beam}=0.46\arcsec$, corresponding to $\sim$3\,400~au.}
    \label{fig:fwhm distribution}
\end{figure}

Figure~\ref{fig:fwhm distribution} displays, for the 208 sources extracted by \textsl{getsf}, histograms of their 1.3~mm physical sizes before and after beam deconvolution\footnote{
    We set a minimum deconvolved size of half the beam, $0.23\arcsec$ or 1\,300~au, to limit deconvolution effects that may give excessively small, and thus unrealistic, sizes.}, 
FWHM$=\sqrt{a_{\rm 1.3mm}\times b_{\rm 1.3mm}}\times d$ and FWHM$^{\rm dec}=\sqrt{a_{\rm 1.3mm}\times b_{\rm 1.3mm}-\Theta_{\rm beam}^2} \times d$, projected at the $d=5.5$~kpc distance of W43. The W43-MM2\&MM3 compact sources have deconvolved sizes ranging from $\sim$1\,300~au to $\sim$10\,000~au with a median value of $\sim$3\,400~au. Given their small physical sizes, these cloud fragments could represent the mass reservoirs, or at least the inner part of those reservoirs, that will undergo gravitational collapse to form a star or a small mutiple system. Following the classical terminology \citep[e.g.,][]{motte2018a} and if they are real cloud fragments (see Sect.~\ref{sect:nature of compact sources}), we hereafter call them cores.

\section{Core nature and core mass estimates} \label{sect:core nature mass estim}

Sources in the W43-MM2\&MM3 protocluster are generally characterized from their measurements in the 1.3~mm \denoised \& \bsens images obtained with the ALMA 12~m array (\cref{appendixtab:core detection table}). Some of them, however, may not correspond to real cores or may have 1.3~mm \denoised \& \bsens fluxes contaminated by line emission; their nature is investigated in Sect.~\ref{sect:nature of compact sources}. When the W43-MM2\&MM3 core sample is cleaned and the 1.3~mm fluxes are corrected, core masses are estimated (see Sect.~\ref{sect:mass estimation}).

\subsection{Core sample of the W43-MM2\&MM3 ridge}
\label{sect:nature of compact sources}

To ensure that the millimeter sources of \cref{appendixtab:core detection table} are indeed dense cloud fragments and to correctly measure their mass, we investigated the contamination of their 1.3~mm and 3~mm continuum fluxes by free-free (see Sect.~\ref{sect:freefree contamination}) and line emission (see Sect.~\ref{sect:line contamination}).
From the 208 sources of \cref{appendixtab:core detection table}, we removed three sources which correspond to structures dominated by free-free emission and corrected the 1.3~mm measurements of 14 cores contaminated by line emission.

\subsubsection{Correction for free-free contamination} \label{sect:freefree contamination}

\begin{figure*}[htbp!]
    \centering
    \begin{minipage}{0.39\textwidth}
      \centering
      \includegraphics[width=1.\textwidth]{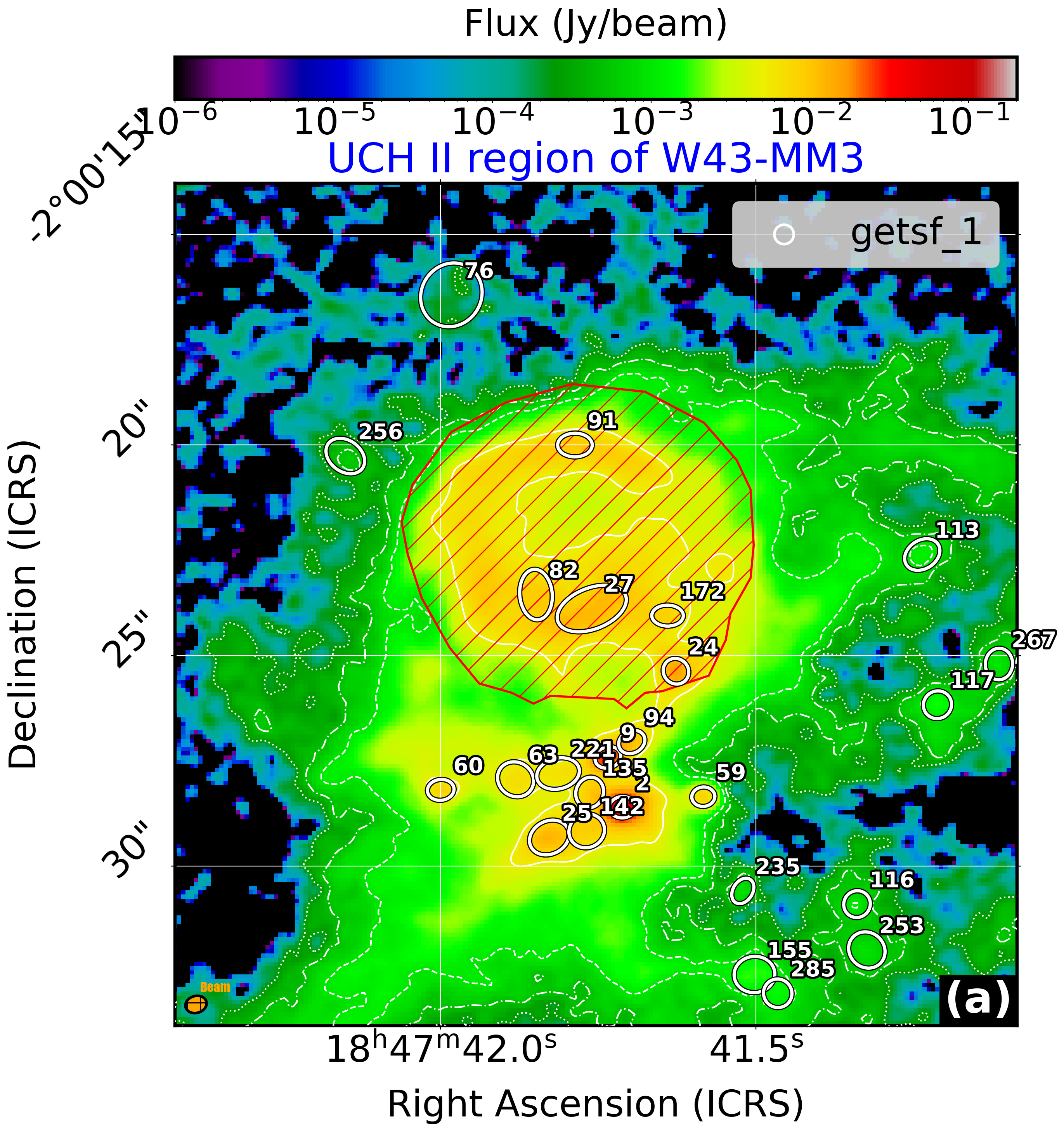}
    \end{minipage}%
    \hskip 0.0198\textwidth
    \begin{minipage}{0.59\textwidth}
      \centering
      \includegraphics[width=1.\textwidth]{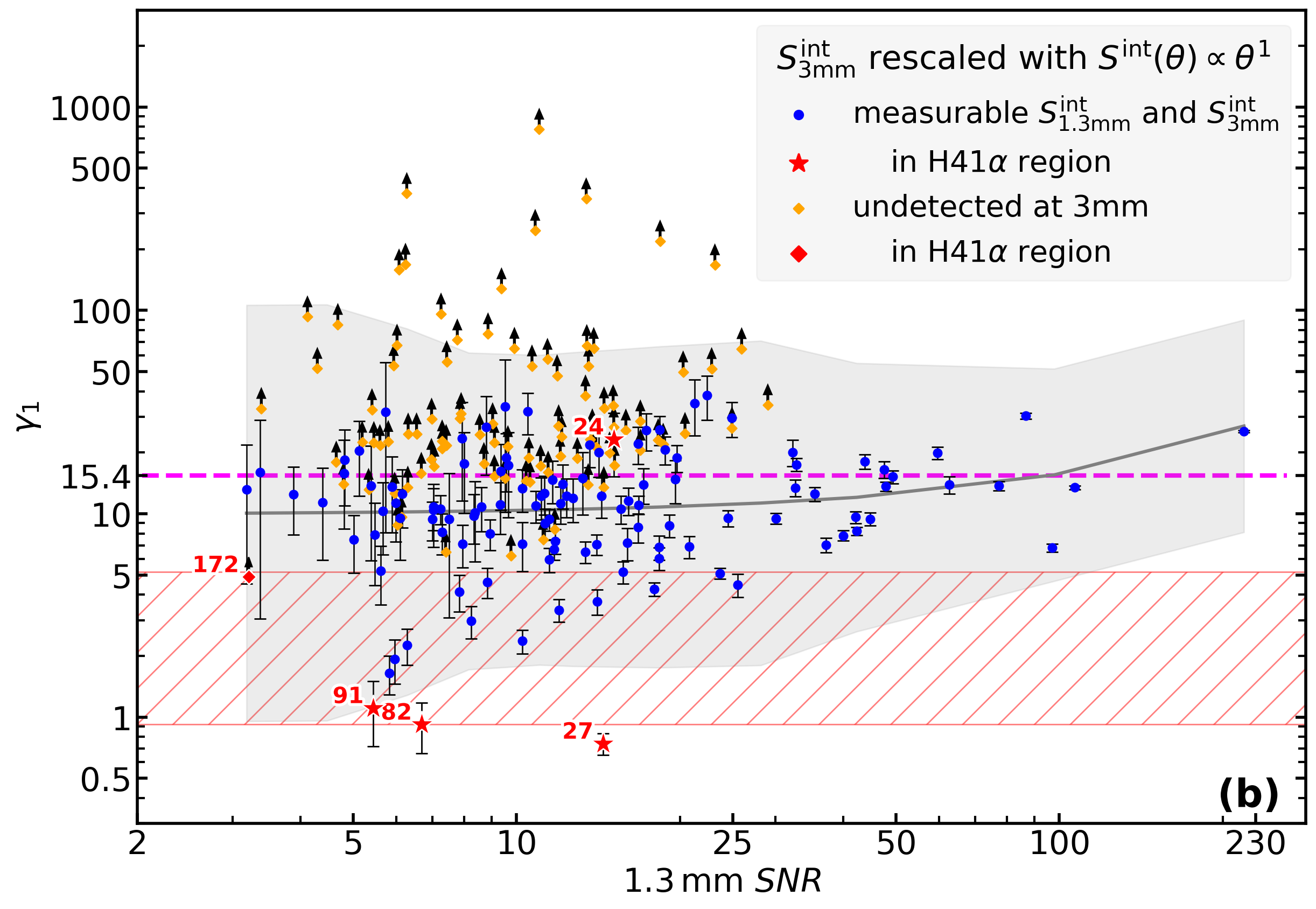}
    \end{minipage}
    \caption{Investigating free-free contaminated sources. Panel \textsl{(a)}: UC\hii region of W43-MM3 and its surrounding cloud imaged by ALMA at 1.3~mm. The red hatched region outlines the H41$\alpha$ recombination line emission of the \hii region. White ellipses outline source boundaries (at half maximum) as defined by \textsl{getsf}. Panel \textsl{(b)}: Thermal dust emission cores separated from free-free emission sources, using their 1.3~mm to 3~mm flux ratios, $\gamma_1$, and shown as a function of the S/N in the 1.3~mm image. Blue points indicate cores with 3~mm thermal dust emission whose flux is rescaled to the source size measured at 1.3~mm (see Eq.~\ref{eq:re-scale}), while orange points locate cores undetected at 3~mm, thus taking the ratio of the 1.3~mm peak flux to the $1\,\sigma$ peak error at 3~mm, corresponding to a lower limit. Red symbols are sources located within the H41$\alpha$ recombination line region displayed in panel \textsl{(a)}. The gray curve indicates the median value of the core ratios, computed over bins of 20 adjacent cores as ranked by their S/N. The shaded gray area indicates the corresponding $3\,\sigma$ dispersion in flux ratio values. The magenta horizontal dashed line represents the theoretical flux ratio of thermal dust emission of 15.4, computed in Eq.~\ref{eq:theo thermal ratio}. The red hatched area locates the theoretical flux ratios of UC\hii or HC\hii regions, whose free-free emission is either optically thin (lower limit) or partly to totally optically thick (upper limit).}
    \label{fig:freefree}
\end{figure*}

Figure~\ref{fig:1.3mm and trichrone}b shows that there is only one localized area associated with free-free emission in the 1.3~mm ALMA-IMF images of W43-MM2 and W43-MM3.
This is the W43-MM3 UC\hii region which is particularly bright at 3~mm. Figure~\ref{fig:freefree}a displays the boundary of this \hii region, as defined by the H41$\alpha$ recombination line emission observed as part of the ALMA-IMF Large Program (Galv\'an-Madrid et al. in prep.). In this area the large-scale continuum emission mainly consists of free-free emission, and the thermal dust emission of cores could only represent a minor part of the total flux at small scales. This calls into question the nature of the five compact sources detected over the extent of the \hii bubble that may not be interpreted as dust cores: \#24, \#27, \#82, \#91, and \#172 (see \cref{fig:freefree}a).

We investigated the free-free contamination of the cores of \cref{appendixtab:core detection table} by measuring the ratio of their 1.3~mm to 3~mm integrated fluxes, $S^{\rm int}_{\rm 1.3mm}$ and $S^{\rm int}_{\rm 3mm}$. To allow a direct comparison of these fluxes, not always integrated over the same area and thus not defining the same parcel of the cloud, we rescaled the 3~mm integrated flux of cores to their deconvolved 1.3~mm sizes, FWHM$^{\rm dec}_{\rm 1.3mm}$.
We assumed a linear relation between the integrated flux and the angular scale, $S^{\rm int}(\Theta)\propto \Theta$, corresponding to the optically thin emission of an isothermal, $T(r)\simeq \rm constant$, protostellar envelope with a $\rho(r)\propto r^{-2}$ density distribution \citep[][]{motte2001, beuther2002}.This flux rescaling was applied in \textit{Herschel} studies that aimed to fit meaningful spectral energy distributions (\citealt{motte2010,nguyen2011a,tige2017}). As discussed in \cite{tige2017}, this correction factor would be larger for starless fragments that have a flatter density distribution, thus leading to a $S^{\rm int}(\Theta)\propto \Theta^m$ relation with $m>1$.
In the case of hyper-compact \hii regions (HCH{\sc ~ii}), potentially optically thick at their center, a larger correction factor would also be necessary.
The rescaled 3~mm fluxes are computed via the following equation:
\begin{equation}
    (S^{\rm int}_{\rm 3mm})^{\rm rescaled}_{m} = S^{\rm int}_{\rm 3mm}\times \left(\frac{{\rm FWHM}^{\rm dec}_{\rm 1.3mm}}{{\rm FWHM}^{\rm dec}_{\rm 3mm}}\right)^{m}.
    \label{eq:re-scale}
\end{equation}

Figure~\ref{fig:freefree}b displays, for the complete catalog of \cref{appendixtab:core detection table}, the ratios of the 1.3~mm to 3~mm fluxes with a rescaling using $m=1$. On average, 3~mm fluxes are corrected by $25\%$, with a maximum of $75\%$, for the cores that have measurable fluxes both at 1.3~mm and 3~mm. For the many cores that remain undetected or that have barely measured fluxes at 3~mm, $S^{\rm int}_{\rm 3mm}\leq \sigma$, we used the $1\,\sigma$ rms noise level to give a lower limit of their 1.3~mm to 3~mm flux ratio, $\frac{S^{\rm int}_{\rm 1.3mm}}{S^{\rm int}_{\rm 3mm}} \geq \frac{S^{\rm peak}_{\rm 1.3mm}}{\sigma_{\rm 3mm}^{\rm peak}}$. In addition, Figs.~\ref{appendixfig:freefree different rescaling}a--b display the same figure without rescaling ($m=0$) and for a rescaling better suited for starless cores ($m=2$). Figures~\ref{fig:freefree}b and \ref{appendixfig:freefree different rescaling}a--b allow a simple separation of sources dominated by thermal dust emission from those dominated by free-free emission. Under the optically thin assumption and arising from the same source area, the 1.3~mm to 3~mm theoretical flux ratio of thermal dust emission is given by
\begin{align}
    \gamma &= \frac{S_{\rm 1.3mm}^{\rm int}}{S^{\rm int}_{\rm 3mm}} \label{eq:gamma}\\
    &= \frac{\kappa_{\rm 1.3mm}}{\kappa_{\rm 3mm}} \frac{B_{\rm 1.3mm}(T_{\rm dust})}{B_{\rm 3mm}(T_{\rm dust})} = \frac{\kappa_{\rm 1.3mm}}{\kappa_{\rm 3mm}} \frac{\nu_{\rm 1.3mm}^3}{\nu_{\rm 3mm}^3} \frac{e^{h\,\nu_{\rm 3mm}/k_{\rm B}\,T_{\rm dust}}-1}{e^{h\,\nu_{\rm 1.3mm}/k_{\rm B}\,T_{\rm dust}}-1} 
    \simeq 15.4,
    \label{eq:theo thermal ratio}
\end{align}
where $k_{\rm B}$ and $h$ are the Boltzmann and Planck constants, and $B_{\rm 1.3mm}(T_{\rm dust})$ and $B_{\rm 3mm}(T_{\rm dust})$ are the Planck function for the mean dust temperature of cores, $T_{\rm dust}=23$~K, at $\nu_{\rm 1.3mm}=228.9$~GHz and $\nu_{\rm 3mm}=100.7$~GHz. These frequency values are taken from Paper~II \citep{ginsburg2021} assuming a spectral index of $\alpha (\nu)=3.5$, which corresponds to a dust opacity spectral index of $\beta=1.5$, suitable for optically thin dense gas at the core scale \citep[see][]{AWB1993, juvela2015}. 
Because the W43-MM2\&MM3 ridge is a dense cloud \citep{nguyen2013}, we adopted a dust opacity per unit (gas $+$ dust) mass adapted for cold cloud structures: $\kappa_{\rm 1.3mm}=0.01\,\rm cm\,g^{-1}$ \citep{OssenkopfHenning1994}. 
The dust opacity mass at 3~mm, $\kappa_{\rm 3mm}$, is computed assuming 
\begin{equation}\label{eq:kappa}
    \kappa_{\lambda} = 0.01 \times \left( \frac{\lambda}{\rm 1.3\,mm} \right)^{-\beta}   
    = 0.01 \times \left( \frac{\nu}{\rm 228.9\,GHz} \right)^{\beta}~\rm cm^2\,g ^{-1}
\end{equation}

with $\beta = 1.5$. For the cores that remain after post-filtering at both wavelengths, we computed their ratio of 1.3~mm flux to 3~mm flux, which is rescaled to the 1.3~mm size with an index of either $m=1$ or $m=2$ (see Eqs.~\ref{eq:re-scale}--\ref{eq:gamma}): $\gamma_1=\gamma^{\rm rescaled}_{m=1}$ and $\gamma_2=\gamma^{\rm rescaled}_{m=2}$, respectively.
They have a median 1.3~mm to 3~mm flux ratio and  associated standard deviation of $\widetilde{\gamma_1}\simeq 11.3\pm 1.8$ (see \cref{fig:freefree}b), which is close to the expected value of 15.4 (see Eq.~\ref{eq:theo thermal ratio}). 
Figure~\ref{fig:freefree}b shows that the 1.3~mm to 3~mm flux ratio tends to increase as the signal-to-noise ratio (S/N) increases, equivalent to the core flux increases. Rescaling the fluxes with an index of $m=2$, rather than $m=1$, removes this unexpected correlation and leads to a median flux ratio of $\widetilde{\gamma_2} \simeq 15.3\pm 2.0$ (see \cref{appendixfig:freefree different rescaling}a), which is closer to the theoretical value (see Eq.~\ref{eq:theo thermal ratio}). 
If confirmed, this result would argue in favor of the pre-stellar rather than protostellar nature of most of the cores extracted in the W43-MM2\&MM3 protoclusters. A companion paper by Nony et al. (in prep.) consistently shows that the protostellar to pre-stellar ratio of the W43-MM2\&MM3 core sample is about $\sim$25\%.

In contrast, the 1.3~mm to 3~mm flux ratio of free-free emission is expected to be much lower than the ratio of thermal dust continuum emission estimated in \cref{eq:theo thermal ratio}. With a spectral index of optically thin and optically thick free-free emission of $\alpha (\nu)=-0.1$ and $\alpha (\nu) \simeq2$ \citep[e.g.,][]{keto2008}, respectively, the theoretical 1.3~mm to 3~mm flux ratios for \hii regions lie within the $\simeq$0.9--5.2 range. As shown in \cref{fig:freefree}a, we found that

\begin{itemize}
    \item three sources have low ratios ($\gamma_1 \simeq 0.9$) and are located along the \hii ring within the free-free continuum bubble of W43-MM3. Sources \#27, \#82, and \#91 most likely correspond to free-free emission fluctuations in the UC\hii region.
    \item source \#24, which is located over the UC\hii region extent, has a high 1.3~mm to 3~mm flux ratio, $\gamma_1 \simeq 18$, and can thus be considered a true core that is dominated by dust emission and lies on the same line of sight as the UC\hii region (see Figs.~\ref{fig:freefree}a--b).
    \item we find 13 sources in \cref{fig:freefree}b that have an intermediate flux ratio, $\gamma_1 \simeq 1.2-5$, and may indicate that they consist of partially optically thick free-free emission. However, only one source (source \#172) lies within the W43-MM3 \hii bubble, and it has a lower-limit ratio of $\gamma_1 \geq 5$. Moreover, none of the sources with $\gamma_1 = 1.2-5$ ratios is associated with strong H41$\alpha$ recombination line emission, as expected for most HC\hii regions. We therefore considered them to be real cores.
\end{itemize}

To confirm this, we developed a methodology that better takes into account the uncertainties of our source extraction and flux measurement process.
For the 121 dust cores detected at 1.3~mm and that have measurable 3~mm fluxes, Figs.~\ref{fig:freefree}b and \ref{appendixfig:freefree different rescaling} locate the $3\,\sigma$ dispersion zone of the logarithm of their flux ratios. 
None of these sources with $\gamma_1= 1.2-5$ ratios lie outside this $3\,\sigma$ zone, suggesting that their flux measurements are too uncertain to securely qualify these sources as being free-free emission peaks. 

In summary, \cref{fig:freefree}b, Figs.~\ref{appendixfig:freefree different rescaling}a--b, and the same figures done for peak fluxes, identified only three sources that likely correspond to free-free emission peaks: \#27, \#82, and \#91.
\cref{appendixtab:core detection table} pinpoints these three sources; they are removed from the core sample of \cref{appendixtab:derived core table} and will not be considered further.

\subsubsection{Correction for line contamination} \label{sect:line contamination}

In order to correctly measure the mass of cores, it is necessary to correct their continuum flux for line contamination. The 1.3~mm and 3~mm \denoised \& \bsens images used to identify sources in Sect.~\ref{sect:extraction of compact sources} indeed provide estimates of their continuum emission, based on all channels of all spectral bands. Some of these bands, however, contain bright emission lines associated with dense gas (see, e.g., Table~3 of \citealt{motte2021}). 
In addition, line forests of complex organic molecules \citep[COMs; e.g.,][]{garrod2006} are expected in all spectral windows when observing hot cores and shocked regions \citep[e.g.,][]{molet2019, bonfand2019}. Investigating the contamination by lines of the \bsens continuum can be done by comparing \bsens fluxes to fluxes measured in the \cleanest images \cite[see][]{motte2018b}.

\begin{figure}[ht]
    \centering
    \includegraphics[width=1.\linewidth]{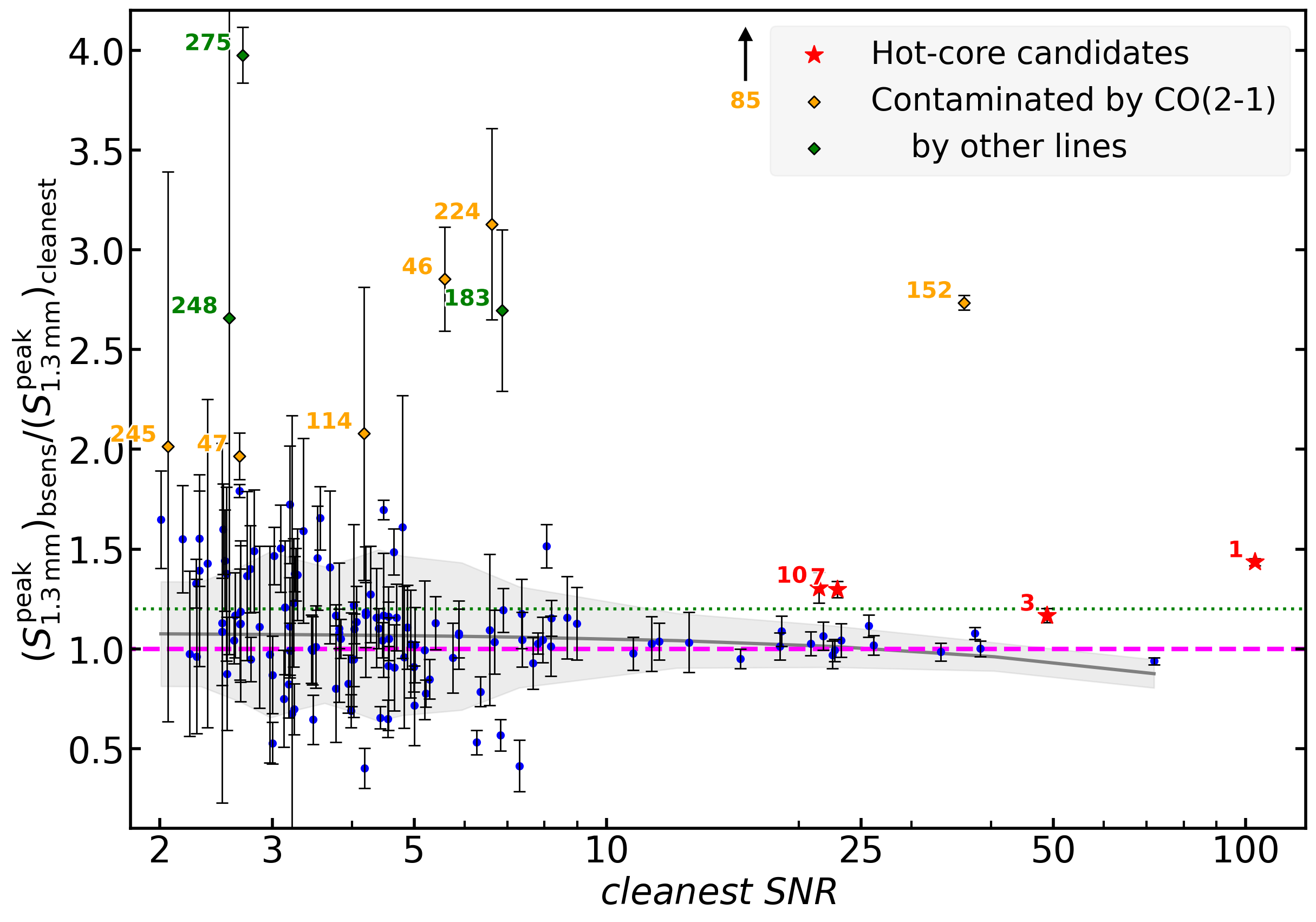}
    \caption{Line contamination of the 1.3~mm continuum fluxes of \textsl{getsf} sources, as estimated from the ratio of \denoised \& \bsens to \cleanest peak fluxes, and shown as a function of the S/N in the cleanest image. The gray curve indicates the median value of the core ratios, computed over bins of 20 adjacent cores as ranked by their S/N. The shaded gray area indicates the corresponding $3\,\sigma$ dispersion in flux ratio values. The red, orange, and green points locate cores with hot-core signatures (Herpin et al. in prep.), cores contaminated by the CO(2-1) line, and cores contaminated by other spectral lines, respectively. The horizontal lines indicate the contamination levels of 0\% (magenta dashed line) and 20\% (green dotted line). By taking only the blue points, the \denoised \& \bsens over \cleanest ratios of \cref{fig:hotcore} have a median value of $\simeq 1.1\pm 0.3$.
    }
    \label{fig:hotcore}
\end{figure}

Figure~\ref{fig:hotcore} presents, for the 155 sources at 1.3~mm with robust \denoised \& \bsens and \cleanest fluxes, the ratios of their \denoised \& \bsens to their \cleanest 1.3~mm peak fluxes. We use the peak rather than the integrated flux because the vast majority of hot cores are expected to be unresolved, and therefore have a higher ratio of \denoised \& \bsens to \cleanest peak fluxes. 
Most sources have ratios that remain close to $1$, with a decrease in the point dispersion as the S/N of the \denoised \& \cleanest fluxes increases (see \cref{fig:hotcore}). As in \cref{fig:freefree}b, we computed the $3\,\sigma$ dispersion zone of the plotted flux ratios and found that
\begin{itemize}
    \item four of the brightest sources (S/N $>20$) that lie above this $3\,\sigma$ zone have been identified as candidates to host a hot core by Herpin et al. (in prep.), namely cores \#1, \#3, \#7, and \#10. The line contamination of their 1.3~mm \denoised \& \bsens peak flux is estimated to range from 20\% to 45\% (see \cref{fig:hotcore}).
    \item ten other sources lie well above the $3\,\sigma$ dispersion zone with high flux ratios, $\frac{ (S_{\rm 1.3mm}^{\rm peak})_{\rm bsens}} { (S_{\rm 1.3mm}^{\rm peak})_{\rm cleanest}} = 2-12$, seven of which (\#46, \#47, \#85, \#114, \#152, \#224, and \#245) correspond to sources contaminated by the $^{12}$CO(2-1) line, which present an excess of flux in the continuum emission of the \denoised \& \bsens image. The three remaining sources (\#183, \#248, and \#275) are most probably contaminated by other lines, undetermined at this stage.
\end{itemize}
As indicated in \cref{appendixtab:derived core table}, the properties of these four and ten cores are derived from their measurements in the \denoised \& \cleanest image. Given that we could only investigate the line contamination of 155 out of 205 sources, we expect to have, in our core catalog of \cref{appendixtab:derived core table}, a maximum of four that have core masses overestimated sources in the $0.1-0.5$~$\Msol$ mass range.

In summary, from the 208 sources of \cref{appendixtab:core detection table}, we removed three sources, that correspond to structures dominated by free-free emission (see contamination tag). 
For the 14 cores contaminated by line emission (see \cref{appendixtab:core detection table}), we corrected their 1.3~mm measurements, including size and fluxes, by taking their \denoised \& \cleanest measurements.

\subsection{Mass estimates}
\label{sect:mass estimation}

We estimate the masses of cores, which are extracted by \textsl{getsf} in Sect.~\ref{sect:extraction of compact sources} and listed in Table~\ref{appendixtab:derived core table}. Because the thermal dust emission of cores is mostly optically thin at 1.3~mm, one generally uses the classical optically thin equation is genrally used to compute their masses. We give it here and provide a numerical application whose dependence on each physical variable is given, for simplicity, in the Rayleigh-Jeans approximation:
\begin{equation}
    \begin{split}
    M_{\rm \tau\ll 1} \: & = \frac{S^{\rm int}_{\rm 1.3\,mm}\; d^2}{ \kappa_{\rm 1.3\,mm}\; B_{\rm 1.3\,mm}(T_{\rm dust})} \\
    &\simeq \: 5\,\Msol \times \left(\frac {S^{\rm int}_{\rm 1.3\,mm}}{\mbox{10~mJy}}\right) \left(\frac {T_{\rm dust}}{\rm 23~K}\right)^{-1} \\
    & ~~~~ \times \left(\frac {d}{\mbox{5.5~kpc}} \right)^2 \left(\frac {\kappa_{\rm 1.3\,mm}}{\rm 0.01\,cm^2\,g^{-1}}\right)^{-1}.
    \end{split}
    \label{eq:optically thin mass}
\end{equation}

We estimated the volume-averaged core temperatures, $T_{\rm dust}$, from a map that combines a moderate angular  resolution dust temperature image with the central heating and self-shielding of protostellar and pre-stellar cores, respectively (see \cref{appendixfig:dust temperature map} and Motte et al. in prep.). The dust temperature image is produced by the Bayesian fit of spectral energy distributions, performed by the \textsl{PPMAP} procedure \citep{marsh2015}. Using the five \textit{Herschel} $70-500~\mu$m images, two APEX 350 and 870~$\mu$m images, and the present ALMA 1.3~mm image, which have a large range of angular resolutions ($0.46\arcsec-36\arcsec$), provides a $2.5\arcsec$-resolution dust temperature image that needs to be extrapolated to the $0.46\arcsec$ resolution of our 1.3~mm ALMA-IMF image. The dust temperature of the immediate background of cores listed in \cref{appendixtab:derived core table} has a mean value of $\overline{T_{\rm dust}}^{\rm core\,bkg}= 24\pm 2$~K. 
Following \cite{motte2018b}, the dust temperature of massive protostellar cores averaged in $0.46\arcsec$-resolution elements is estimated from the total luminosity of the W43-MM2 cloud \citep[$\sim$$2\times 10^4~L_\odot$,][]{motte2003} divided between cores, in proportion to their associated line contamination in the 1.3~mm band (see Motte et al. in prep.). This leads to volume-averaged temperatures, $T_{\rm dust}$, between 20~K and 65~K. In addition, the mean core temperature of lower-mass cores driving outflows (see Nony et al. in prep.) is increased by $4\pm4$~K compared to the core background temperature. The temperature of candidate pre-stellar cores is itself decreased by $2\pm2$~K compared to their background temperature. The resulting estimates of the mass-averaged temperature of cores range from 19~K to 65~K, with uncertainties ranging from $\pm 2$~K to $\pm10$~K (see \cref{appendixtab:derived core table}).

For the cores that reach sufficiently high densities ($\gtrsim5\times 10^7$~cm$^{-3}$, see Eq.~\ref{eq:density}), in other words the most massive ones, we expect them to be optically thick \citep[e.g.,][]{cyganowski2017, motte2018a}. To partly correct for this opacity, \cite{motte2018a} proposed an equation, which is given below and fully explained in \cref{appendixsect:detailed approach for the mass calculation}:
\begin{equation} \label{eq:optically thick mass}
M_{\rm \tau\gtrsim 1} \: = -\, \frac{\Omega_{\rm beam} \;d^2} {\kappa_{1.3{\rm mm}}}\, \frac{S^{\rm int}_{1.3{\rm mm}}} {S^{\rm peak}_{1.3{\rm mm}}} \,
	\ln\left(1\,-\,\frac{S^{\rm peak}_{1.3{\rm mm}}}{\Omega_{\rm beam}\;B_{1.3{\rm mm}}(T_{\rm dust})}\right). 
\end{equation}
Here $\Omega_{\rm beam}$ is the solid angle of the beam. This correction is significant for two cores (cores \#1 and \#2), whose masses estimated with the optically thin assumption would have been underestimated by $\sim$15\%. With this correction of optical thickness and the temperatures estimated in \cref{appendixfig:dust temperature map}, the core mass range is $0.1-70~\Msol$ (see \cref{tab:cmf and cores}). 
To start estimating which of these cores are gravitationally bound, we compared the measured masses with virial masses.
The core virial masses were calculated from their FWHM sizes measured at 1.3~mm and their estimated temperatures, $T_{\rm dust}$, given in  \cref{appendixtab:derived core table}. All the W43-MM2\&MM3 cores could be gravitationally bound because their virial parameter, $\alpha_{\rm vir}=M_{\rm vir}/M_{\rm \tau\gtrsim 1}$, is always smaller than the factor 2 chosen by \cite{bertoldi1992} to define self-gravitating objects. 
Their dynamical state, however, requires further study of the non-thermal motions of the cores, which will be measured in part by future ALMA-IMF studies of spectral lines.

We estimated the absolute values of the core masses to be uncertain by a factor of a few, and the relative values between cores to be uncertain by $\sim$50\%. Dust opacity should indeed evolve as the core grows and the protostar heats up \citep{OssenkopfHenning1994} and may also have a radial dependence from the core surroundings to its center. We therefore assumed a $1\,\sigma$ uncertainty for the dust opacity that should cover its variations with gas density and temperature; divided or multiplied by a factor of 1.5 it becomes $\kappa_{\rm 1.3mm}=0.01\pm^{0.005}_{0.0033} \,\rm cm\,g^{-1}$.

\cref{appendixtab:derived core table} lists the physical properties of the 205 cores derived from their 1.3~mm \denoised \& \bsens measurements and the analysis made in Sect.~\ref{sect:core nature mass estim}:  deconvolved size, FWHM$^{\rm dec}$; mass corrected for optical depth, $M_{\rm \tau\gtrsim 1}$; dust temperature, $T_{\rm dust}$; volume density, $n_{\rm H_2}$. 
Volume densities are computed assuming a spherical core: 
\begin{equation}
    \begin{split}
    n_{\rm H_2} &= \frac{M_{\rm \tau\gtrsim 1}}{\frac{4}{3}\pi\,\mu\,m_{\rm H}\,\left({\rm FWHM}^{\rm dec}_{\rm 1.3mm}\right)^3} \\
     &\simeq 7.8\times 10^7\,{\rm cm}^{-3} \times \left( \frac{M_{\rm \tau\gtrsim 1}}{\mbox{70~\Msol}}\right) \left( \frac{{\rm FWHM}^{\rm dec}_{\rm 1.3mm}}{\mbox{3\,000~au}}\right)^{-3}.
    \end{split}
    \label{eq:density}
\end{equation}

\begin{figure*}[htbp!]
    \centering
    \begin{minipage}{0.48\textwidth}
      \centering
      \includegraphics[width=1.\textwidth]{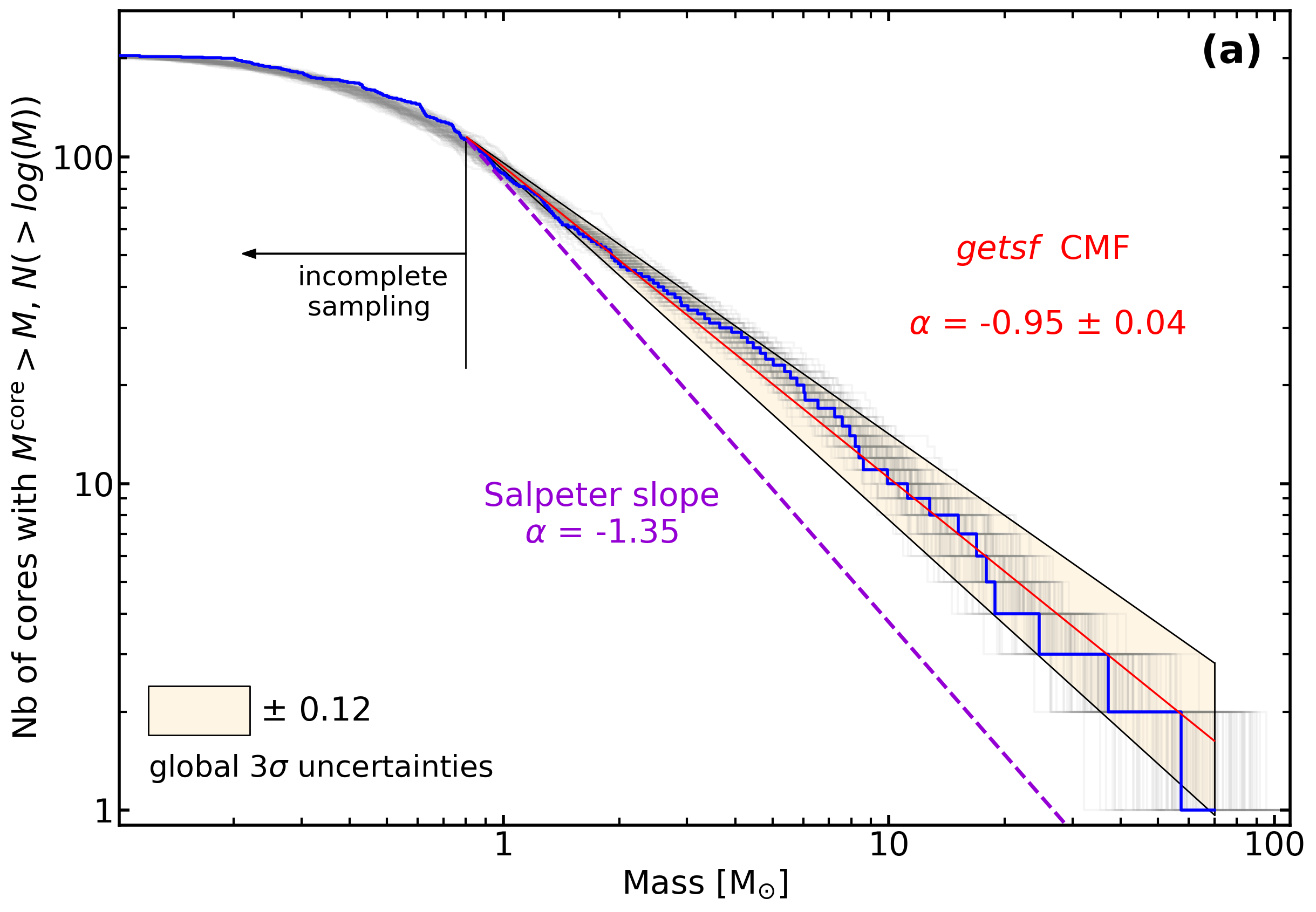}
    \end{minipage}%
    \hskip 0.0199\textwidth
    \begin{minipage}{0.48\textwidth}
      \centering
      \includegraphics[width=1.\textwidth]{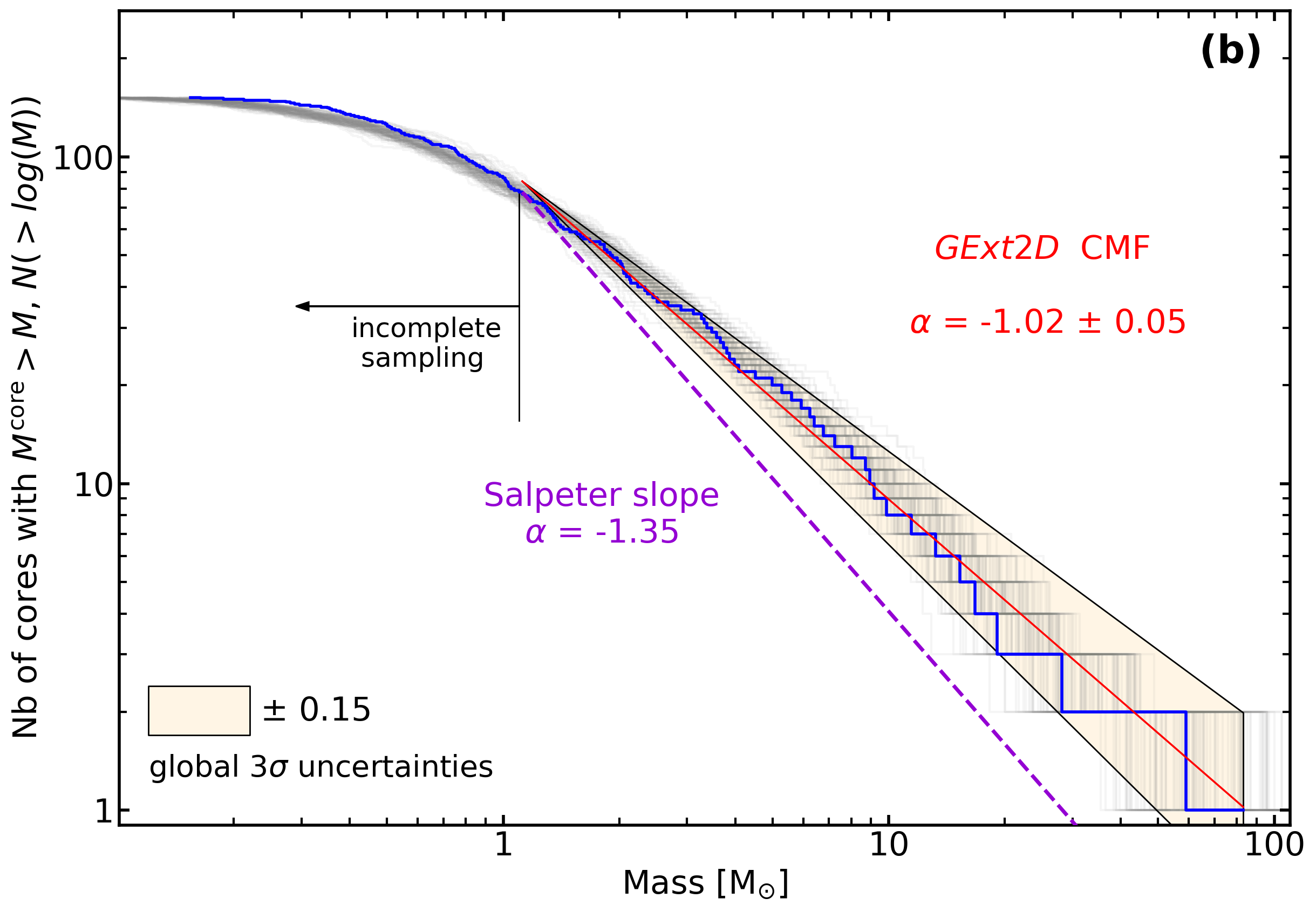}
    \end{minipage}
    \caption{Top-heavy CMF of the W43-MM2\&MM3 ridge, with cores extracted by the \textsl{getsf} (panel \textsl{a}) and \textsl{GExt2D} (panel \textsl{b}) software packages, in the \denoised \& \bsens and \original \& \bsens images, respectively. The cumulative forms of CMFs (blue histograms) are fitted above their 90\% completeness levels (black vertical lines) by single power-laws of the form $N(>\log M)\propto M^{\alpha}$, with $\alpha = -0.95 \pm 0.04$ (\textsl{a}) and $\alpha = -1.02 \pm 0.05$ (\textsl{b}) (red lines and $1\,\sigma$ global uncertainties). The global $3\,\sigma$ uncertainties are computed from 2\,000 CMFs that are uniformly randomly generated (light gray histograms) and from the fit uncertainty (see Sect.~\ref{sect:top heavy cmf}). The W43-MM2\&MM3 CMF slope is clearly shallower than the high-mass end of the canonical IMF, which has a power-law index of $\alpha = -1.35$ (\citealt{salpeter1955}, dashed magenta lines).}
    \label{fig:cmfs software}
\end{figure*}

\begin{figure*}[htbp!]
    \centering
    \begin{minipage}{0.49\textwidth}
      \centering
      \includegraphics[width=1.\textwidth]{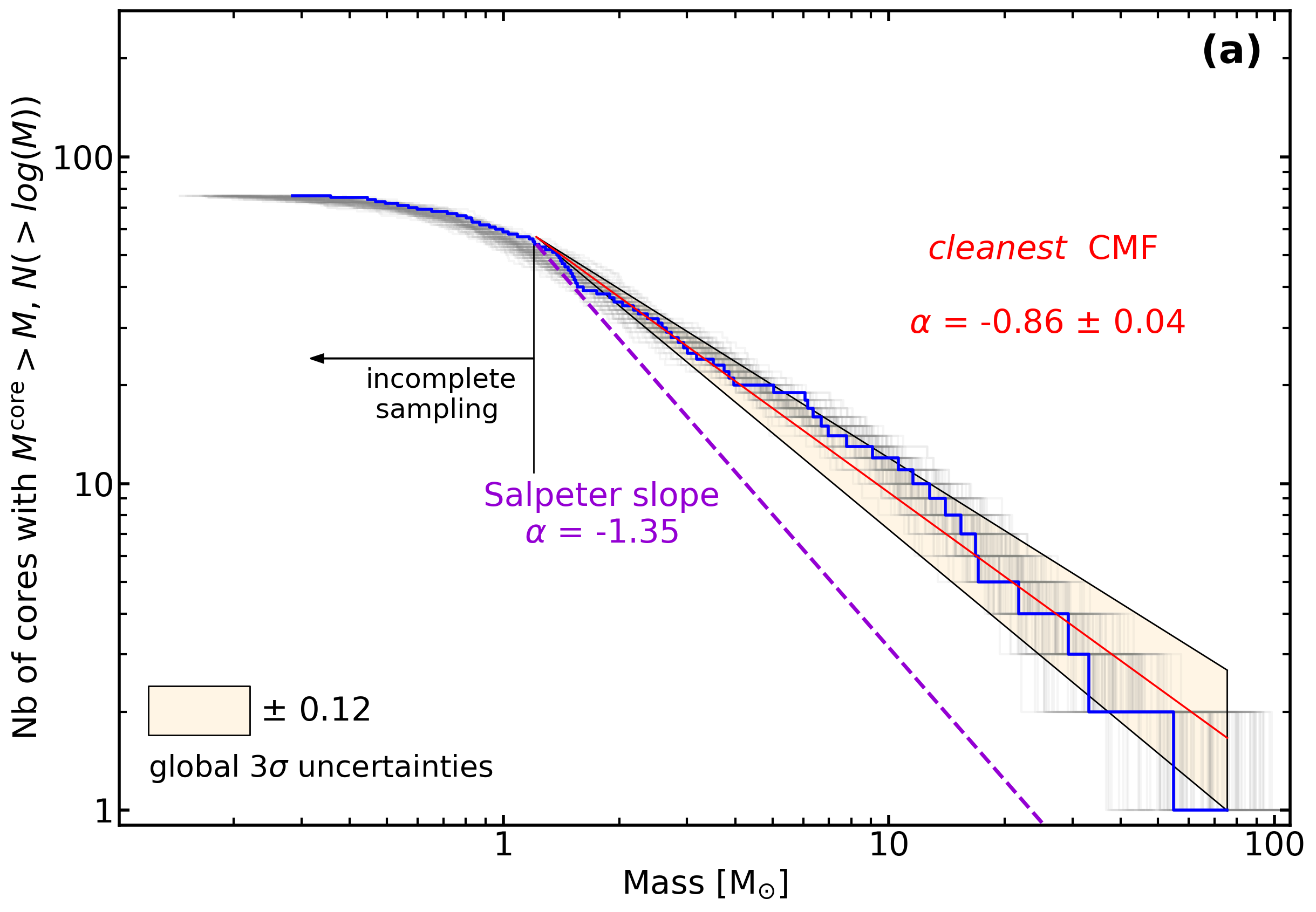}
    \end{minipage}%
    \begin{minipage}{0.49\textwidth}
      \centering
      \includegraphics[width=1.\textwidth]{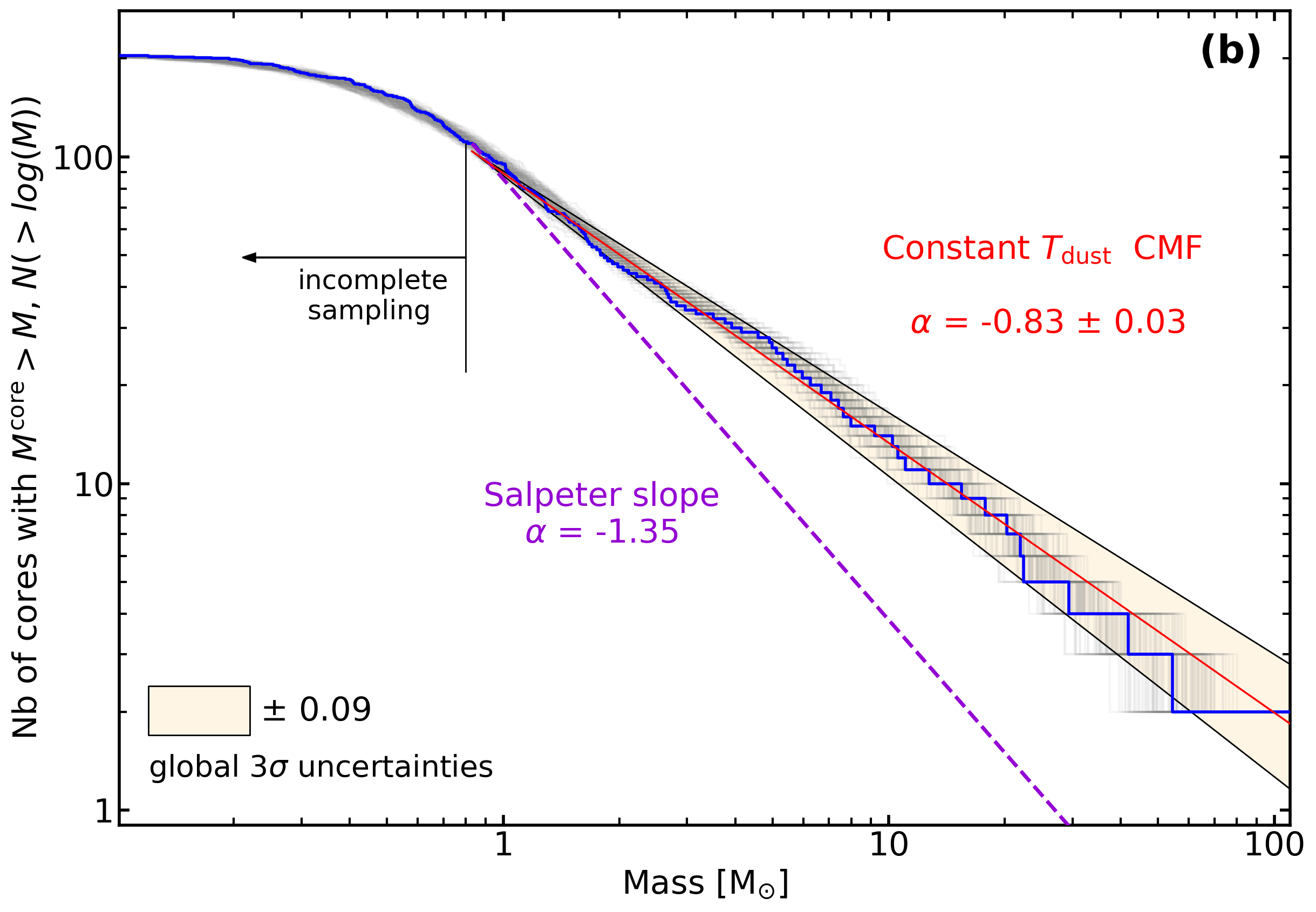}
    \end{minipage}
    \begin{minipage}{0.49\textwidth}
      \centering
      \includegraphics[width=1.\textwidth]{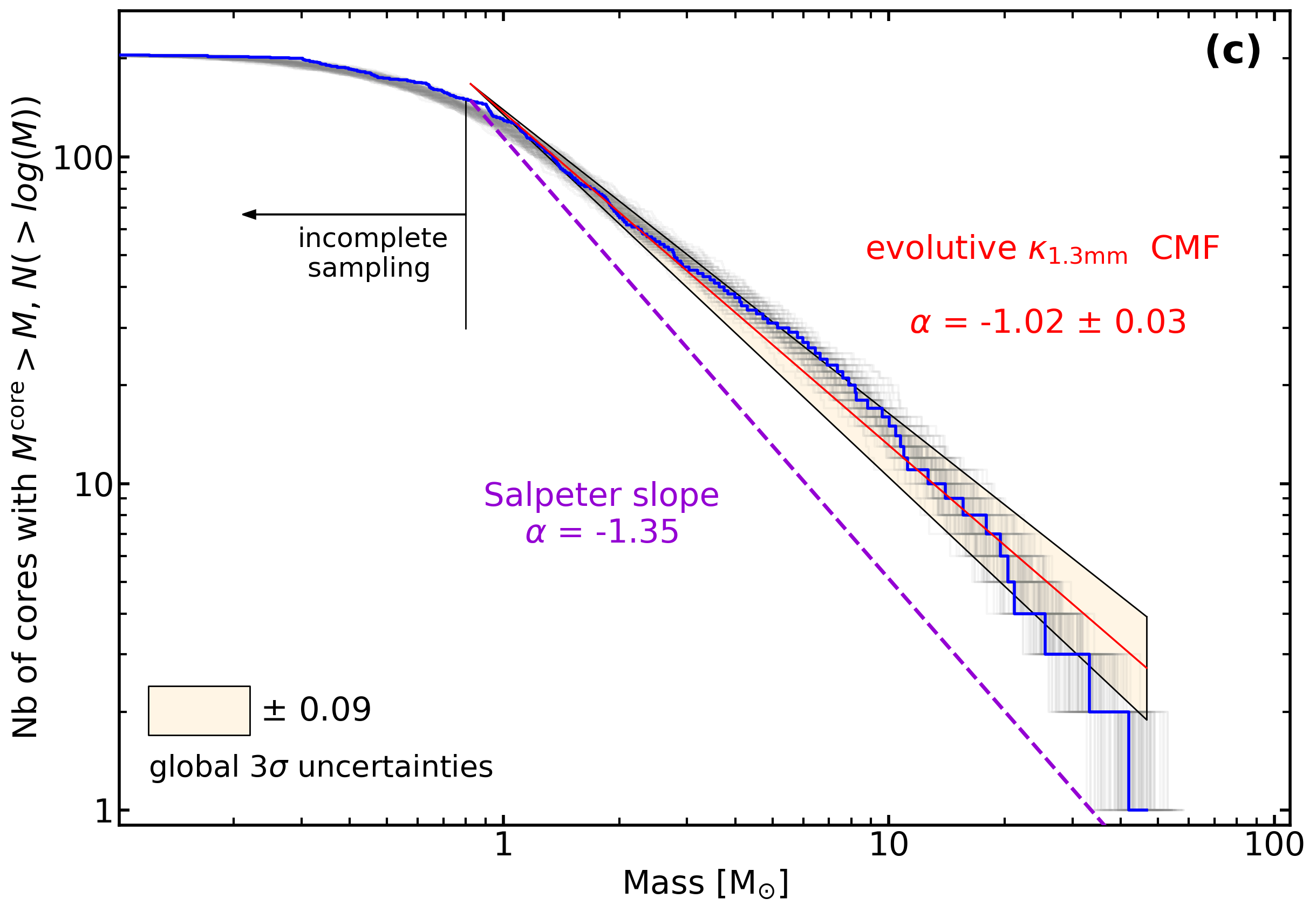}
    \end{minipage}%
    \begin{minipage}{0.49\textwidth}
      \centering
      \includegraphics[width=1.\textwidth]{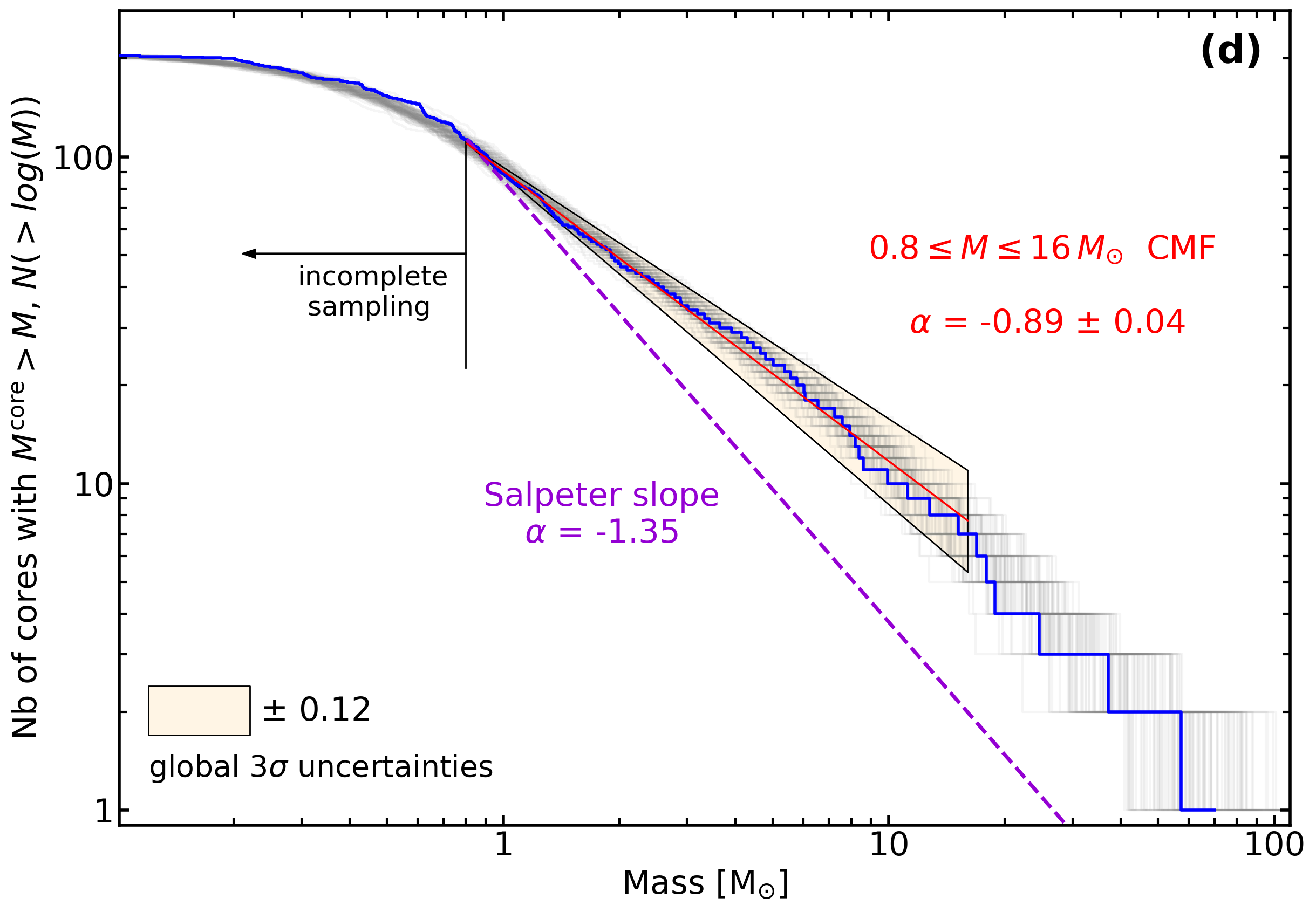}
    \end{minipage}
    \caption{\textsl{getsf} CMFs of the W43-MM2\&MM3 ridge built for a different core catalog (\textsl{a}), under different assumptions of dust temperature and emissivity (\textsl{b} and \textit{c}), and fit over a different mass range (\textsl{d}). The cumulative CMFs, their completeness levels, power-law fits, global $3\,\sigma$ uncertainties (explained in Sect.~\ref{sect:top heavy cmf}), and the Salpeter slope of the canonical IMF are represented as in \cref{fig:cmfs software}. Panel \textsl{(a)}: CMF derived from the core catalog of Paper~V (Louvet et al. in prep.), itself obtained by \textsl{getsf} extraction in the \original \& \cleanest images of W43-MM2 and W43-MM3, showing a similar but slightly shallower slope of $\alpha = 0.86\pm0.04$. Panel \textsl{(b)}: CMF obtained with a mean $T_{\rm dust}=23$~K dust temperature for all cores, instead of T$_{\rm dust}$ in \cref{appendixfig:dust temperature map}, displaying a similar but slightly shallower slope of $\alpha=-0.83\pm0.03$. Panel \textsl{(c)}: CMF derived assuming a linear relation for the dust opacity with core mass (see Sect.~\ref{sect:robustness against our assumptions}) showing a steeper slope of $\alpha=-1.02\pm0.03$. Panel \textsl{(d)}: Fitting the CMF of \cref{fig:cmfs software}a in the low- to intermediate-mass range,$0.8-16~\Msol$. This leads to a similar but slightly shallower slope of $\alpha=-0.89\pm0.04$.}
    \label{fig:cmf tests}
\end{figure*}

{\renewcommand{\arraystretch}{1.5}%
\begin{table}[ht]
\centering
\tiny
\begin{threeparttable}[c]
\caption{W43-MM2\&MM3 core populations and CMF parameters, as derived by two core extraction algorithms.}
\label{tab:cmf and cores}
\begin{tabular}{ccccc}
    \hline\hline
    Extraction & Number & 
    \multirow{2}{*}{$\sum M_{\rm \tau\lesssim 1}$} & 
    \multirow{2}{*}{Mass range} & 
    \multirow{2}{*}{$\alpha$} \\
    packages & of cores & & & \\
     & & [\Msol] & [\Msol] & \\
          (1)       & (2) &    (3)      &     (4)      &      (5)       \\ \hline
                    &     &             & $0.8-69.9$ & $-0.95\pm0.04$ \\
    \textsl{getsf}  & 205 & $541\pm29$  &   $0.8-16$ & $-0.89\pm0.04$ \\
                    &     &             &  $2.0-69.9$ & $-1.05\pm0.06$ \\ \hline
                    &     &             &  $1.1-83.1$ & $-1.02\pm0.05$ \\
    \textsl{GExt2D} & 152 & $468\pm35$  &    $1.1-16$ & $-0.98\pm0.06$ \\
                    &     &             &  $2.0-83.1$ & $-1.07\pm0.07$ \\ \hline
    \end{tabular}
\begin{tablenotes}[flushleft]
\item (3) Cumulative mass of cores, listed in \cref{appendixtab:derived core table}. Uncertainties arise from those associated with individual core mass estimates.
\item (4) Mass range used to fit a power-law to the cumulative form of the CMFs. The lower limit of this mass range is the 90\% completeness limit (see \cref{appendixsect:completeness simulation} and Sect.~\ref{sect:top heavy cmf}) of $2~\Msol$; its upper limit corresponds to the maximum core mass detected or $16~\Msol$.
\item (5) Power law index of the CMFs in their cumulative form, $N(>\log M)\propto M^{\alpha}$. Uncertainties are estimated by varying dust temperature and emissivity and by taking into account the fit uncertainty notably associated with a completeness limit uncertainty of $\pm0.2~\Msol$ (see Sect.~\ref{sect:robustness against our assumptions}).
\end{tablenotes}
\end{threeparttable}
\end{table}}

{\renewcommand{\arraystretch}{1.5}%
\begin{table*}[ht]
\centering
\begin{threeparttable}[c]
\caption{
CMFs and predicted IMFs of the W43-MM2\&MM3 protocluster: Uncertainty evaluation and predicted evolution.}
\label{tab:tests cmf}
\begin{tabular}{ll|ccc}
    \hline\hline
     & & Mass range & $\alpha$ & Associated figure \\ 
     & & [\Msol] & & \\ \hline
    \multicolumn{2}{l|}{\textbf{Reference CMF} (using \textsl{getsf} cores from the \denoised \& \bsens image)} & $0.8-69.9$ & $-0.95\pm0.04$ & \cref{fig:cmfs software}a \\
    CMF for & cores extracted in the \original \& \cleanest image                                        &  $1.2-75.6$ & $-0.86\pm0.04$ & \cref{fig:cmf tests}a      \\ 
            & masses computed with a constant $T_{\rm dust}$                                             &  $0.8-492$  & $-0.83\pm0.03$ & \cref{fig:cmf tests}b      \\
            & masses computed with a linear function of $\kappa_{\rm 1.3mm}$ with the mass               &  $0.8-46.6$ & $-1.02\pm0.03$ & \cref{fig:cmf tests}c      \\ 
    \hline 
    IMF for & a constant mass conversion efficiency, $\epsilon_{\rm core} = 50$\%                        &  $0.4-35.0$ & $-0.95\pm0.04$ & \cref{fig:fragmentation}a  \\
            & a linear function with the mass of $\epsilon_{\rm core} \propto M$                         & $0.44-69.9$ & $-0.59\pm0.04$ & \cref{fig:fragmentation}a  \\
            & a dependence on core density of $\epsilon_{\rm core} \propto (n_{\rm H_2})^{0.9}$          & $0.44-44.6$ & $-0.67\pm0.06$ & \cref{fig:fragmentation}a  \\
    IMF for & thermal Jeans fragmentation with $\epsilon_{\rm core} = 50$\%                              & $0.4-1.6$   & $-3.46\pm0.55$ & \cref{fig:fragmentation}b  \\
            & an analytical function of $N_{\rm frag} \propto M^{0.4}$ with $\epsilon_{\rm core} = 50$\% &  $0.3-3.9$  & $-1.42\pm0.10$ & \cref{fig:fragmentation}b  \\
            & a fractal hierarchical cascade with $\epsilon_{\rm core} = 50$\%                           & $0.27-23.3$ & $-1.00\pm0.04$ & \cref{fig:fragmentation}c  \\
            & a fractal hierarchical cascade with $\epsilon_{\rm core} \propto M$                        &$0.004-46.6$ & $-0.49\pm0.06$ & \cref{fig:fragmentation}c  \\
    \hline
    \end{tabular}
\begin{tablenotes}[para,flushleft]
Notes: Cumulative CMFs and predicted IMFs are fitted by power-laws of the form $N(>\log M) \propto M^{\alpha}$. Mass ranges of the CMF and IMF fits are limited by the estimated completeness level.
\end{tablenotes}
\end{threeparttable}
\end{table*}}

\section{CMF results} \label{sect:cmf results}

We use the core masses estimated in Sect.~\ref{sect:core nature mass estim} to build the CMF of the W43-MM2\&MM3 ridge in Sect.~\ref{sect:top heavy cmf} and discuss its robustness in Sect.~\ref{sect:robustness against our assumptions}. 
Tables~\ref{tab:cmf and cores} and \ref{tab:tests cmf} list the parameters of the W43-MM2\&MM3 CMFs derived from different catalogs and under different assumptions.

\subsection{Top-heavy CMF in the W43-MM2\&MM3 ridge} \label{sect:top heavy cmf}

Figure~\ref{fig:cmfs software} displays the W43-MM2\&MM3 CMFs as derived from the \textsl{getsf} and \textsl{GExt2D} samples of 205 and 152 cores, respectively. The 90\% completeness limits for \textsl{getsf} and \textsl{GExt2D} are estimated to be $0.8\pm0.2~\Msol$ and $1.1\pm 0.2~\Msol$, respectively (see \cref{appendixsect:completeness simulation}). 
Following the recommendations of \cite{maiz2005} and \cite{reid2006} for improving the measurement statistics, we chose to analyze the complementary cumulative distribution form (hereafter called cumulative form) rather than the differential form of these CMFs. The \textsl{getsf} and \textsl{GExt2D} CMFs are least-squares fitted above their completeness limits by single power-laws of the form $N(>\log M)\propto M^{\rm \alpha}$ with $\alpha=-0.95 \pm 0.04$ for \textsl{getsf} and $\alpha = -1.02\pm 0.05$ for \textsl{GExt2D} (see Figs.~\ref{fig:cmfs software}a--b). 

A slope uncertainty driven by uncertainties on the core masses, referred to below as mass-driven uncertainty, is computed from two thousand randomly generated CMFs, taking for each core a uniformly random mass in the range $[M_{\rm min}-M_{\rm max}]$. 
For each core, $M_{\rm max}$ and $M_{\rm min}$ are the maximum and minimum masses, respectively, computed from its measured flux, estimated temperature, and dust opacity, plus or minus the associated $1\,\sigma$ uncertainties (see Tables~\ref{appendixtab:core detection table}--\ref{appendixtab:derived core table}, and Sect.~\ref{sect:mass estimation}). The mass-driven uncertainties of the power-law indices range from $\sigma\simeq 0.03$ to $0.06$. 
In addition, we estimated a slope uncertainty due to the power-law fit, referred to as the fit uncertainty, from the $\chi^2$ uncertainty and by varying the initial point of the slope fit using the 90\% completeness level and its uncertainty (see \cref{tab:cmf and cores} and \cref{appendixfig:completeness}). The fit uncertainty of the power-law indices is about $\sigma\simeq 0.03$.
The global uncertainties of the power-law indices are finally taken to be the quadratic sum of the mass-driven uncertainties and the fit uncertainties (see Tables~\ref{tab:cmf and cores} and \ref{tab:tests cmf}).

When taking into account these global uncertainties, the CMF slopes measured in \cref{fig:cmfs software} are much shallower than the high-mass end, $>$1~$\Msol$, of the canonical IMF that is often represented by a power-law function close to $N(>\log M)\propto M^{\rm -1.35}$ \citep{salpeter1955, kroupa2001, chabrier2005}. 
Using the shape of the IMF as a reference, these CMFs are qualified as top-heavy. 
They are overpopulated by high-mass cores compared to intermediate-mass cores and are overpopulated by intermediate-mass cores compared to low-mass cores (see \cref{fig:cmfs software}). 

\begin{figure}[ht]
    \centering
    \includegraphics[width=1.\linewidth]{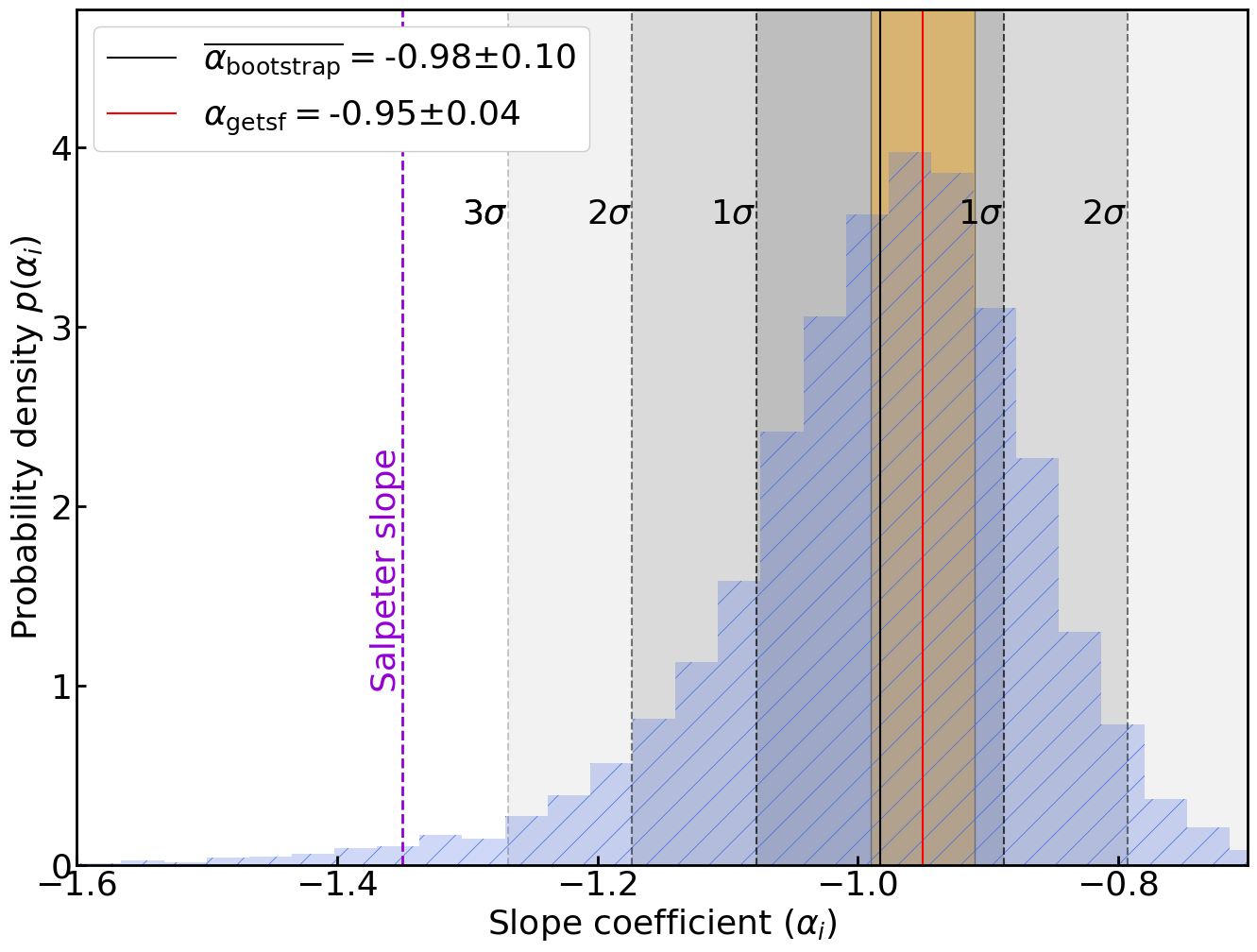}
    \caption{Bootstrapping probability density histogram of the $N=10^4$ slopes fitted using the \cite{alstott2014} KS metric, measured for data sets generated using the metric parameters obtained for our sample of 205 cores. The black and red vertical lines indicate the resulting slope coefficient of $\alpha=-0.98\pm0.10$ and the \textsl{getsf} fitted slope of $\alpha=-0.95\pm0.04$ (orange area corresponding to 1$\sigma$, see Sect.~\ref{sect:top heavy cmf}). 1, 2, and 3$\sigma$ dispersions are estimated from the bootstrapping (shaded gray areas). The Salpeter slope (dashed vertical magenta line) is rejected with a probability of 99.98\%.}
    \label{fig:bootstrap powerlaw}
\end{figure}

We use statistical tests to compare the \textsl{getsf} CMF with either the \textsl{GExt2D} CMF or the Salpeter IMF.
A two-sample Kolmogorov-Smirnov (KS) test is used to assess the likelihood that two distributions are drawn from the same parent sample (null hypothesis). In the case of the \textsl{getsf} and \textsl{GExt2D} CMFs, above the \textsl{GExt2D} completeness level we found no significant evidence that the core samples are drawn from different populations (with a KS statistic of $0.09$ and a p-value of $0.91$). 
We also used a statistical library based on the KS metric and dedicated to probability laws fitted by power-laws \citep{clauset2009, alstott2014} to estimate the robustness of our linear regression fit. Run on the \textsl{getsf} CMF of W43-MM2\&MM3 shown in \cref{fig:cmfs software}a, this toolbox suggests that if fitted by a power-law, its best-fit parameters would be a slope coefficient of $\alpha=-0.95\pm0.08$ above a minimum mass of $0.61~\Msol$. This result is in good agreement with the regression fit performed on the \textsl{getsf} sample of cores, above our completeness level of $0.8~\Msol$. 
Figure~\ref{fig:bootstrap powerlaw} presents the bootstrapping probability density histogram of the $N=10^4$ slopes fitted using the KS metric of \cite{alstott2014}, measured for data sets generated using the metric parameters obtained for our sample of 205 cores. The resulting slope coefficient is slightly steeper, $\alpha=-0.98\pm0.1$, but still consistent with those found by the KS metric alone and the fitted value of \cref{fig:cmfs software}a. Moreover, the sigma value obtained with this bootstrapping allows the Salpeter slope to be rejected with a probability of 99.98\%, that is further than the $3.5\sigma$ level.

\subsection{Robustness against our assumptions}
\label{sect:robustness against our assumptions}

Figure~\ref{fig:cmf tests} shows various W43-MM2\&MM3 ridge CMFs built for a different core catalog, under different assumptions of dust temperature and emissivity, and fit over a different mass range. 
For each CMF, we introduced randomly generated CMFs by varying core fluxes, dust temperatures, and opacities and computed the associated $3\,\sigma$ global uncertainty of their fit.
We discuss below the robustness of the observed CMF slope against the chosen extraction strategy and assumptions behind the measurements of core masses.

Comparing Figs.~\ref{fig:cmfs software}a--b shows that the CMF of the W43-MM2\&MM3 ridge is top-heavy regardless of the source extraction technique, either \textsl{getsf} or \textsl{GExt2D} (see \cref{tab:cmf and cores}). Of the 100 cores detected by both software packages (see \cref{appendixtab:core detection table}), 90\% have no significant differences in their integrated fluxes.
Above the \textsl{GExt2D} completeness limit, they constitute the $1.1-69.9~\Msol$ range of the CMF. This striking similarity argues for the robustness of core fluxes measured with different extraction methods, as long as they have a similar core definition.

Furthermore, when comparing \cref{fig:cmf tests}a and \cref{fig:cmfs software}a, our extraction strategy, which is based on \bsens images denoised by \textsl{MnGSeg}, does not seem to impact quantitatively the W43-MM2\&MM3 CMF. Figure~\ref{fig:cmf tests}a indeed presents the CMF of cores extracted in a companion paper (Paper~V; Louvet et al. in prep.) from the \original \& \cleanest images of the W43-MM2 and W43-MM3 protoclusters\footnote{
    For consistency, we applied our filtering and analysis methods to the \original \& \cleanest core catalog (see Sects.~\ref{sect:extraction of compact sources} and \ref{sect:nature of compact sources})
    and made the same assumptions for the mass estimates (see Sect.~\ref{sect:mass estimation}). The resulting catalog of $\sim$75 cores is thus slightly different from that obtained in Paper~V (Louvet et al. in prep.)}.
In agreement with the noise level of the \cleanest images at 1.3~mm (see \cref{tab:sensivity stat}), the 90\% completeness limit of the \cleanest core catalog is two times larger than that of the \denoised \& \bsens CMF. The power-law index of the high-mass end of the \original \& \cleanest CMF is close to, but even shallower than, that of the \denoised \& \bsens CMF (see \cref{tab:tests cmf}). 
The $\sim$75 cores detected in the \cleanest image are in fact among the most massive cores listed in \cref{appendixtab:derived core table}. Moreover, the consistency between the two CMFs comes from the fact that, on average, the \original \& \cleanest cores have fluxes within 15\% of their corresponding flux in \cref{appendixtab:core detection table} and are at worst within 50\% of each other. 

Beyond the uncertainty of flux measurements used to compute the core masses, the main uncertainties of CMFs arise from the mass-averaged dust temperature and dust opacity used to convert fluxes into masses (see Eq.~\ref{appendixeq:core mass}, \cref{appendixfig:dust temperature map}, and \cref{appendixtab:core detection table}). If we do not take into account the central heating by protostars and self-shielding of pre-stellar cores, the core temperatures would homogeneously be $\overline{T_{\rm dust}}\simeq23\pm 2$~K. 
The CMF of \textsl{getsf}-extracted cores with a constant temperature (\cref{fig:cmf tests}b) has a slightly shallower slope than when the individual dust temperature estimates are used (\cref{fig:cmfs software}a, see \cref{tab:tests cmf}). 
We also determined that the CMF flattening is robust against dust opacity variations. As the dust opacity is expected to increase with core density \citep[e.g.,][]{OssenkopfHenning1994}, we made a test assuming a linear relation with mass, starting at $\kappa_{\rm 1.3mm}= 0.007$~cm$^2$\,g$^{-1}$ for the lowest-density core ($0.12~\Msol$) and ending at $\kappa_{\rm 1.3mm}= 0.015$~cm$^2$\,g$^{-1}$ for the highest-density core ($69.9~\Msol$). The resulting CMF has a power-law index lower than the CMF index found in \cref{fig:cmfs software}a, but still greater than the Salpeter slope (see \cref{fig:cmf tests}c and \cref{tab:tests cmf}).

With all tests summarized in Tables~\ref{tab:cmf and cores} and \ref{tab:tests cmf}, we can state that the W43-MM2\&MM3 CMF is top-heavy with a power-law index within the $\alpha=[-1.02;-0.83]$ range. The resulting 1$\,\sigma$ uncertainty is estimated to be about $\pm 0.08$, still excluding the Salpeter slope. 

\section{Discussion on the origin of stellar masses} \label{sect:discussion on the origin of stellar masses}

In Sect.~\ref{sect:classical interpretation}, we compare the CMF of the W43-MM2\&MM3 mini-starburst to published CMF studies. In the framework of several scenarios, we then predict the IMF that would result from the observed W43-MM2\&MM3 CMF. In particular, we apply various mass conversion efficiencies (Sects.~\ref{sect:classical interpretation}--\ref{sect:mass conversion eff}) and various subfragmentation scenarios (Sect.~\ref{sect:core subfrag}), and mention the other processes to consider (Sect.~\ref{sect:other processes}). \cref{tab:tests cmf} lists the parameters of the W43-MM2\&MM3 IMFs derived and fitted under these various assumptions.

\begin{figure}[ht]
    \centering
    \includegraphics[width=1.\linewidth]{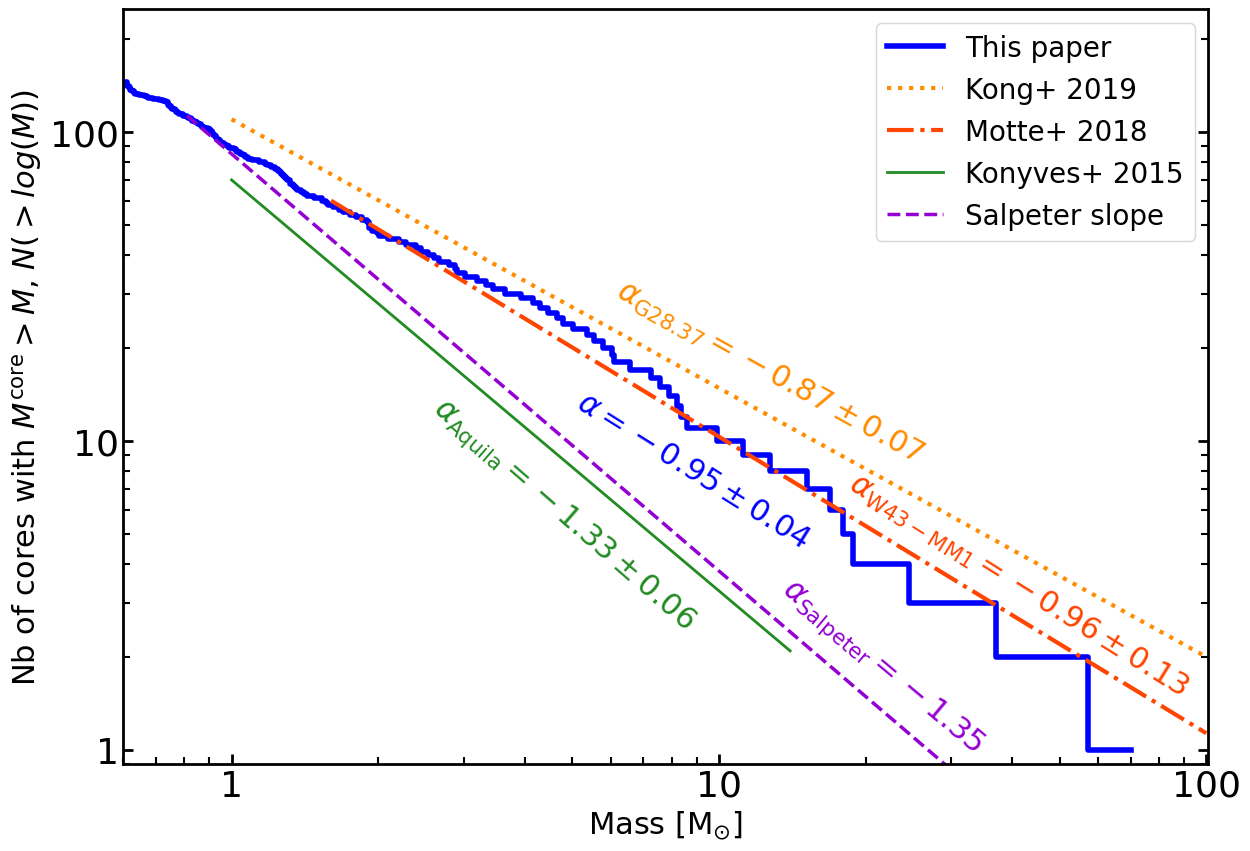}
    \caption{Comparison of the W43-MM2\&MM3 CMF (blue histogram, see \cref{fig:cmfs software}a) with power-laws fitted to the high-mass end, $>$1~$\Msol$, of CMFs measured in three star-forming regions. The proto-typical CMF of low-mass star-forming regions, derived in Aquila \citep[green line,][]{konyves2015}, resembles the Salpeter slope of the canonical IMF \citep[dashed magenta line,][]{salpeter1955}. In contrast, the CMFs in W43-MM2\&MM3 and in the two high-mass star-forming protoclusters W43-MM1 and G28.37+0.07 \citep[red dot-dashed and orange dotted lines,][]{motte2018b,kong2019} are top-heavy.}
    \label{fig:CMF comparison}
\end{figure}

\subsection{In the framework of the classical interpretation}
\label{sect:classical interpretation}

CMFs measured in low-mass star-forming regions are generally strikingly similar to the IMF \cite[e.g.,][]{motte1998, enoch2008, konyves2015}.
In contrast, CMFs of Figs.~\ref{fig:cmfs software}a--b are much shallower than the high-mass end of the canonical IMF. 
The usual methodology to compare observed CMFs to the IMF is to assume a one-to-one correspondence between cores and stars and a given mass conversion efficiency of core mass into star mass. CMF studies of low-mass, low-density cores, $10^5-10^7$~cm$^{-3}$, often derived mass conversion efficiencies of $\epsilon_{\rm core}\sim 30-40\%$ \citep[e.g.,][]{alves2007, konyves2015}. We could expect a larger mass conversion efficiency for our extreme-density cores, $\gtrsim5\times 10^7$~cm$^{-3}$ (see \cref{appendixtab:derived core table}). Therefore, we assume here a mass conversion efficiency of $\epsilon_{\rm core} =50\%$, following \cite{motte2018b}. With this efficiency, the mass range of $0.8-69.9$~\Msol, where the \textsl{getsf} sample is 90\% complete, covers the progenitors of low- to high-mass stars, $0.4-35~\Msol$. Fitting the CMF high-mass end, which would then formally start above $1~\Msol$ or $2~\Msol$, would lead to a slightly steeper slope, $\alpha$ values between $-0.98\pm 0.06$ and $-1.07\pm 0.07$, still shallower than the Salpeter slope of the canonical IMF (see \cref{tab:cmf and cores} for a fit above $2~\Msol$). 
As shown in Figs.~\ref{fig:cmfs software}a and \ref{fig:cmf tests}d, the \textsl{getsf} CMFs for all cores and for those that should form low- to intermediate-mass stars are similarly flat (see \cref{tab:cmf and cores}). We refrain from fitting the CMF of high-mass cores alone because it has too few cores to be statistically robust.
The flattening observed for the W43-MM2\&MM3 CMF is a general trend in all mass regimes. Therefore, it cannot solely be attributed to high-mass stars that could form by processes different from those of low-mass stars \citep[e.g.,][]{motte2018a}. 

Figure~\ref{fig:CMF comparison} compares the high-mass end, $>$1~$\Msol$, of the W43-MM2\&MM3 CMF with a few reference CMF studies obtained in one low-mass star-forming region, Aquila \citep{konyves2015}, and two high-mass protoclusters, W43-MM1 and G28.37+0.07 \citep{motte2018b, kong2019}.
All published studies of core populations found in the nearby, low-mass star-forming regions have argued for the interpretation that the shape of the IMF can simply be derived directly from the CMF  \citep[e.g.,][]{motte1998, andre2014}. We here use a similar definition for cores and very similar tools to extract them to those used in these studies. In particular, \textsl{getsf} \citep{men2021getsf} has the same philosophy as the software used to extract cores from \textit{Herschel} images, \textsl{getsources} and \textsl{CuTEx} \citep{men2012multi, molinari2011}, and ground-based images, \textsl{MRE-GCL} \citep{motte2007}. 
Even so, the CMF measured for the W43-MM2\&MM3 ridge is different from the CMF found in low-mass star-forming regions, including Aquila, which was studied in detail with \textsl{Herschel} \citep[][see \cref{fig:CMF comparison}]{konyves2015}.
It has a high-mass end shallower than most published CMFs, and thus shallower than the IMF of \cite{salpeter1955}. It only resembles, for now, the CMFs observed for the W43-MM1 mini-starburst ridge \citep{motte2018b} and the G28.37+0.07 filament \citep{kong2019} (see \cref{fig:CMF comparison}).

The CMF results obtained for both the W43-MM2\&MM3 and W43-MM1 ridges indicate that either their IMF will be abnormally top-heavy and/or that the  mapping between their core and star masses will not be direct. In the framework of the first interpretation, we assume that the shape of the IMF is directly inherited from the CMF. The results from these two mini-starbursts would thus put into question the IMF universality, which is now being debated \citep[e.g.,][]{hopkins2018}. 
In the framework of the second interpretation, several processes could, in principle, help reconcile the top-heavy CMF observed in the W43-MM2\&MM3 ridge with a Salpeter-like IMF. We investigate below the effect of several of them: mass conversion efficiency, core subfragmentation, star formation history, and disk fragmentation.

\begin{figure}[htbp!]
    \centering
    \begin{minipage}{0.49\textwidth}
      \centering
      \includegraphics[width=1.\textwidth]{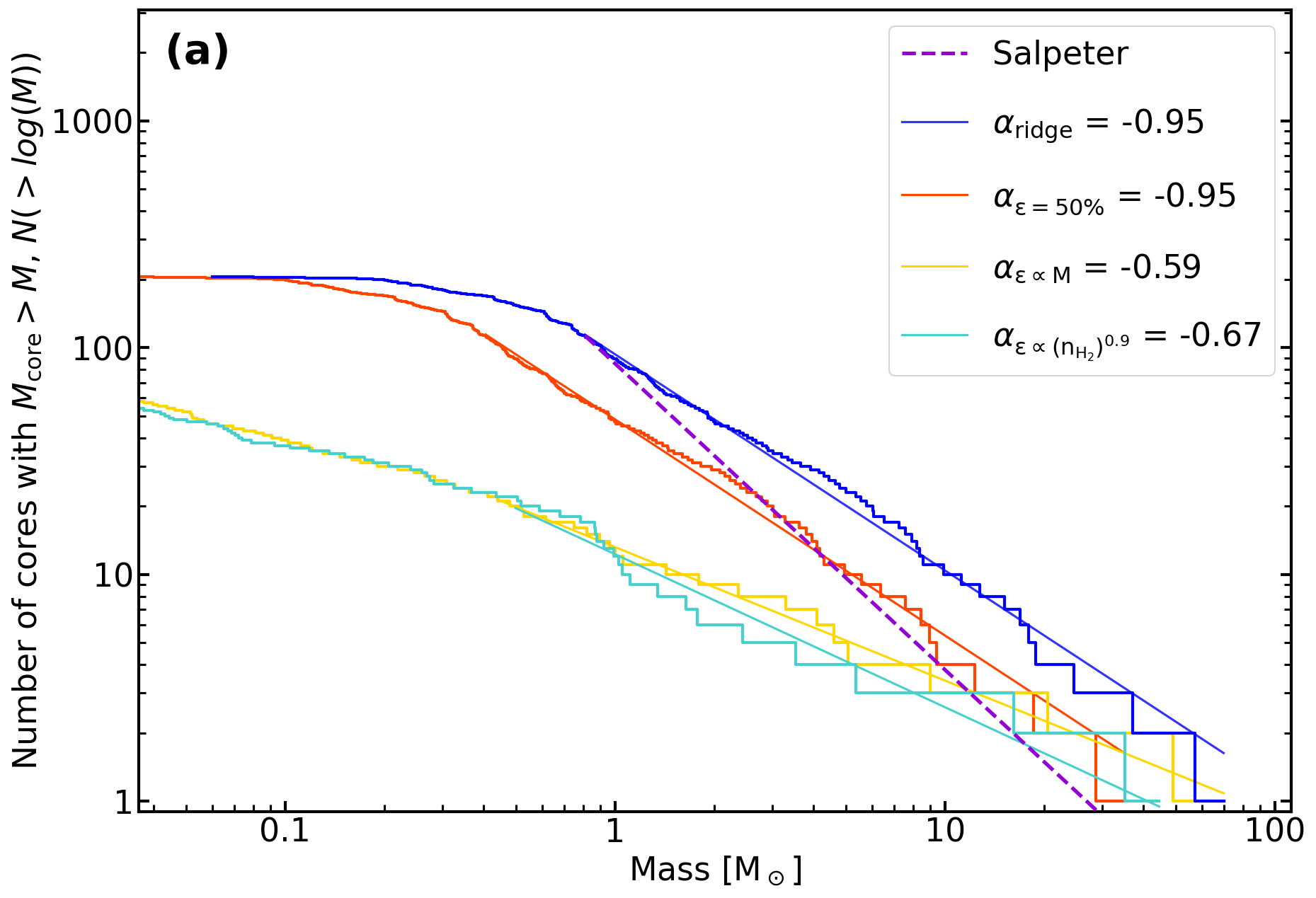}
    \end{minipage}
    \begin{minipage}{0.49\textwidth}
      \centering
      \includegraphics[width=1.\textwidth]{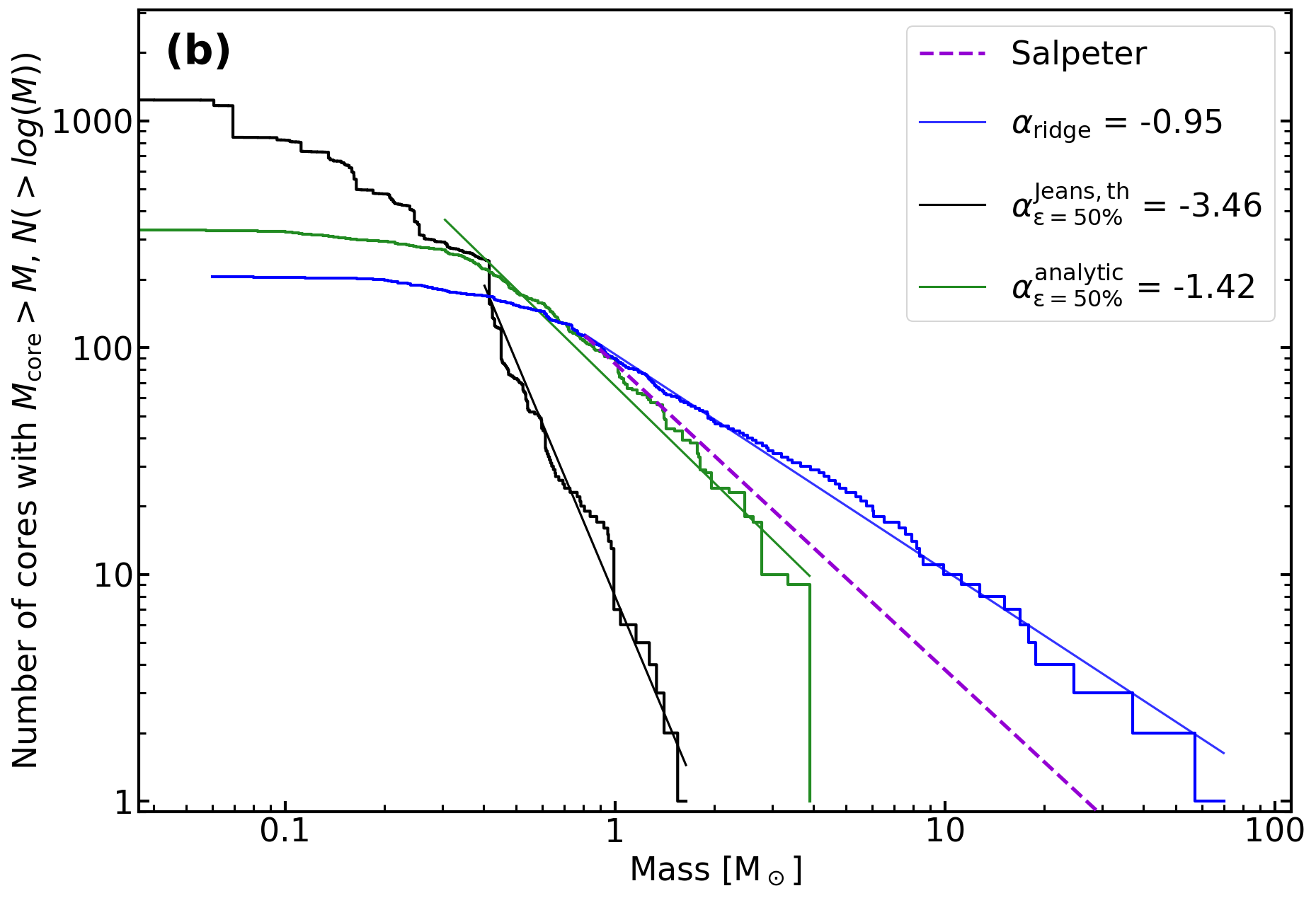}
    \end{minipage}
    \begin{minipage}{0.49\textwidth}
      \centering
      \includegraphics[width=1.\textwidth]{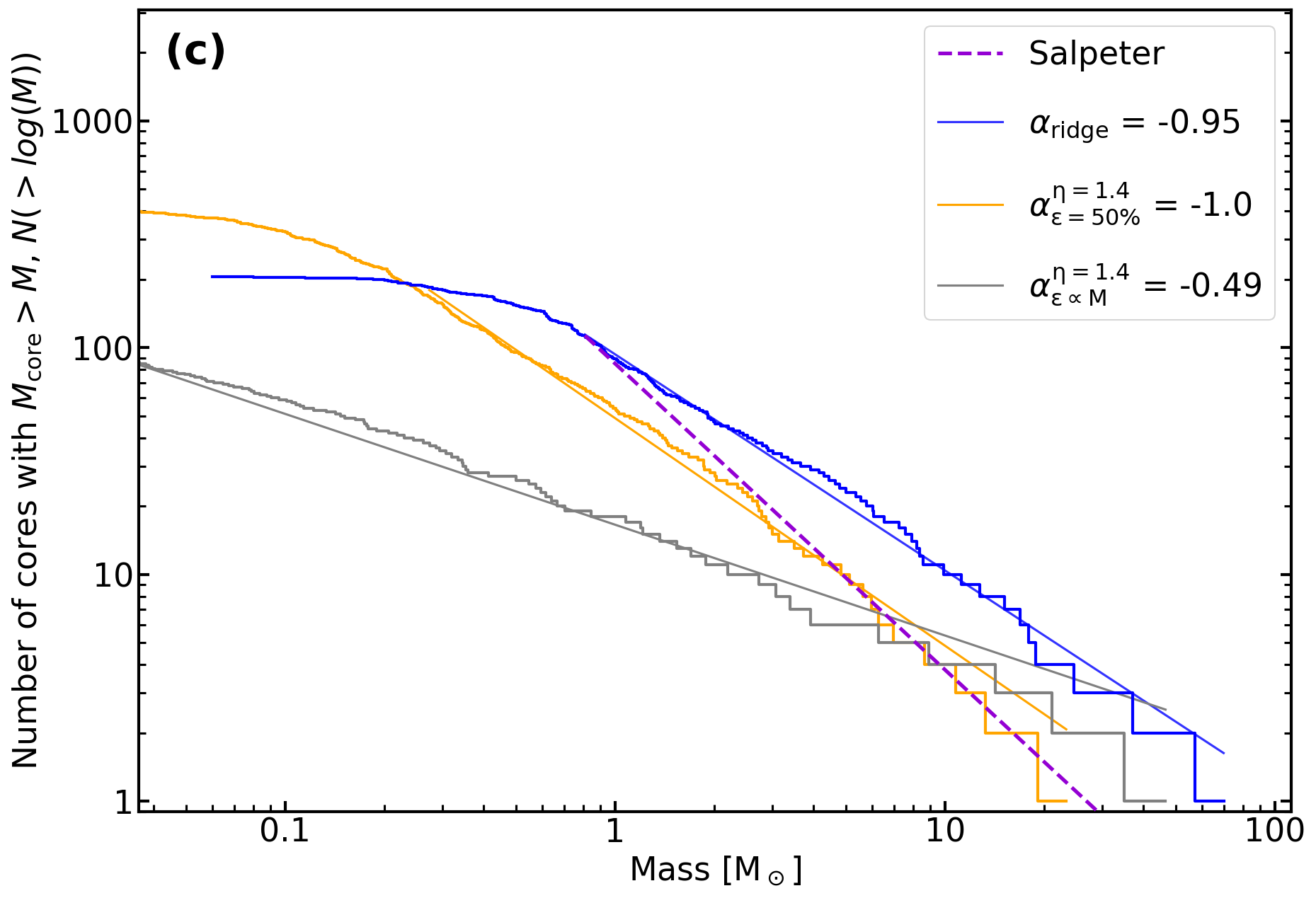}
    \end{minipage}
    \caption{IMFs resulting from various mass conversion efficiencies and fragmentation scenarios, all applied to the W43-MM2\&MM3 CMF of \cref{fig:cmfs software}a (blue histogram). Panel \textsl{(a)}: IMFs predicted for a constant mass conversion efficiency of 50\% (red histogram), linear with the mass (yellow histogram), and dependent on the core density \citep[][cyan histogram]{louvet2014} (see Sect.~\ref{sect:mass conversion eff}). Panel \textsl{(b)}: IMFs predicted by the two extreme fragmentation scenarios in Sect.~\ref{sect:core subfrag}: thermally supported Jeans fragmentation (black) and the analytic fragmentation function leading to a Salpeter slope (green histogram). Panel \textsl{(c)}: IMFs predicted by the hierarchical cascade scenario of Thomasson et al. (subm.), leading to binary fragments. A 2:1 mass partition and two constant mass conversion efficiencies, 50\% (orange histogram) or linear with the mass (gray histogram), are assumed. The number of fragments is taken as the lower integer, with a minimum of 1.}
    \label{fig:fragmentation}
\end{figure}

\subsection{Using different mass conversion efficiencies}
\label{sect:mass conversion eff}
In the present paper we define cores as emission peaks whose sizes are limited by their structured background and neighboring cores. In dynamical clouds, however, the mass, the structure, and even the existence of these cores will evolve over time, as they are expected to accrete or dissolve gas from their background and split into several components or merge with their neighbors \citep[e.g.,][]{smith2014, motte2018a, vazquez2019}.
To account for both these static and dynamical views of cores, we used different functions for the conversion efficiency of core mass into star mass and predict the resulting IMF. 

We first assume a mass conversion efficiency that accounts for the mass loss associated with protostellar outflows in the core-collapse model \citep{matznerMckee2000}. With a constant mass conversion efficiency with core mass, the IMF has the same shape as the CMF as it is simply shifted to lower masses. 
As mentioned in Sect.~\ref{sect:classical interpretation}, we choose a mass conversion efficiency of $\epsilon_{\rm core} =50\%$. Figure~\ref{fig:fragmentation}a displays the IMF resulting from cores whose distribution is shown in \cref{fig:cmfs software}a. The predicted IMF presents, as expected, the same high-mass end slope above $\sim$0.4$~\Msol$ (see \cref{tab:tests cmf}). 

In the case of dynamical clouds, the competitive or gravitationally driven accretion process allows high-mass cores to more efficiently accrete gas mass from their surroundings than low-mass cores \citep[e.g.,][]{bonnell2006, clarkWhitworth2021}. This generally leads to efficiencies of the core formation and mass conversion that depend, to the first order, on the clump and core masses, respectively. We use two analytical models for the mass conversion efficiency.

Since the gravitational force scales linearly with mass, as a first toy model we assumed a linear relation between the mass conversion efficiency and the core mass, normalized by its maximum value: $\epsilon_{\rm core} = \frac{M}{69.9\,\Msol}\times 100\%$. The IMF resulting from this relation applied to the CMF of \cref{fig:cmfs software}a presents a much shallower high-mass end slope (see \cref{fig:fragmentation}a and \cref{tab:tests cmf}).

As a second toy model, we assumed a mass conversion efficiency depending on the mean volume density of cores, normalized by its maximum value: $\epsilon_{\rm core} = \left(\frac{n_{\rm {H_2}}}{2.2\times 10^8{\rm cm}^{-3}}\right)^{0.9}\times 100\%$. This quasi-linear relation is an extrapolation at 3\,400~au scales (the typical size of our cores) of the relation observed in W43-MM1 for large cloud structures, $\sim$1~pc \citep{louvet2014}. The IMF resulting from this toy model has a high-mass end slope which is slightly shallower than the CMF of \cref{fig:cmfs software}a (see \cref{fig:fragmentation}a and \cref{tab:tests cmf}).

Therefore, the fact that a core is no longer considered a static and isolated cloud structure, but rather a cloud structure that accretes its mass from its surrounding cloud at a rate depending on its mass and location in the cloud, tends to flatten the high-mass end of the predicted IMF relative to the observed CMF of cores. This result is in qualitative agreement with analytical models following the evolution of the CMF through the growth of core mass expected in dynamical clouds \citep{dib2007, hatchellFuller2008, clarkWhitworth2021}.

\subsection{Using different scenarios of core subfragmentation}
\label{sect:core subfrag}

The definition of a core is also closely associated with the angular resolution of the observed (or simulated) images of a protocluster \citep[see][]{leeHennebelle2018a, pelkonen2021, louvet2021simu}. The turbulent subfragmentation within these core entities cannot be neglected, but fragmentation functions are barely constrained. We therefore assumed three extreme fragmentation scenarios after applying a 50\% mass conversion efficiency to the W43-MM2\&MM3 CMF displayed in \cref{fig:cmfs software}a. Figure~\ref{fig:fragmentation}b presents the resulting distribution of fragment masses, here called core fragmentation mass function, as in \cite{elmegreen2011}, and sometimes also called system mass function \citep{clarkWhitworth2021}. Since a mass conversion efficiency is applied beforehand, the core fragmentation mass function could directly correspond to the IMF.

The first most extreme fragmentation scenario is the Jeans fragmentation of a core only supported by its thermal pressure. Under this hypothesis and with a mass conversion efficiency of $\epsilon_{\rm core}=50\%$, we assume a mass equipartition between fragments; the number of fragments is thus half the ratio of the core mass to its Jeans mass, $N_{\rm frag}(M) = 0.5 \times \frac{M}{M_{\rm Jeans}}$. We took the measured temperature and FWHM size of our cores (see Tables~\ref{appendixtab:core detection table}-\ref{appendixtab:derived core table}) and computed the Jeans mass of fragments within cores with masses ranging from $2~\Msol$ and $\sim$70$~\Msol$.
In the W43-MM2\&MM3 ridge, most cores are super-Jeans and the most massive cores, in the $16-70~\Msol$ range, would fragment into $50-85$ objects. The resulting IMF is much steeper than the CMF of \cref{fig:cmfs software}a and even steeper than the Salpeter slope of the canonical IMF (see \cref{fig:fragmentation}b and \cref{tab:tests cmf}).

As a second extreme scenario, we found that a fragmentation function of the form $N_{\rm frag}(M) = \left(\frac{\epsilon_{\rm core}\times M}{0.12\;\Msol}\right)^{0.4}$ is necessary to steepen the high-mass end slope of the CMF to a core fragmentation mass function and IMF with a slope close to Salpeter (see \cref{fig:fragmentation}b). This analytical function predicts a single star in $0.24~\Msol$ cores, about five stars in $16~\Msol$ cores and about ten stars in the $\sim$70~$\Msol$ core of W43-MM2. 
These fragmentation prescriptions may apply to evolved cores (referred to as IR-bright), which are observed with a high level of fragmentation \citep{broganHunter2016, palau2018, tang2021}.
However, most of the W43-MM2\&MM3 cores are expected to be much younger \citep{motte2003}. Since the subfragmentation of young cores (referred to as IR-quiet or IR-dark) is rarely observed and therefore very poorly constrained, we simply assume similar levels of fragmentation from young massive clumps to cores, that is from $\sim$20\,000~au to $\sim$2\,000~au scales, and from cores to fragments, that is from $\sim$2\,000~au to $\sim$500~au scales. If we follow studies by \cite{bontemps2010}, \cite{palau2013}, \cite{busquet2016} and \cite{louvet2019} showing that high-mass clumps generally fragment in two cores at most, the two extreme fragmentation scenarios proposed above are both unlikely to be taking place in the W43-MM2\&MM3 ridge.

A third fragmentation scenario is derived from a new type of model aimed at constraining the hierarchical cascade, also called the fragmentation cascade, in observed and simulated clouds (e.g., Thomasson et al. subm.). These studies are based on the finding that the density structure of molecular clouds is hierarchical, and more precisely multi-fractal \citep{elmegreen2001fract, robitaille2020}, and that the spatial distribution of stars is also hierarchical \citep{joncour2017, joncour2018}. Thomasson et al. (subm.) studied the fractal hierarchical cascade of the intermediate-mass star-forming region NGC~2264, using {\it Herschel}-based column density maps. 
The authors found, for clustered clumps, a fractal fragmentation index of $\eta \simeq 1.4\pm0.1$, from the clump to the core scales and more precisely from 13\,000~au to 5\,000~au. A fractal index of $\eta=1.4$ means that for every factor of $2$ decrease in physical scale, the number of fragments multiplies by $1.4$. If we use this fractal index to extrapolate to scales ranging from 2\,500~au to 500~au and generally apply it to all of our cores, we expect to find about two fragments at 500~au resolution within our $0.12-69.9~\Msol$ cores.
Below this 500~au scale, we assume that disk fragmentation dominates turbulent fragmentation and that therefore the hierarchical cascade stops. The distribution of the core mass between subfragments, hereafter called mass partition, is not yet well constrained; we assume below two different cases.

The simplest case assumes a uniformly random mass distribution. As shown by \cite{swift2008}, among others, with this mass partition the high-mass end slopes of the core fragmentation mass function of fragments and the resulting IMF cannot change much from that of the CMF of their parental cores.

For the second case we can assume a very unbalanced mass partition. A preliminary study of 11 W43-MM2\&MM3 core systems\footnote{
    At a $2\,\Theta_{\rm beam}$ distance, paired systems are cores [\#1, \#7], [\#9, \#94], [\#12, \#28], [\#35, \#217], [\#80, \#103], [\#112, \#131], [\#135, \#142], [\#157, \#171], [\#155, \#285]. At a $4\Theta_{\rm beam}$ distance, multiple systems are cores [\#2, \#135, \#142], [\#3, \#43], [\#86, \#98], and [\#112, \#131, \#204].}
identified within $<\,2\,\Theta_{\rm beam}$ distances (or 5000~au in \cref{fig:1.3mm and trichrone}a) suggests mass partition fractions close to 2:1. Interestingly, this is consistent with observations of other high-mass core systems \citep{busquet2016, motte2018b}. Such an unbalanced mass partition is also predicted in the competitive accretion model of \cite{clarkWhitworth2021}, which shows that the large majority of the core mass is used to increase the masses of existing fragments. This unbalanced mass partition and a mass conversion efficiency of $\epsilon_{\rm core} = 50\%$, applied to the W43-MM2\&MM3 CMF, slightly steepens the high-mass end slope (see \cref{fig:fragmentation}c and \cref{tab:tests cmf}).

As the last and most complex test, we assumed the third fragmentation scenario with a 2:1
mass partition and a mass conversion efficiency depending on the core mass, $\epsilon_{\rm core} \propto M$. The resulting IMF is top-heavy with a slope even shallower than that in \cref{fig:cmfs software}a. Interestingly, these assumptions tend to agree with the model of \cite{clarkWhitworth2021}, which combines turbulent fragmentation and competitive accretion. The high-mass end of the the predicted core fragmentation mass functions is broadly invariant over time because the formation of new multiple cores balances the accretion of the gas mass onto existing cores.

\subsection{In the framework of other processes}
\label{sect:other processes}

Beyond the turbulent fragmentation discussed in Sect.~\ref{sect:core subfrag}, disk fragmentation and N-body interactions could further alter the shape of the core fragmentation mass function and thus of the resulting IMF of single stars. Stellar multiplicity studies of low- to intermediate-mass systems have generally revealed mass equipartition \citep{duquennoy1991}, which would not impact the slope of the IMF high-mass end \citep[e.g.,][]{swift2008}. In contrast, given the low number statistics of high-mass star studies, the mass partition of stellar systems that contain high-mass stars is poorly constrained \citep{ducheneKraus2013}. Because of the lack of constraints on disk fragmentation and on N-body interactions, we did not apply a model to the core fragmentation mass function to determine the IMF of single stars.

The other process used to reconcile the observed top-heavy CMF high-mass end with a Salpeter-like CMF is the continuous formation of low-mass cores versus short bursts of formation of high-mass stars. In the case of dense clumps or ridges, most high-mass cores could indeed form in short bursts of $\sim$10$^5$~years, while lower-mass cores would more continuously form over longer periods of time. 
We recall that the IMF of young stellar clusters of a few $10^6$~years is the sum of several instantaneous CMFs built over one to two free-fall times with $\tau_{\rm free-fall} \simeq 10^5$~years. Before and after a burst with a single top-heavy CMF, about ten star formation events of more typical CMFs could develop, diluting the top-heavy IMF resulting from the star formation burst into an IMF with a close-to-canonical shape. Studying the evolution of the CMF shape over time is necessary to quantify this effect, and is one of the goals of the ALMA-IMF survey \citep[see Paper~I and Paper~V;][Louvet et al. in prep.]{motte2021}.

In conclusion, it is difficult to predict the resulting IMF from the observed CMF in the W43-MM2\&MM3 ridge. However, the various mass conversion efficiencies and fragmentation scenarios discussed here suggest that the high-mass end of the IMF could remain top-heavy. 
This will have to go through the sieve of more robust functions of the mass conversion efficiency and core subfragmentation, and of better constrained disk fragmentation and burst-versus-continuous star formation scenarios. 
If it is confirmed that the predicted IMF of W43-MM2\&MM3 is top-heavy, this result will clearly challenge the IMF universality. If we dare to generalize, the IMFs emerging from starburst events could inherit their shape from that of their parental CMFs and could all be top-heavy, disproving the IMF universality.


\section{Summary and conclusion} \label{sect:conclusions}

We used ALMA images of the W43-MM2\&MM3 mini-starburst to make an extensive census of cores and derive its CMF. Our main results and conclusions can be summarized as follows:

\begin{itemize}
    \item We combined the 12~m array images of the W43-MM2 and W43-MM3 protoclusters that were individually targeted by the ALMA-IMF Large Program \citep[see Sect.~\ref{sect:obs and DR} and \cref{tab:observation table};][]{motte2021,ginsburg2021}. At 1.3~mm, the resulting $\rm 4.2~pc\times 3.2~pc$ mosaic has a spatial resolution of $\sim$0.46\arcsec, or 2\,500~au. The 3~mm mosaic is wider, $\rm 7.3~pc\times 5.3~pc$, with a similar angular resolution but a mass sensitivity about three times lower (see \cref{appendixfig:3mm image with cores}).
    \item To have the most complete and most robust sample of cores possible, we used both the best-sensitivity and the line-free  ALMA-IMF images and removed part of the cirrus noise with \textsl{MnGSeg} (see Sect.~\ref{sect:extraction of compact sources}). This new strategy proved to be efficient both in increasing the number of sources detected and in improving the accuracy of their measurements, when applied to present observations and synthetic images (see \cref{tab:sensivity stat} and \cref{appendixsect:mngseg}). In the end, it allows the $5\,\sigma$ detection of point-like cores with gas masses of $\sim$0.20~$\Msol$ at 23~K (see \cref{fig:1.3mm and trichrone}a).
    \item We extracted 1.3~mm compact sources using both the \textsl{getsf} and \textsl{GExt2D} software packages. \textsl{getsf} provides a catalog of 208 objects, which have a median FWHM size of 3\,400~au (see \cref{appendixtab:core detection table} and Figs.~\ref{fig:1.3mm and trichrone}--\ref{fig:fwhm distribution}). The 100 cores extracted by both \textsl{getsf} and \textsl{GExt2D} have sizes and thus fluxes, on average, consistent to within 30\%.
    \item The nature of the W43-MM2\&MM3 sources is investigated to exclude free-free emission peaks and correct source fluxes from line contamination (see Figs.~\ref{fig:freefree}--\ref{fig:hotcore} and Sects.~\ref{sect:freefree contamination}--\ref{sect:line contamination}). The resulting catalog contains 205 \textsl{getsf} cores (see \cref{appendixtab:derived core table}).
    Their masses are estimated and, for the most massive cores, they are corrected for their optically thick thermal dust emission (see Eq.~\ref{eq:optically thick mass} in Sect.~\ref{sect:mass estimation} and \cref{appendixsect:detailed approach for the mass calculation}). The core mass range is $0.1-70~\Msol$ and the \textsl{getsf} catalog is 90\% complete down to $0.8~\Msol$ (see \cref{appendixsect:completeness simulation}).
    \item The W43-MM2\&MM3 CMFs derived from the \textsl{getsf} and \textsl{GExt2D} core samples are both top-heavy with respect to the Salpeter slope of the canonical IMF (see Sect.~\ref{sect:top heavy cmf} and \cref{fig:cmfs software}). The high-mass end of the \textsl{getsf} CMF is well fitted, above its 90\% completeness limit, by a power-law of the form $N(>\log M)\propto M^{\alpha}$, with $\alpha = -0.95 \pm 0.04$ (see \cref{tab:cmf and cores}). The error bars include the effect of uncertainties on core mass, fit, and completeness level. The CMF high-mass end thus cannot be represented by a function resembling the Salpeter IMF (see also \cref{fig:bootstrap powerlaw}). We showed that the shape of the CMF is robust against flux differences arising from the map or software chosen to extract cores, and against variations of the dust emissivity and temperature variations (see Sect.~\ref{sect:robustness against our assumptions}, \cref{fig:cmf tests} and \cref{tab:tests cmf}). Our result, in striking contrast with most CMF studies, argues against the universality of the CMF shape.
    \item We used different functions of the conversion efficiency from core to stellar masses to predict the IMF resulting from the W43-MM2\&MM3 CMF (see Sect.~\ref{sect:discussion on the origin of stellar masses}). While in the framework of the core-collapse model the slope of the IMF high-mass end remains unchanged, it becomes shallower for competitive accretion or hierarchical global collapse models (see \cref{fig:fragmentation}a). We explored several fragmentation scenarios, which all slightly steepen the high-mass end of the predicted IMF (see \cref{fig:fragmentation}b--c). It is possible to set an artificial analytical model that predicts an IMF with the Salpeter slope. However, the best-constrained fragmentation model, which is a hierarchical cascade with 2:1 mass partition, predicts an IMF slope which does not reconcile with the canonical value (see \cref{fig:fragmentation}c). 
Most scenarios tested here suggest that the resulting IMF could remain top-heavy. More constrained functions of the mass conversion efficiency, core subfragmentation, disk fragmentation, and burst development are required to provide a more definitive prediction. However, if this result is confirmed, the IMFs emerging from starburst events could inherit their shape from that of their parental CMFs and be top-heavy, thus challenging the IMF universality.
\end{itemize}


\begin{acknowledgements}
    This paper makes use of the ALMA data ADS/JAO.ALMA\#2017.1.01355.L. ALMA is a partnership of ESO (representing its member states), NSF (USA) and NINS (Japan), together with NRC (Canada), MOST and ASIAA (Taiwan), and KASI (Republic of Korea), in cooperation with the Republic of Chile. The Joint ALMA Observatory is operated by ESO, AUI/NRAO and NAOJ.
    This project has received funding from the European Research Council (ERC) via the ERC Synergy Grant \textsl{ECOGAL} (grant 855130), from the French Agence Nationale de la Recherche (ANR) through the project \textsl{COSMHIC} (ANR-20-CE31-0009), and the French Programme National de Physique Stellaire and Physique et Chimie du Milieu Interstellaire (PNPS and PCMI) of CNRS/INSU (with INC/INP/IN2P3).
    YP acknowledges funding from the IDEX Universit\'e Grenoble Alpes under the Initiatives de Recherche Strat\'egiques (IRS) “Origine de la Masse des \'Etoiles dans notre Galaxie” (OMEGa).
    YP 
    and GB acknowledge funding from the European Research Council (ERC) under the European Union’s Horizon 2020 research and innovation programme, for the Project “The Dawn of Organic Chemistry” (DOC), grant agreement No 741002.
    RGM and TN acknowledge support from UNAM-PAPIIT project IN104319. RGM is also supported by CONACyT Ciencia de Frontera project ID 86372. TN acknowledges support from the postdoctoral fellowship program of the UNAM.
    SB acknowledges support from the French Agence Nationale de la Recherche (ANR) through the project \textsl{GENESIS} (ANR-16-CE92-0035-01).
    FL acknowledges the support of the Marie Curie Action of the European Union (project \textsl{MagiKStar}, Grant agreement number 841276).
    AGi acknowledges support from the National Science Foundation under grant No. 2008101.
    PS and BW were supported by a Grant-in-Aid for Scientific Research (KAKENHI Number 18H01259) of the Japan Society for the Promotion of Science (JSPS). P.S. and H.-L.L. gratefully acknowledge the support from the NAOJ Visiting Fellow Program to visit the National Astronomical Observatory of Japan in 2019, February.
    %
    AS gratefully acknowledges funding support through Fondecyt Regular (project code 1180350), from the ANID BASAL project FB210003, and from the Chilean Centro de Excelencia en Astrof\'isica y Tecnolog\'ias Afines (CATA) BASAL grant AFB-170002. 
    TB acknowledges the support from S. N. Bose National Centre for Basic Sciences under the Department of Science and Technology, Govt. of India.
    GB also acknowledges funding from the State Agency for Research (AEI) of the Spanish MCIU through the AYA2017-84390-C2-2-R grant and from the PID2020-117710GB-I00 grant funded by MCIN/ AEI  /10.13039/501100011033 . 
    TCs 
    has received financial support from the French State in the framework of the IdEx Universit\'e de Bordeaux Investments for the future Program.
    %
    %
    LB gratefully acknowledges support by the ANID BASAL projects ACE210002 and FB210003. 
    K.T. was supported by JSPS KAKENHI (Grant Number 20H05645).
    DW gratefully acknowledges support from the National Science Foundation under Award No. 1816715.
\end{acknowledgements}

\bibliographystyle{aa}
\bibliography{ALMA-IMF-PaperIII}

\begin{appendix}

\section{Quality of the core extraction carried out using images denoised by MnGSeg}\label{appendixsect:mngseg}

\begin{figure*}[ht!]
\centering
    \includegraphics[width=1.\linewidth]{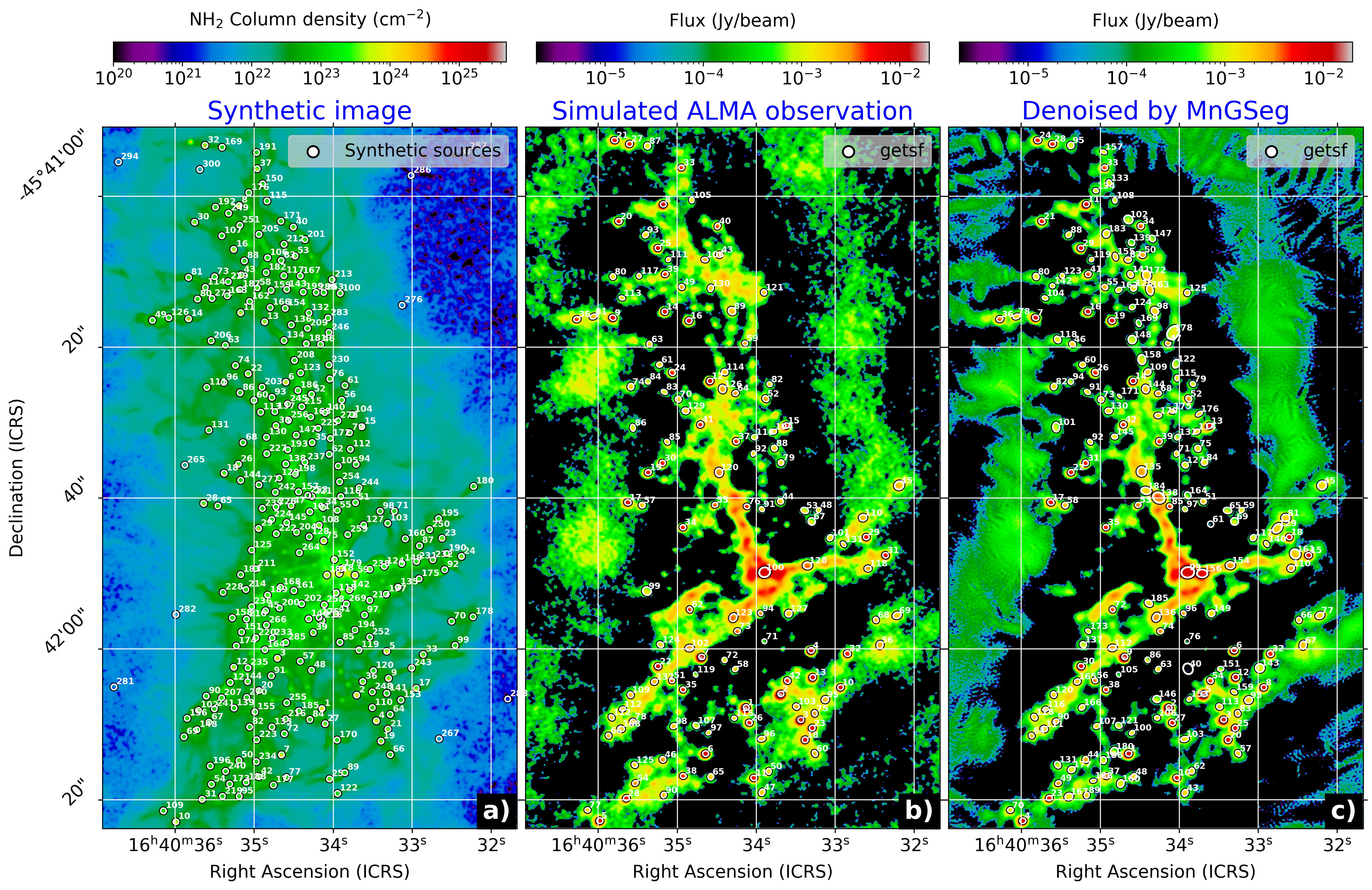}
    \caption{Synthetic column density images used to qualify the extraction of cores in images denoised by \textsl{MnGSeg}. Panel \textsl{a)}: Projected column density map of a numerical simulation by \cite{ntormousi2019}, from which simulated cores have been removed and to which synthetic sources have been added ($0.2-3.2~\Msol$ cores with a $I(\theta)\propto \theta^{-1}$ intensity profile up to $0.5\arcsec$). Panel \textsl{b)}: Simulation of the ALMA imaging of the column density map of \textsl{a} at 1.3~mm, with a 240-minutes integration time and a beam size of $0.81\arcsec\times0.76\arcsec$. Panel \textsl{c)}: Simulated ALMA image from \textsl{b)}, with noise reduced by $\sim$30\% using \textsl{MnGSeg}. Cores are outlined by ellipses and labeled by numbers according to the truth table (in \textsl{a)}) and \textsl{getsf} identifiers in the \original and \denoised catalogs (in \textsl{b)} and \textit{c)}).}
    \label{appendixfig:3tiles}
\end{figure*}

Core extractions are limited by the noise level of input images. The noise measured in any molecular cloud image can schematically be seen as the sum of the white instrumental noise at scales smaller than the beam, interferometric artifacts (when applicable), and some structural noise. \textsl{MnGSeg} is an efficient technique to separate the hierarchical cloud structures into coherent structures associated with star formation and incoherent Gaussian structures \citep{robitaille2019}. The latter consist of structures that do not persist from scale to scale including white noise and low-density cirrus clouds on the line of sight to our star formation sites. We used \textsl{MnGSeg} to remove Gaussian structures of our image at scales larger than the beam, thus removing structural noise like our low-density cirrus. In practice, we keep coherent structures, point-like structures that correspond in particular to isolated cores, interferometric artifacts, and the white instrumental noise, which is a flux component needed to reliably extract cores. We hereafter call this image the \denoised image since the noise level decreases (see \cref{tab:observation table}).

To characterize the quality of the core extraction performed on images denoised by \textsl{MnGSeg}, it is necessary to quantify the gain of the catalog in terms of number of detected cores, fraction of spurious sources, and quality of the flux measurements. To do so, we studied core extractions done on a synthetic image, consisting of a background image plus well-characterized synthetic sources. For the background image, we used a column density image created by \cite{louvet2021simu} from a three-dimensional numerical simulation of an Orion-sized turbulent molecular cloud \citep{ntormousi2019}. 
For our purpose, we assumed it represents a typical cloud complex of the ALMA-IMF Large Program like G338.93, located at 3.9~kpc. The simulated density cube was projected along a single axis and a core extraction was performed with \textsl{getsf} to remove the flux contribution of simulated cores, thus creating an image only made of their background. These cores are not suitable for our purpose due to their ill-constrained sizes and masses. A total of 306 synthetic sources were then added where the column density exceeds $4\times10^{22}~\textrm{cm}^{-2}$. They have masses arranged in ten logarithmic bins spanning $0.2-3.2~\Msol$ and, for simplicity, they are set to be unresolved by the simulated ALMA beam (here $0.8\arcsec$). The resulting column density image is presented in \cref{appendixfig:3tiles}a. All the characteristics of synthetic sources are stored in a reference table hereafter called the truth table.

Afterward, we simulated the ALMA observation of this synthetic protocluster using the CASA task \textsl{simobserve} (see \cref{appendixfig:3tiles}b). We first created an image of the 1.3~mm, or more precisely 224.55~GHz, flux arising from the synthetic column density image of \cref{appendixfig:3tiles}a, assuming a dust temperature of 20~K and a dust $+$ gas mass opacity of $\kappa_{\rm 1.3mm} = \rm 0.01~cm^2\,g^{-1}$. By specifying a typical array configuration of ALMA-IMF images and a given integration time, \textsl{simobserve} uses an incomplete UV coverage to smooth the synthetic image to a $0.81\arcsec\times0.76\arcsec$ beam and creates interferometric artifacts, including filtering of the extended emission. It then adds some white noise, which is characteristic of submillimeter observations. Finally, we applied \textsl{MnGSeg} to this \original image and removed all incoherent structures with sizes larger than the beam (see definition above and \cref{appendixfig:3tiles}c).

{\renewcommand{\arraystretch}{1.5}
\begin{table*}[t]
\centering
\begin{threeparttable}[c]
\caption{Quality of core extractions done using the \original and \denoised images of \cref{appendixfig:3tiles}b-c for either detection or measurement or both.}
\label{appendixtab:simalma stat}
\begin{tabular}{cc|cccccc}
    \hline\hline
    \multicolumn{2}{c|}{Extraction strategy} & Cores & \multicolumn{5}{c}{Cores (number and their proportion in the \textsl{getsf} catalog)} \\
    \multirow{2}{*}{Detection} & \multirow{2}{*}{Measurement} & 
    extracted
    & correctly & with bad & with rough & with bad & with rough \\
     & & 
    by \textsl{getsf}
    & extracted \hspace{2mm} &
    \multicolumn{2}{c}{detection} &
    \multicolumn{2}{c}{measurement} \\
    (1) & (2) & (3) & (4) & (5) & (6) & (7) & (8) \\
    \hline
    \original & \original & 132 & 128 (97\%) & 0 & 1 (1\%) & 4  (3\%)  & 21 (16\%) \\
    \denoised & \original & 183 & 163 (90\%) & 0 & 5 (3\%) & 18 (10\%) & 27 (15\%) \\
    \denoised & \denoised & 183 & 173 (95\%) & 0 & 6 (3\%) & 10 (5\%)  & 32 (17\%) \\
    \hline
\end{tabular}
\begin{tablenotes}
\item (4) Extracted cores, whose peak position is at worst partly inaccurate, and whose flux measurement is inaccurate by at worst a factor of 2 (see definitions below). 
\item (5) Detected cores, whose peak position is very inaccurate: $\Theta_\textrm{offset-position} > \Theta_{\textrm{beam}}/2$.
\item (6) Detected cores whose peak position is partly inaccurate: $\Theta_{\textrm{beam}}/4 < \Theta_\textrm{offset-position} \leq
\Theta_{\textrm{beam}}/2$.  
\item (7) Correctly extracted cores, whose flux measurement is inaccurate by at least a factor of 2, $\frac{S^{\rm int}_{\rm measured}}{S^{\rm int}_{\rm true}} < \frac{1}{2}$ or $> 2$. These sources are not considered correctly extracted cores (see Col.~4), and as such are excluded when estimating completeness levels in \cref{appendixfig:completeness}.
\item (8) Correctly extracted cores, whose flux measurement is inaccurate by a factor between 1.5 and 2, $\frac{1}{2} \leq \frac{S^{\rm int}_{\rm measured}}{S^{\rm int}_{\rm true}} < \frac{2}{3}$ or $\frac{3}{2} < \frac{S^{\rm int}_{\rm measured}}{S^{\rm int}_{\rm true}} \leq 2$. Roughly detected cores generally have rough measurements.
\end{tablenotes}
\end{threeparttable}
\end{table*}}

We compared the core extractions done by \textsl{getsf} \citep{men2021getsf} on the \original and \denoised images of Figs.~\ref{appendixfig:3tiles}b-c. We applied the post-selection criteria recommended by \textsl{getsf} and described in Sect.~\ref{sect:extraction of compact sources} to remove sources that are not reliable enough.
\cref{appendixtab:simalma stat} lists the number of cores, detected by \textsl{getsf}, which correspond to real synthetic cores. Since the noise level is $\sim$30\% smaller in the \denoised image (see \cref{tab:observation table}), \textsl{getsf} detected an increased number of sources, $40\%$ more than in the \original image (see \cref{appendixtab:simalma stat}). 

\cref{appendixtab:simalma stat} also quantifies the quality of core extractions, from their detection to their flux measurement.
To identify the ``correctly detected'' cores, we cross-matched the \textsl{getsf} catalogs and the truth table. Their peak positions are considered accurate when they lie at less than $\Theta_{\textrm{beam}}/2\simeq0.4\arcsec$ from the position of a synthetic core in the truth table. 
A ``badly detected'' source refers to a source whose peak position is inaccurate by more than this value, and consequently not present in the synthetic core population. 
The \textsl{getsf} software was adjusted by \cite{men2021getsf} not to return any badly detected sources, also called spurious source. All cores identified in both in the \original and \denoised images of Figs.~\ref{appendixfig:3tiles}b-c qualify as correctly detected, without spurious detections. 
To go beyond this binary description of core detection, \cref{appendixtab:simalma stat} lists the number of ``roughly detected'' cores, which are those with peak position offset by more than $\Theta_{\textrm{beam}}/4\simeq0.2\arcsec$, but less than $\Theta_{\textrm{beam}}/2\simeq0.4\arcsec$. 
These cores consist of the merging of a synthetic core and part of its surrounding background cloud, which inevitably has consequences on the flux measurement, as shown below. The number of roughly detected cores barely increased, from 1\% to 3\%, when applying the \textsl{MnGSeg} technique to denoise the simulated ALMA image (see \cref{appendixtab:simalma stat}).

\begin{figure*}[htbp]
\centering
\begin{minipage}{.45\textwidth}
  \centering
  \includegraphics[width=1.\textwidth]{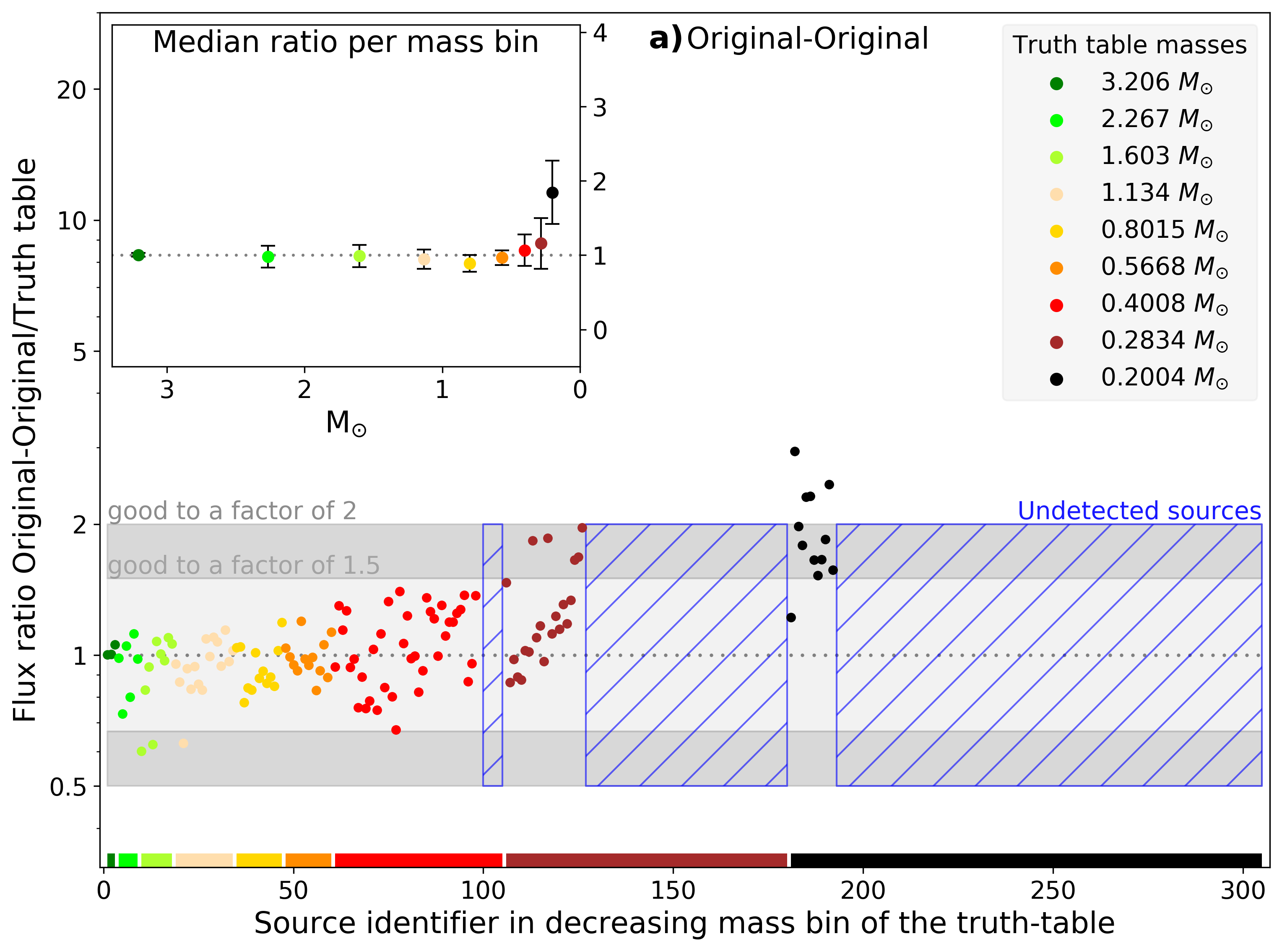}
  \label{appendixfig:ratio-original}
\end{minipage}%
\hskip 0.5cm
\begin{minipage}{.45\textwidth}
  \centering
  \includegraphics[width=1.\textwidth]{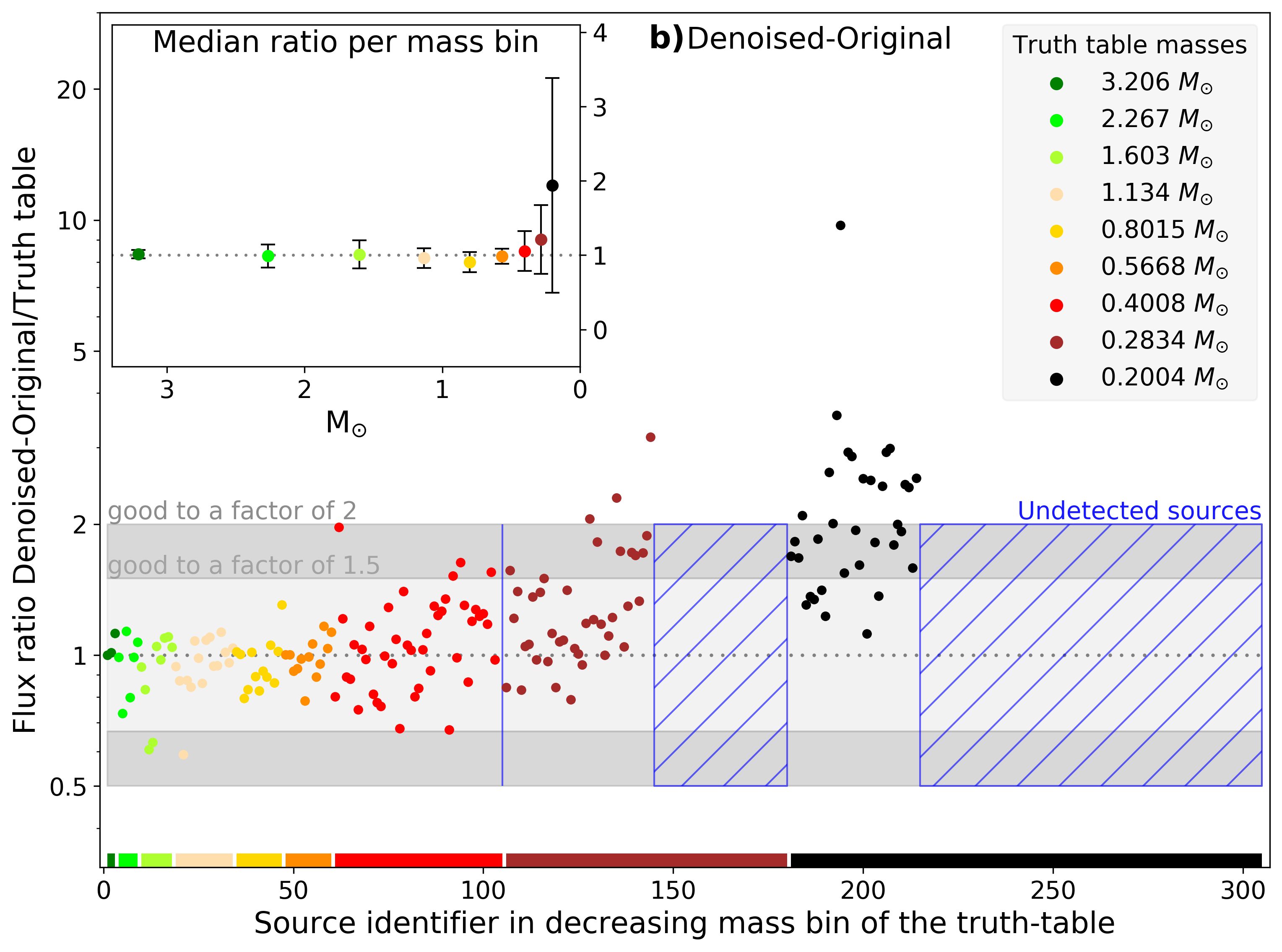}
  \label{appendixfig:ratio-mngseg}
\end{minipage}
\begin{minipage}{.45\textwidth}
  \centering
  \includegraphics[width=1.\textwidth]{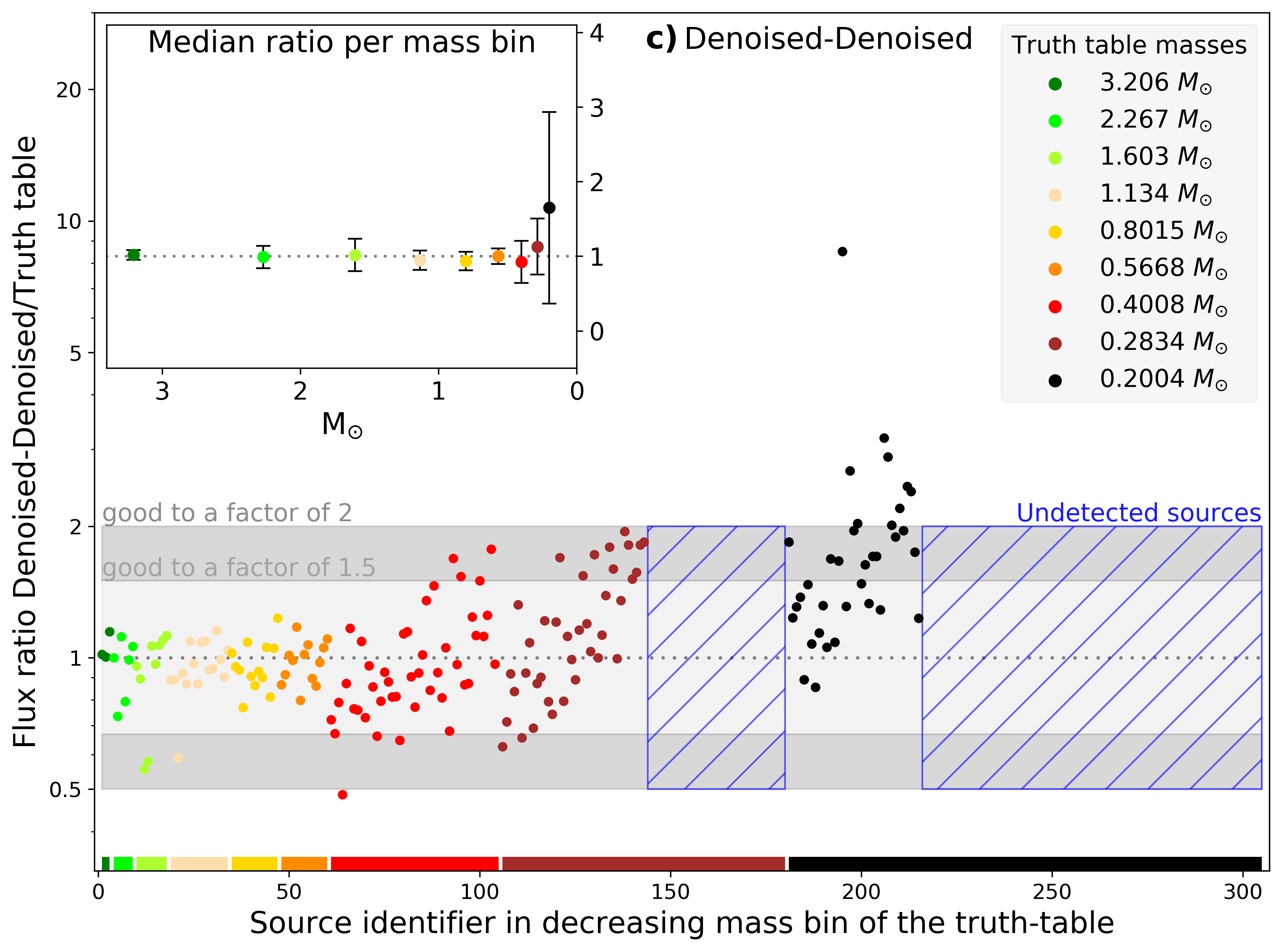}
  \label{appendixfig:ratio-mngseg-only}
\end{minipage}
\vskip -0.3cm
\caption{Quality of the flux measurements made by \textsl{getsf} when sources are both detected and extracted in the \original image (in \textsl{a}), detected in the \denoised and measured in the \original images (in \textsl{b}), and both detected and measured in the \denoised image (in \textsl{c}). The ratios of the \textsl{getsf} integrated fluxes to the true fluxes are plotted for individual sources, arranged in decreasing mass bin (main panel) and median values for the nine mass bins (enclosed plots). Blue hatched regions represent fraction of the truth table sources undetected by the algorithm, for each mass bin. For cores down to the $0.28~\Msol$ bin, fluxes are correct by $\sim$10\% for the three extraction runs. The \denoised image provides a catalog with 40\% more sources whose flux measurements are, for 90\% of them, better than a factor of 2.}
\label{appendixfig:ratio}
\end{figure*}

We investigated the quality of the flux measurements of cores extracted both in the \original and \denoised images by computing the ratios of the \textsl{getsf} integrated fluxes over the true fluxes. 
Given the complexity of the process of extracting cores in molecular clouds, we qualify a core as correctly extracted when its flux measurement is correct within a factor of 2, $\frac{1}{2} \leq \frac{S^{\rm int}_{\rm measured}}{S^{\rm int}_{\rm true}} \leq 2$.
An inaccuracy of the flux ratio larger than 2 indicates a source with a ``badly measured'' flux, while a flux ratio between 1.5 and 2 indicates it is ``roughly measured''. \cref{appendixtab:simalma stat} shows that measurements in the \denoised image are as accurate as those made in the \original image. When cores are detected in the \denoised image and their flux measured in the noisier \original image, the latter are less accurate but not by large factors.
Figures~\ref{appendixfig:ratio}a-c, display the ratios of the \textsl{getsf} fluxes over the true fluxes for the different detection and measurement images. Undetected sources of a given bin mass lie in the hatched regions of Figs.~\ref{appendixfig:ratio}a--c, leading to non-continuous source groups. For cores down to the $0.28~\Msol$ bin, which is below the 90\% completion level of $0.37~\Msol$ (see \cref{appendixfig:completeness simu}), median fluxes are correct by $\sim$10\% for the three extraction runs. 
The additional sources of the \denoised catalog (out of the hatched region of \cref{appendixfig:ratio}c, contrary to \cref{appendixfig:ratio}a), whose detection and flux measurement were estimated as not reliable enough in the original image, have less accurate fluxes, but for 90\% of them they remain correct to within a factor of 2 and on average are better than a factor of 1.55.

These synthetic simulations strongly suggest that denoising real images with \textsl{MnGSeg} provide the opportunity to extract a larger number of cores with good flux measurements. The parameters of the sources, such as sizes and fluxes, appear more accurate when measured directly in the \denoised image (see enclosed plots of \cref{appendixfig:ratio}). Completeness levels of the \denoised versus \original core extractions can also be measured to quantify the gain when denoising the simulated ALMA image using \textsl{MnGSeg}. Figure~\ref{appendixfig:completeness simu} shows that the 90\% completeness level improved by $\sim$16\%. While this value depends on the chosen background and core shape, such an improvement of the completeness level is a definite asset for studies of the mass and spatial distributions of cores. We therefore expect that the denoising process proposed by \textsl{MnGSeg} will improve the statistics of core catalogs and their resulting studies.

\begin{figure}[hbtp]
    \centering
    \includegraphics[width=1.\linewidth]{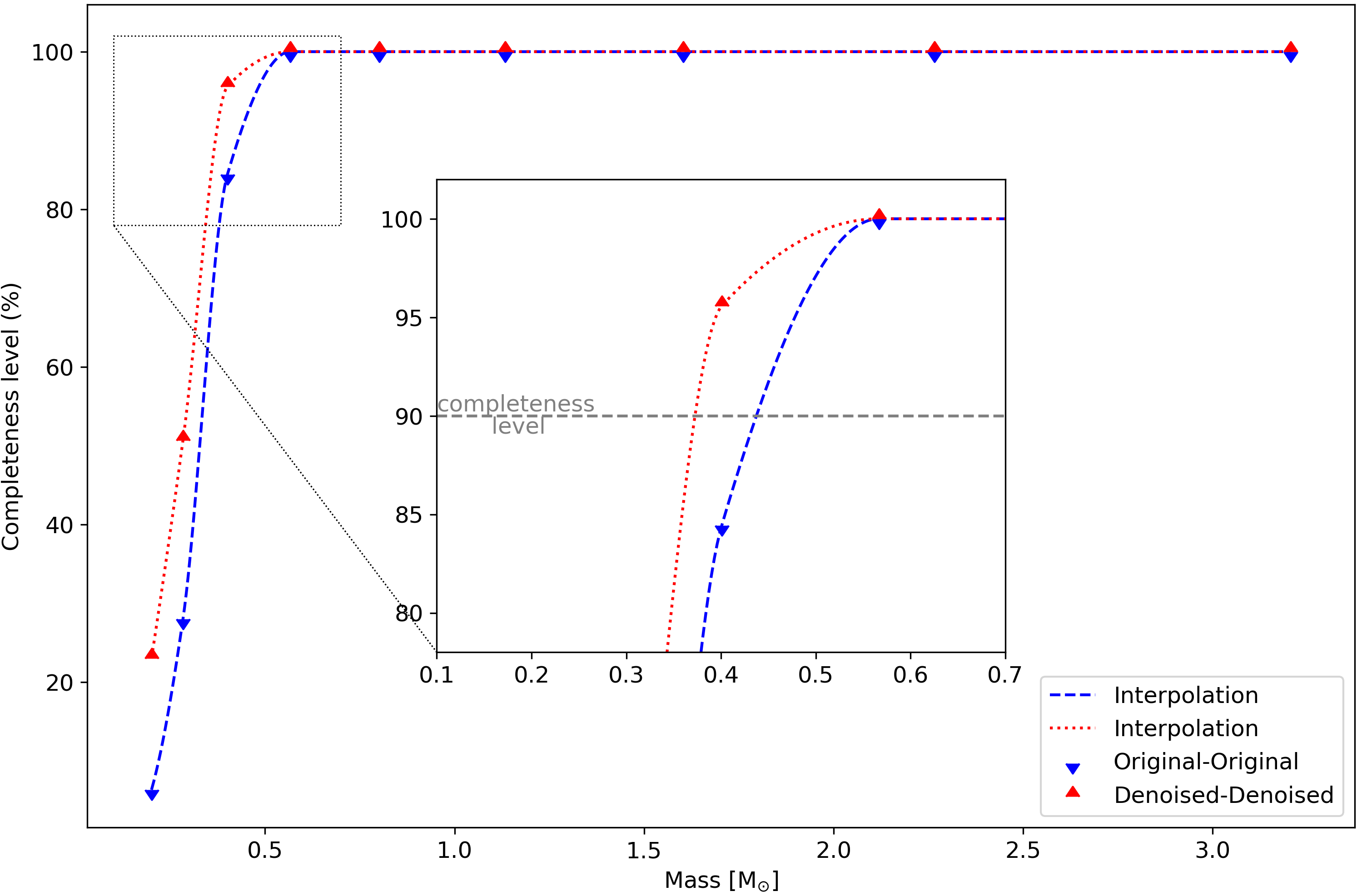}
    \caption{Completeness levels of the core samples of the \original and \denoised catalogs obtained by \textsl{getsf}, excluding badly detected and badly measured sources. Data points were interpolated using the Piecewise Cubic Hermite Interpolating Polynomial method. The core content is 90\% complete down to $\sim$$0.44~\Msol$ and $\sim$$0.37~\Msol$ for the \original and \denoised images, respectively, which correspond to an improvement of $\sim$16\% in mass completeness.}
    \label{appendixfig:completeness simu}
\end{figure}

\clearpage
\section{Method proposed to correct for the optical depth of the continuum emission of compact sources}
\label{appendixsect:detailed approach for the mass calculation}

One of the difficulties in calculating masses is the optical depth of the emission. In general, at 1.3 mm, the emission is largely optically thin. Studies of high-mass star-forming regions revealed, however, that some cores can reach sufficiently high densities ($n_{\rm H2}= 10^7-10^8$~cm$^{-3}$) to become optically thick \cite[e.g.,][]{cyganowski2017,motte2018a}. We propose here a method to correct, at first order, for the optical depth of cores observed at submillimeter wavelengths.

A point-like source that is at a distance $d$ and subtends a solid angle $\Omega_{\rm beam}$ has a cross-sectional area $A=\Omega_{\rm beam}\;d^2$. Its optical depth at wavelength $\lambda$ is $\tau_{\lambda} = \Sigma\;\kappa_{\lambda}$, where $\Sigma$ is the surface density and $\kappa_{\lambda}$ the dust opacity of the source. The mass measured in a telescope beam, $M^{\rm peak}_{\rm \tau\gtrsim 1}$, is then
\begin{equation}\label{appendixeq:mass w/ optical depth}
    \centering
    M^{\rm peak}_{\rm \tau\gtrsim 1} = A\;\Sigma 
    = \Omega_{\rm beam}\;d^2\;\Sigma 
    = \frac{\Omega_{\rm beam}\;d^2}{\kappa_{\lambda}}\tau_{\lambda}.
\end{equation}

Since the monochromatic intensity from the source is given by $I_{\lambda} = B_{\lambda}\left(T_{\rm dust}\right) \left[1\,-\,e^{-\tau_{\lambda}}\right]$, the monochromatic flux measured in a telescope beam, $S_{\lambda}^{\rm peak}$, is
\begin{equation} \nonumber
    \centering
    S_{\lambda}^{\rm peak} = \Omega_{\rm beam}\;I_{\lambda} = \Omega_{\rm beam}\;B_{\lambda}\left(T_{\rm dust}\right) \left[1\,-\,e^{-\tau_{\lambda}}\right].
\end{equation}
It follows that the optical depth at wavelength $\lambda$ is
\begin{equation} \label{appendixeq:optical depth}
    \tau_{\lambda} = -\ln{\left(1\,-\,\frac{S_{\lambda}^{\rm peak}}{\Omega_{\rm beam}\; B_{\lambda}\left(T_{\rm dust}\right)}\right)}. 
\end{equation}
When substituting \cref{appendixeq:optical depth} in \cref{appendixeq:mass w/ optical depth}, we obtain
\begin{equation}\label{appendixeq:peak mass}
    M^{\rm peak}_{\rm \tau\gtrsim 1} =\,-\,\frac{\Omega_{\rm beam}\;d^2}{\kappa_{\lambda}}\; \ln{\left(1\,-\,\frac{S^{\rm peak}_{\lambda}}{\Omega_{\rm beam}\; B_{\lambda}\left(T_{\rm dust}\right)}\right)}. \nonumber
\end{equation}
A compact source like our cores in \cref{appendixtab:core detection table}, which have deconvolved sizes about 1.5 times the beam (see \cref{fig:fwhm distribution} and Sect.~\ref{sect:extraction of compact sources}), has most of its flux in an area of size equal to the synthesize beam. Its mass, $M_{\rm \tau\gtrsim 1}$, can therefore be estimated applying the optical depth of \cref{appendixeq:optical depth}, measured over the beam assuming point-like sources, to its whole solid angle, $\Omega_{\rm core}$, using the following equation:
\begin{equation}\label{appendixeq:core mass1}
    M_{\rm \tau\gtrsim 1} \approx \,-\,\frac{\Omega_{\rm beam}\;d^2}{\kappa_{\lambda}} \sum_{\Omega_{\rm core}}\; \ln{\left(1\,-\,\frac{S^{\rm peak}_{\lambda}}{\Omega_{\rm beam}\; B_{\lambda}\left(T_{\rm dust}\right)}\right)}.
\end{equation}
With similar solid angles, $\Omega_{\rm core}$ and $\Omega_{\rm beam}$, we approximate \cref{appendixeq:core mass1} by
\begin{align} \label{appendixeq:core mass}
    M_{\rm \tau\gtrsim 1} &\approx \,-\,\frac{\Omega_{\rm beam}\;d^2}{\kappa_{\lambda}}\; \ln{\left(1\,-\,\frac{S^{\rm peak}_{\lambda}}{\Omega_{\rm beam}\; B_{\lambda}\left(T_{\rm dust}\right)}\right)} \times \frac{\Omega_{\rm core}}{\Omega_{\rm beam}} \nonumber \\
     &\approx\,-\,\frac{\Omega_{\rm core}\;d^2}{\kappa_{\lambda}}\; \ln{\left(1\,-\,\frac{S^{\rm peak}_{\lambda}}{\Omega_{\rm beam}\; B_{\lambda}\left(T_{\rm dust}\right)}\right)}.
\end{align}

Since for a Gaussian source $\frac{S^{\rm int}}{\Omega_{\rm core}} = \frac{S^{\rm peak}}{\Omega_{\rm beam}}$, it follows that the core mass of \cref{appendixeq:core mass} can be estimated by
\begin{align} \label{appendixeq:optically thick mass detailled}
    M_{\rm \tau\gtrsim 1} &\simeq\,-\,\frac{\Omega_{\rm beam}\;d^2}{\kappa_{\lambda}} \frac{S^{\rm int}}{S^{\rm peak}}\; \ln{\left(1\,-\,\frac{S^{\rm peak}_{\lambda}}{\Omega_{\rm beam}\;B_{\lambda}\left(T_{\rm dust}\right)}\right)} \\
     &\simeq\,-\,\frac{S^{\rm int}\; d^2}{\kappa_{\lambda}\;B_{\lambda}\left(T_{\rm dust}\right)} \nonumber \\
     &\qquad \times \frac{\Omega_{\rm beam}\;B_{\lambda}\left(T_{\rm dust}\right)}{S^{\rm peak}}\;\ln{\left(1\,-\,\frac{S^{\rm peak}_{\lambda}}{\Omega_{\rm beam}\; B_{\lambda}\left(T_{\rm dust}\right)}\right)} \nonumber\\
     &\simeq\,-\, M_{\rm \tau\ll 1} \nonumber\\
     &\qquad \times \frac{\Omega_{\rm beam}\;B_{\lambda}\left(T_{\rm dust}\right)}{S^{\rm peak}}\;\ln{\left(1\,-\,\frac{S^{\rm peak}_{\lambda}}{\Omega_{\rm beam}\; B_{\lambda}\left(T_{\rm dust}\right)}\right)}. \nonumber
\end{align}

We can retrieve the optically thin equation for the mass estimates of \cref{eq:optically thin mass} applying the optically thin medium assumption at 1.3~mm:
\begin{equation} \nonumber
    \frac{S^{\rm peak}_{1.3{\rm mm}}}{\Omega_{\rm beam}} \ll B_{\rm 1.3{\rm mm}}\left(T_{\rm dust}\right). 
\end{equation}


\section{Completeness of the core catalogs} \label{appendixsect:completeness simulation}

\begin{figure}[h!]
    \centering
    \includegraphics[width=0.48\textwidth]{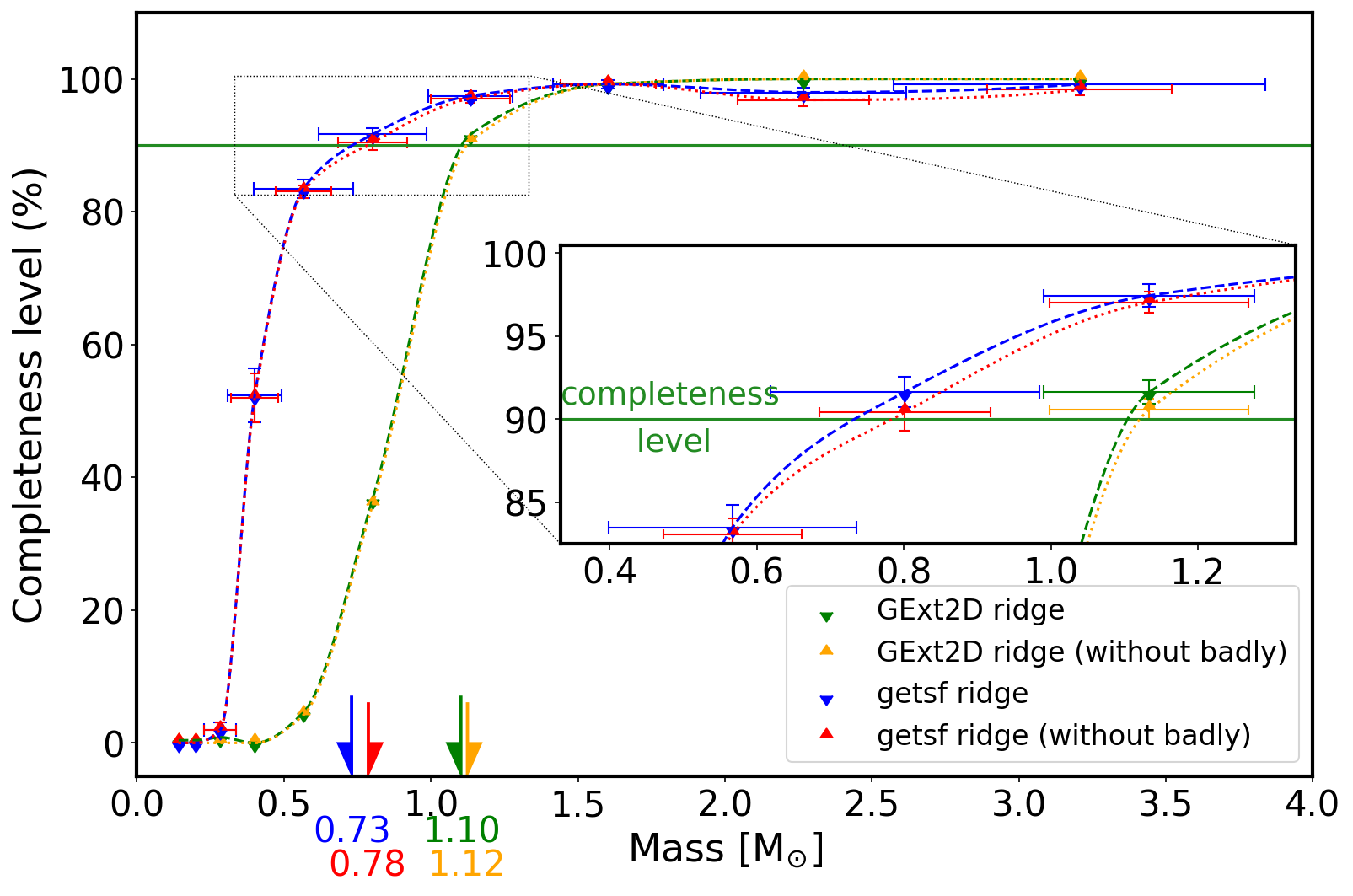}
    \caption{Completeness levels of the $\sim$900 synthetic cores added on the background image of W43-MM2\&MM3. The core content is 90\% complete down to $0.8\pm0.2~\Msol$ and $1.1\pm0.2~\Msol$ for \textsl{getsf} and \textsl{GExt2D}, respectively. The error bars represent the $\pm1\sigma$ uncertainties for mass estimations across each bin (\textit{x-axis}) and total of cores retrieved per bin (\textit{y-axis}). Data points were interpolated using the Piecewise Cubic Hermite Interpolating Polynomial method. Blue and green points represent the full sample of cores detected by \textsl{getsf} and \textsl{GExt2D}, respectively, while red and orange points measure the bin completeness for cores that have mass measurements at worst larger or smaller than a factor of 2 compared to the truth table, thus excluding badly detected sources (see \cref{appendixsect:mngseg} and \cref{appendixtab:simalma stat} for complementary information).}
    \label{appendixfig:completeness}
\end{figure}

We estimated the completeness level of each of the \textsl{getsf} and \textsl{Gext2D} core catalogs by injecting synthetic populations of $\sim$1900 sources over the background image of W43-MM2\&MM3. Background images are produced by \textsl{getsf} during the source extraction process in the \denoised \& \bsens and \bsens images at 1.3~mm, respectively (see Sect.~\ref{sect:extraction of compact sources}). Synthetic sources were split into ten bins logarithmically spaced between $0.1~\Msol$ and $3.2~\Msol$, with a constant number of about 190 sources per bin in order to properly sample the bins useful for defining the completeness level. Synthetic sources more massive than $3.2~\Msol$ have a flux contrast to their background that allows their detection in all test cases. 
The density profile chosen for synthetic cores is Gaussians with FWHM of $0.7\arcsec$ (or 3\,400~au at 5.5~kpc), equal to the median size of extracted sources (see \cref{fig:fwhm distribution}), and with an outer diameter of 2.5\arcsec. Following \cref{appendixsect:mngseg} (see \cref{appendixfig:3tiles}), sources are randomly injected in a regular grid, not allowing cores to overlap. 
We focus on the $\sim$900 synthetic sources located within the central part of the W43-MM2\&MM3 ridge, corresponding to the location of its detected cores (see \cref{fig:1.3mm and trichrone}a). This method allows us to estimate a level of completeness as close as possible to that of our core catalog. With a source grid covering the entire image, the completeness level would be 1.4 times smaller.
We performed five series of completeness simulations, varying the location of synthetic sources to mitigate the effects of the chosen grid relative to the background and allow estimation of the error bars of \cref{appendixfig:completeness}. We ran the extraction algorithms \textsl{getsf} and \textsl{Gext2D} on all these synthetic images with the same parameters as for the observations (see Sect.~\ref{sect:extraction of compact sources}). 

Figure~\ref{appendixfig:completeness} shows the detection rates of synthetic sources injected on the W43-MM2\&MM3 background image versus the synthetic source mass. We use it to estimate a global 90\% completeness level (excluding badly measured sources) of $\sim$0.8$\pm0.2~\Msol$ for the \textsl{getsf} and $\sim$1.1$\pm0.2~\Msol$ for the \textsl{Gext2D} catalogs, respectively. Uncertainties are estimated from the error bars shown in \cref{appendixfig:completeness} in the mass bins located near the point of intersection with the 90\% completeness level. 
75\% of the sample of \cref{appendixtab:derived core table} lie above the \textsl{getsf} completeness level.

\section{Complementary figures} 
\label{appendixsect:complementary figure 3mm and freefree}

\begin{figure*}[htbp!]
    \centering
    \includegraphics[width=1.\textwidth]{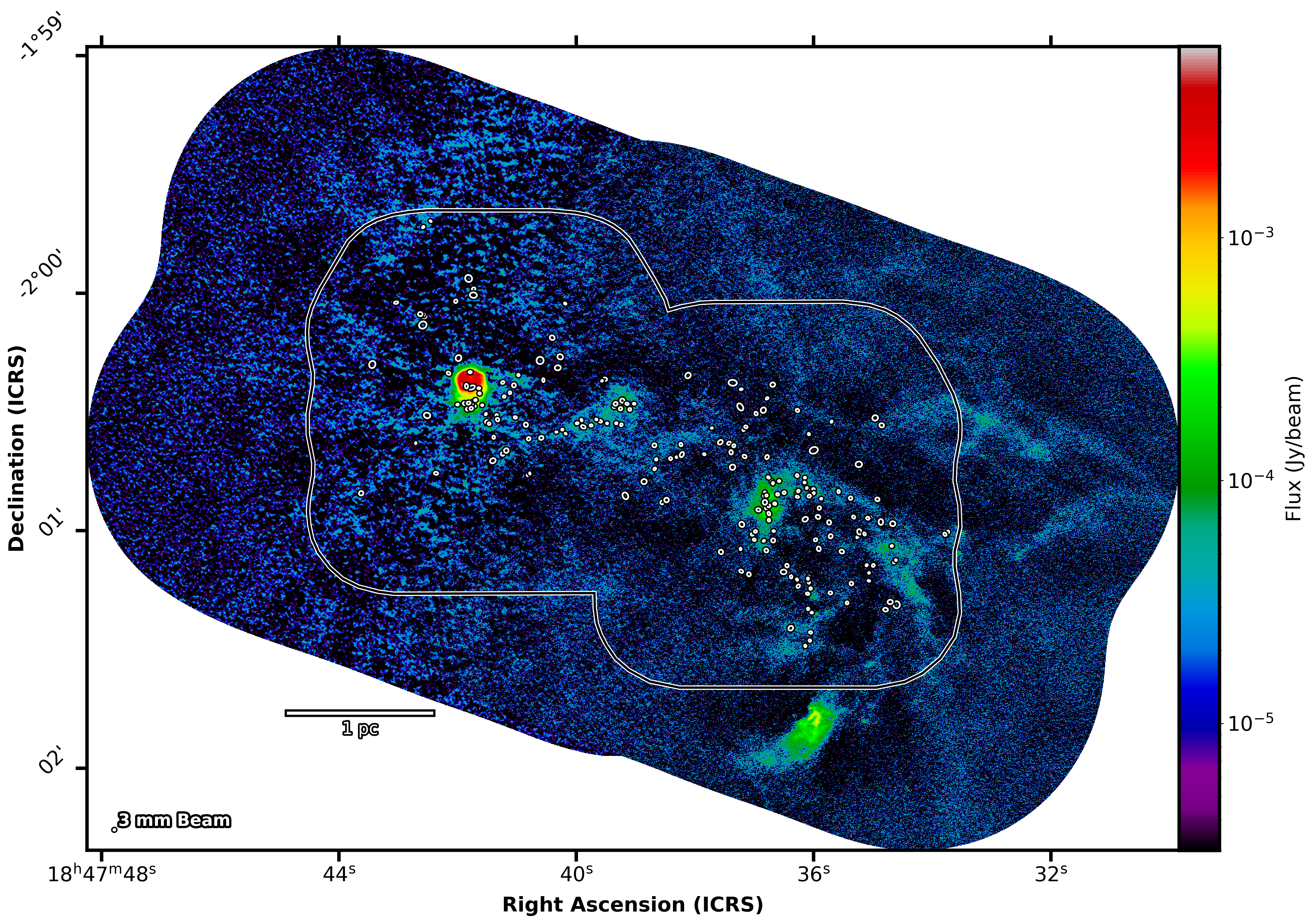}
    \caption{W43-MM2\&MM3 protocluster cloud imaged at 3~mm by the ALMA 12~m array (best-sensitivity image prior to primary-beam correction). White ellipses outline the FWHM size of compact cores extracted by \textsl{getsf} at 1.3~mm and whose 3~mm flux is measurable (see Sect.~\ref{sect:extraction of compact sources}). the ellipse in the lower left corner represents the angular resolution of the \bsens 3~mm image and the scale bar indicates the size in physical units.}
    \label{appendixfig:3mm image with cores}
\end{figure*}

\begin{figure*}[htbp!]
    \centering
    \begin{minipage}{0.49\textwidth}
      \centering
      \includegraphics[width=1.\textwidth]{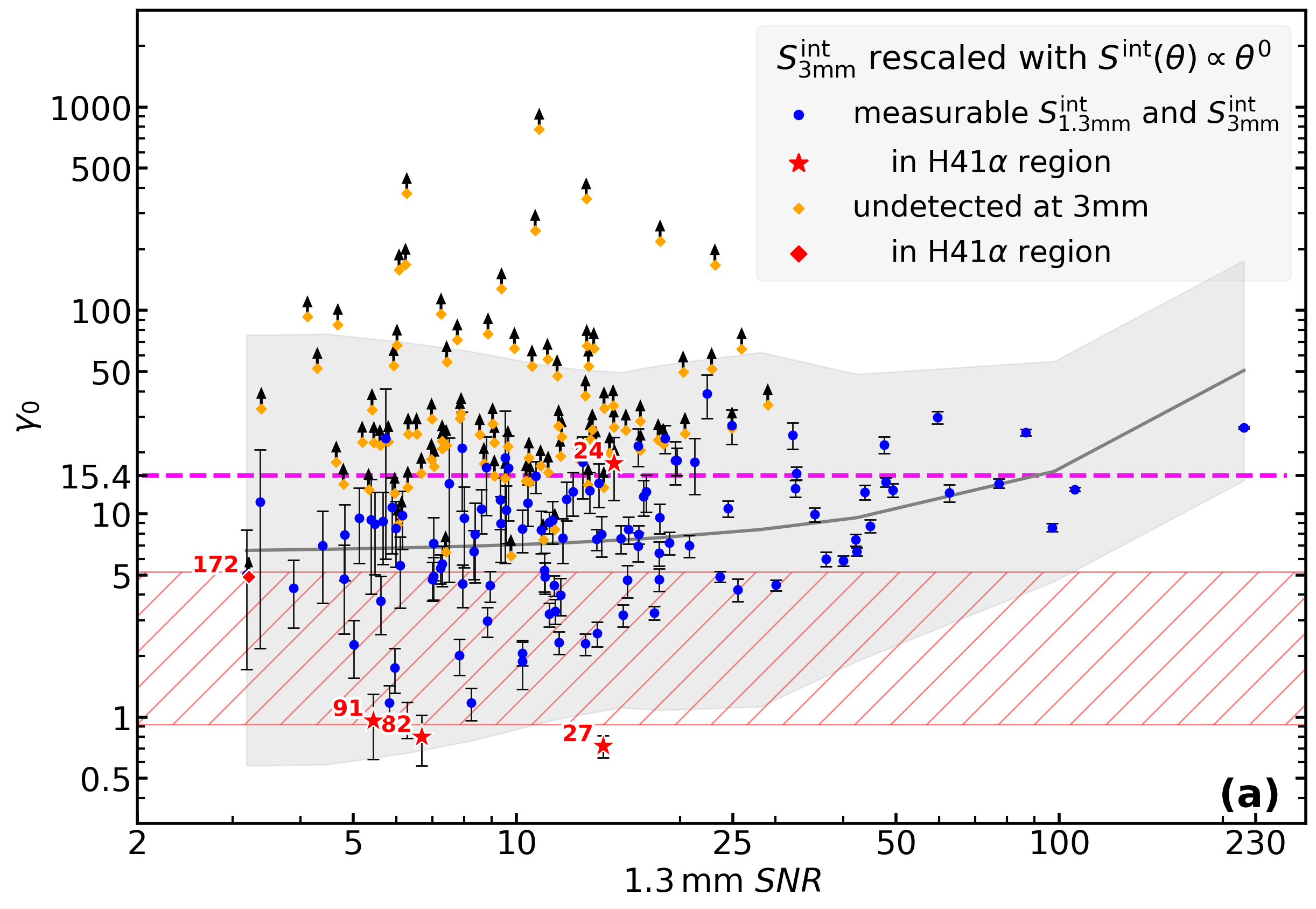}
    \end{minipage}%
    \begin{minipage}{0.49\textwidth}
      \centering
      \includegraphics[width=1.\textwidth]{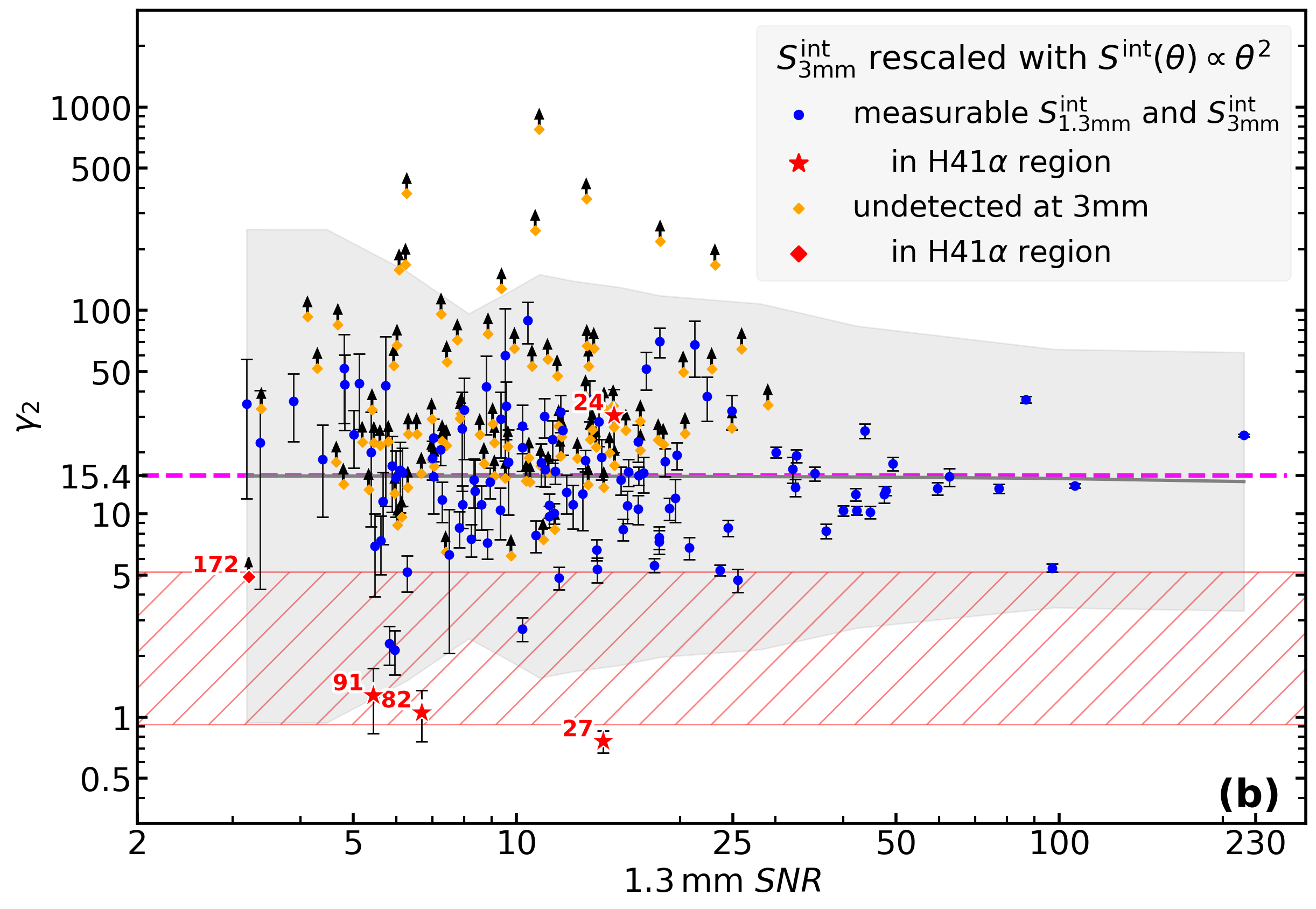}
    \end{minipage}
    \caption{Thermal dust emission cores separated from free-free emission sources, using their 1.3 mm to 3 mm flux ratios without re-scaling, $\gamma_0$ (in \textsl{a}), and with a stronger re-scaling suited for the flatter density distribution of starless cores, $\gamma_2$ (in \textsl{b}). Plots are shown as a function of the S/N in the 1.3~mm image. The magenta horizontal dashed line represents the theoretical flux ratio of thermal dust emission of 15.4, computed in \cref{eq:theo thermal ratio}. The gray segments correspond to the median values of the core flux ratios and the shaded gray area is their $3\,\sigma$ dispersion. Blue points indicate cores with 3~mm thermal dust emission whose flux is re-scaled to the source size measured at 1.3~mm (see Eq.~\ref{eq:re-scale}), while orange points locate cores undetected at 3~mm, thus taking the ratio between the 1.3~mm peak flux and the $1\,\sigma$ peak error at 3~mm, corresponding to a lower limit. Red symbols are sources located within the H41$\alpha$ recombination line region of \cref{fig:freefree}a. The gray curve indicates the median value of the core ratios, computed over bins of 20 adjacent cores as ranked by their S/N. The shaded gray area indicates the corresponding $3\,\sigma$ dispersion in flux ratio values. The magenta horizontal dashed line represents the theoretical flux ratio of thermal dust emission of 15.4, computed in Eq.~\ref{eq:theo thermal ratio}. The red hatched area locates the theoretical flux ratios of UC\hii or HC\hii regions, whose free-free emission is either optically thin (lower limit) or partly to totally optically thick (upper limit). The median value of the blue points is $8.4\pm2.1$ (in \textsl{a}) and $15.3\pm2.0$ (in \textsl{b}).}
    \label{appendixfig:freefree different rescaling}
\end{figure*}

\begin{figure*}[htbp!]
    \centering
    \includegraphics[width=1.\textwidth]{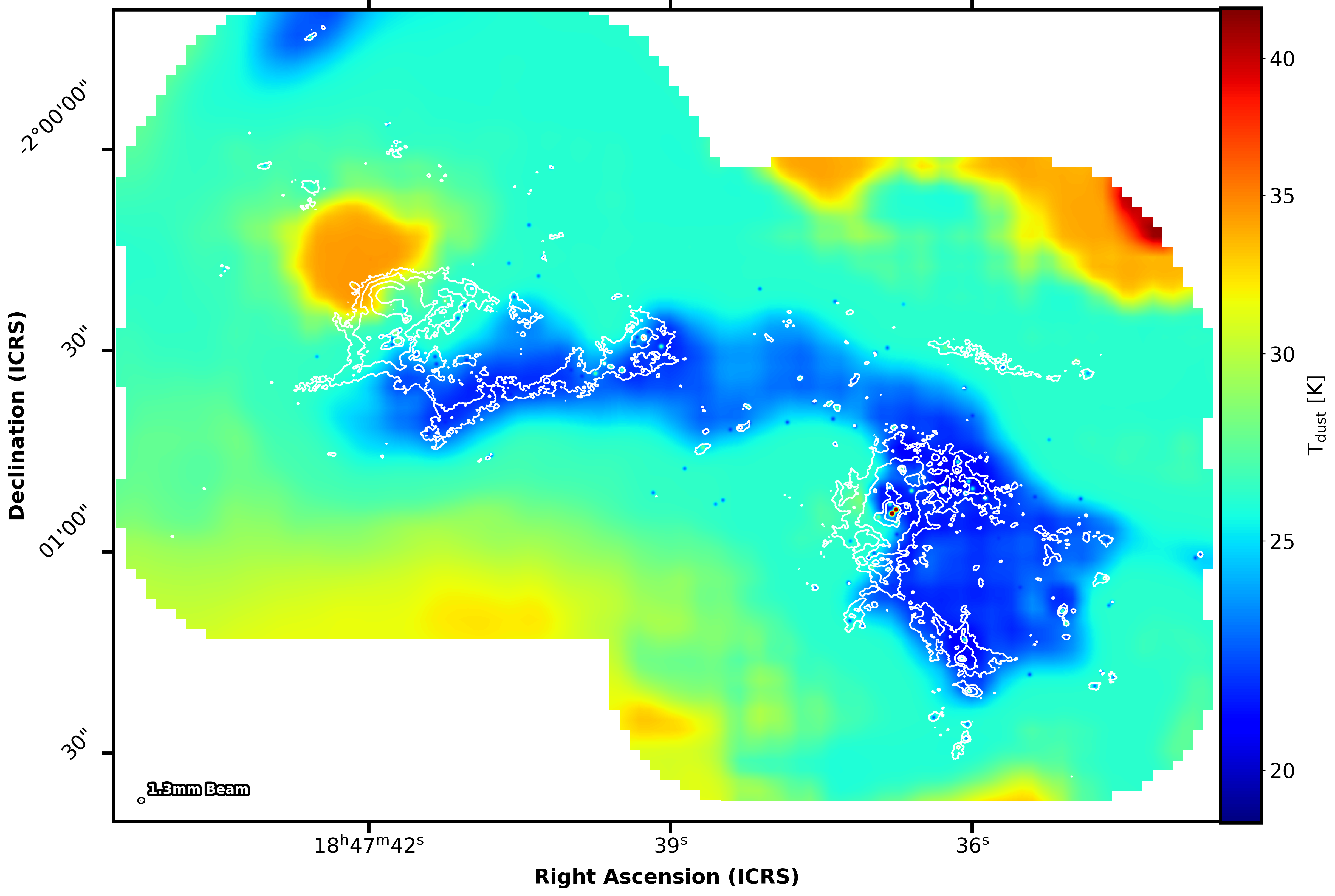}
    \caption{Background dust temperature of the W43-MM2\&MM3 protocluster cloud from Motte et al. (in prep.). It combines a $2.5\arcsec$-resolution dust temperature image computed by \textsl{PPMAP} with the central heating and self-shielding of protostellar and pre-stellar cores, respectively, at $0.46\arcsec$ resolution (see Sect.~\ref{sect:mass estimation}). White contours correspond to 4, 15, 75, and $150\,\sigma$ of the ALMA 12~m array \bsens image, at 1.3~mm.}
    \label{appendixfig:dust temperature map}
\end{figure*}

\cref{appendixsect:complementary figure 3mm and freefree} presents the 3~mm continuum image of the W43-MM2\&3 ridge (see \cref{appendixfig:3mm image with cores}), complementary figures used to identify sources associated with free-free emission peaks (see \cref{appendixfig:freefree different rescaling}), and the dust temperature background image of Motte et al. (in prep.) (see also Sect.~\ref{sect:mass estimation}).

\section{Online tables} \label{appendixsect:supplementary tables}
\cref{appendixsect:supplementary tables} presents Tables~\ref{appendixtab:core detection table} and \ref{appendixtab:derived core table}. The first lists the sources detected by \textsl{getsf} at 1.3~mm and the second gives their physical properties (see Sects.~\ref{sect:extraction of compact sources} and \ref{sect:core nature mass estim}).

{\renewcommand{\arraystretch}{1.5}%
\begin{sidewaystable*}[ht]
\centering
\tiny
\begin{threeparttable}[c]
\caption{Catalog of dense cores identified by \textsl{getsf} (v210403) in the ALMA-IMF images of the W43-MM2\&MM3 mini-starburst.}
\label{appendixtab:core detection table}
\begin{tabular}{cccccccccccccc}
\hline\hline
n & Core name & RA & Dec & $a_{\rm 1.3mm} \times b_{\rm 1.3mm}$ & PA$_{\rm 1.3mm}$ & $S_{\rm 1.3mm}^{\rm peak}$ & $S_{\rm 1.3mm}^{\rm int}$ & $a_{\rm 3mm} \times b_{\rm 3mm}$ & PA$_{\rm 3mm}$ & $S_{\rm 3mm}^{\rm peak}$ & $S_{\rm 3mm}^{\rm int}$ & \textsl{GExt2D} tag & Contamination \\
 & W43-MM2\&3\_ALMAIMF$\ast$ & [J2000] & [J2000] & [$\arcsec \times \arcsec$] & [deg] & [mJy\,beam$^{-1}$] & [mJy] & [$\arcsec \times \arcsec$] & [deg] & [mJy\,beam$^{-1}$] & [mJy] &  & \\
\hline
\hline
  1 & 184736.80-20054.27  & 18:47:36.80 & -2:00:54.27 & $0.8\times0.6$ &  5  & $140.00\pm1.35$ & $398.70\pm3.07$ & $0.8\times0.5$ & 25  & $  9.24\pm0.14$ & $ 23.16\pm0.20$ & $ \star\star $ &   COMs    \\ 
  2 & 184741.71-20028.60  & 18:47:41.71 & -2:00:28.60 & $0.6\times0.5$ & 94  & $ 53.73\pm1.15$ & $100.70\pm1.60$ & $0.7\times0.5$ & 117 & $  4.12\pm0.37$ & $  7.94\pm0.66$ & $ \star\star $ &           \\ 
  3 & 184739.26-20028.10  & 18:47:39.26 & -2:00:28.10 & $0.7\times0.6$ & 23  & $ 17.06\pm0.35$ & $ 42.18\pm0.60$ & $0.8\times0.7$ & 173 & $  0.86\pm0.04$ & $  2.31\pm0.06$ & $ \star\star $ &   COMs    \\ 
  5 & 184736.10-20115.98  & 18:47:36.10 & -2:01:15.98 & $0.9\times0.6$ & 75  & $ 15.11\pm0.40$ & $ 42.42\pm0.55$ & $0.8\times0.7$ & 72  & $  1.25\pm0.09$ & $  3.00\pm0.11$ & $ \star\star $ &           \\ 
  6 & 184736.03-20120.73  & 18:47:36.03 & -2:01:20.73 & $0.7\times0.5$ & 75  & $ 11.98\pm0.20$ & $ 20.78\pm0.21$ & $0.6\times0.5$ & 57  & $  1.77\pm0.09$ & $  2.43\pm0.09$ & $ \star\star $ &           \\ 
  7 & 184736.75-20053.75  & 18:47:36.75 & -2:00:53.75 & $0.8\times0.8$ & 19  & $ 33.85\pm1.47$ & $ 83.72\pm1.48$ & $0.8\times0.8$ & 121 & $  3.63\pm0.13$ & $  9.46\pm0.13$ & $            $ &   COMs    \\ 
  9 & 184741.73-20027.42  & 18:47:41.73 & -2:00:27.42 & $0.6\times0.5$ & 108 & $ 26.32\pm1.02$ & $ 39.18\pm0.90$ & $0.7\times0.6$ & 115 & $  1.75\pm0.21$ & $  3.07\pm0.18$ & $ \star\star $ &           \\ 
 10 & 184736.28-20050.75  & 18:47:36.28 & -2:00:50.75 & $0.6\times0.5$ &  1  & $ 10.00\pm0.47$ & $ 15.78\pm0.45$ & $0.6\times0.5$ &  5  & $  1.06\pm0.06$ & $  1.74\pm0.06$ & $   \star    $ &   COMs    \\ 
 11 & 184740.97-20020.73  & 18:47:40.97 & -2:00:20.73 & $0.6\times0.6$ & 34  & $  6.77\pm0.25$ & $ 14.63\pm0.30$ & $0.7\times0.6$ & 69  & $  0.51\pm0.05$ & $  1.12\pm0.06$ & $ \star\star $ &           \\ 
 12 & 184736.70-20047.55  & 18:47:36.70 & -2:00:47.55 & $0.7\times0.5$ & 109 & $ 15.83\pm0.69$ & $ 26.98\pm0.82$ & $0.6\times0.6$ & 102 & $  1.10\pm0.08$ & $  1.70\pm0.07$ & $ \star\star $ &           \\ 
 13 & 184736.15-20047.87  & 18:47:36.15 & -2:00:47.87 & $0.6\times0.5$ & 69  & $  9.53\pm0.31$ & $ 13.22\pm0.29$ & $0.6\times0.5$ & 45  & $  1.13\pm0.08$ & $  1.52\pm0.07$ & $ \star\star $ &           \\ 
 14 & 184735.10-20108.77  & 18:47:35.10 & -2:01:08.77 & $0.6\times0.5$ & 134 & $  5.85\pm0.19$ & $ 10.60\pm0.25$ & $0.7\times0.6$ & 137 & $  0.82\pm0.03$ & $  1.63\pm0.04$ & $ \star\star $ &           \\ 
 15 & 184736.84-20102.61  & 18:47:36.84 & -2:01:02.61 & $0.6\times0.4$ & 128 & $  8.18\pm0.38$ & $ 11.66\pm0.39$ & $0.9\times0.6$ & 156 & $  0.99\pm0.06$ & $  2.61\pm0.07$ & $   \star    $ &           \\ 
 16 & 184735.69-20032.50  & 18:47:35.69 & -2:00:32.50 & $0.6\times0.5$ & 152 & $  3.71\pm0.11$ & $  6.45\pm0.15$ & $0.6\times0.5$ & 104 & $  0.44\pm0.03$ & $  0.87\pm0.04$ & $ \star\star $ &           \\ 
 18 & 184740.23-20034.51  & 18:47:40.23 & -2:00:34.51 & $0.6\times0.5$ & 103 & $  3.83\pm0.17$ & $  5.68\pm0.17$ & $0.6\times0.5$ & 114 & $  0.29\pm0.03$ & $  0.43\pm0.03$ & $ \star\star $ &           \\ 
 20 & 184736.06-20127.82  & 18:47:36.06 & -2:01:27.82 & $0.7\times0.6$ & 56  & $  4.75\pm0.21$ & $  9.56\pm0.24$ & $0.8\times0.7$ & 110 & $  0.57\pm0.04$ & $  1.62\pm0.06$ & $ \star\star $ &           \\ 
 21 & 184741.39-20036.43  & 18:47:41.39 & -2:00:36.43 & $0.9\times0.7$ &  6  & $  0.45\pm0.18$ & $  1.14\pm0.18$ &       --       & --  & $   \leq 0.003$ & $   \leq 0.003$ & $   \star    $ &           \\ 
 22 & 184736.65-20053.23  & 18:47:36.65 & -2:00:53.23 & $0.8\times0.7$ & 85  & $ 11.67\pm1.36$ & $ 31.92\pm1.62$ & $0.8\times0.7$ & 25  & $  0.71\pm0.15$ & $  1.75\pm0.18$ & $   \star    $ &           \\ 
 24 & 184741.63-20025.37  & 18:47:41.63 & -2:00:25.37 & $0.6\times0.6$ & 44  & $ 11.09\pm1.28$ & $ 19.65\pm1.30$ & $0.7\times0.7$ & 164 & $  0.53\pm0.31$ & $  1.11\pm0.31$ & $ \star\star $ &           \\ 
 25 & 184741.83-20029.32  & 18:47:41.83 & -2:00:29.32 & $1.0\times0.8$ & 117 & $  9.18\pm0.96$ & $ 36.75\pm1.47$ & $1.1\times0.8$ & 101 & $  0.42\pm0.18$ & $  1.35\pm0.21$ & $ \star\star $ &           \\ 
 27 & 184741.76-20023.88  & 18:47:41.76 & -2:00:23.88 & $1.7\times1.0$ & 112 & $  5.03\pm1.33$ & $ 33.15\pm2.30$ & $1.6\times1.1$ & 116 & $  6.53\pm1.46$ & $ 45.93\pm2.47$ & $            $ & Free-free \\ 
 28 & 184736.68-20048.06  & 18:47:36.68 & -2:00:48.06 & $0.6\times0.5$ & 166 & $  9.54\pm0.70$ & $ 15.02\pm0.61$ & $0.6\times0.5$ & 167 & $  0.96\pm0.08$ & $  1.41\pm0.07$ & $ \star\star $ &           \\ 
 30 & 184737.17-20034.45  & 18:47:37.17 & -2:00:34.45 & $1.0\times0.7$ & 142 & $  0.43\pm0.08$ & $  1.27\pm0.11$ & $1.6\times1.6$ & 98  & $  0.05\pm0.01$ & $  0.38\pm0.02$ & $            $ &           \\ 
 32 & 184738.30-20041.47  & 18:47:38.30 & -2:00:41.47 & $0.6\times0.4$ & 98  & $  2.42\pm0.09$ & $  3.56\pm0.10$ & $0.6\times0.5$ & 81  & $  0.25\pm0.02$ & $  0.36\pm0.02$ & $ \star\star $ &           \\ 
 33 & 184736.82-20052.88  & 18:47:36.82 & -2:00:52.88 & $1.2\times1.1$ & 170 & $ 12.63\pm1.01$ & $ 69.47\pm1.17$ & $1.0\times0.7$ &  9  & $  0.60\pm0.14$ & $  2.33\pm0.13$ & $            $ &           \\ 
 34 & 184736.97-20030.47  & 18:47:36.97 & -2:00:30.47 & $1.0\times0.8$ & 30  & $  0.37\pm0.09$ & $  1.24\pm0.10$ &       --       & --  & $   \leq 0.013$ & $   \leq 0.020$ & $            $ &           \\ 
 35 & 184733.73-20100.38  & 18:47:33.73 & -2:01:00.38 & $0.8\times0.6$ & 178 & $  3.37\pm0.17$ & $  7.51\pm0.20$ & $0.9\times0.6$ & 98  & $  0.43\pm0.05$ & $  1.25\pm0.07$ & $ \star\star $ &           \\ 
 37 & 184739.48-20032.93  & 18:47:39.48 & -2:00:32.93 & $0.5\times0.5$ & 98  & $  3.75\pm0.33$ & $  4.92\pm0.29$ & $0.6\times0.5$ & 106 & $  0.38\pm0.03$ & $  0.62\pm0.03$ & $ \star\star $ &           \\ 
 38 & 184735.05-20056.75  & 18:47:35.05 & -2:00:56.75 & $0.8\times0.8$ & 13  & $  0.27\pm0.10$ & $  0.79\pm0.11$ &       --       & --  & $   \leq 0.016$ & $   \leq 0.018$ & $            $ &           \\ 
 39 & 184736.14-20046.65  & 18:47:36.14 & -2:00:46.65 & $0.6\times0.4$ & 104 & $  3.56\pm0.26$ & $  4.22\pm0.23$ & $0.5\times0.5$ & 98  & $  0.55\pm0.06$ & $  0.66\pm0.05$ & $   \star    $ &           \\ 
 40 & 184734.99-20108.83  & 18:47:34.99 & -2:01:08.83 & $0.8\times0.7$ & 136 & $  0.40\pm0.11$ & $  0.91\pm0.12$ &       --       & --  & $   \leq 0.019$ & $   \leq 0.026$ & $   \star    $ &           \\ 
 41 & 184736.14-20129.16  & 18:47:36.14 & -2:01:29.16 & $0.6\times0.6$ & 12  & $  2.57\pm0.26$ & $  5.22\pm0.30$ & $0.7\times0.6$ & 106 & $  0.20\pm0.05$ & $  0.43\pm0.06$ & $ \star\star $ &           \\ 
 43 & 184739.22-20027.20  & 18:47:39.22 & -2:00:27.20 & $1.1\times1.0$ & 45  & $  3.43\pm0.27$ & $ 12.90\pm0.27$ & $0.9\times0.8$ & 153 & $  0.21\pm0.04$ & $  0.59\pm0.04$ & $ \star\star $ &           \\ 
\hline
\end{tabular}
\begin{tablenotes}[para,flushleft]
Notes: RA, right ascension; Dec, declination; $a$ and $b$, major and minor sizes at half maximum; PA, counterclockwise ellipse orientation from north to east; $S^{\rm peak}$ and $S^{\rm int}$, peak and integrated fluxes; $\star$, detected by \textsl{GExt2D}; $\star\star$, detected by \textsl{GExt2D} and with a 1.3~mm integrated flux at worst 30\% larger or smaller than \textsl{getsf} fluxes; Contamination, tag to indicate source with partial contamination (COMs, Lines) or almost fully contamined fluxes (free-free). The full table is available in electronic form through CDS.
\end{tablenotes}
\end{threeparttable}
\end{sidewaystable*}}

{\renewcommand{\arraystretch}{1.5}%
\begin{table*}[ht]
\centering
\begin{threeparttable}[c]
\caption{Derived properties of cores identified by \textsl{getsf} (v210403) in the ALMA-IMF images of the W43-MM2\&MM3 mini-starburst.}
\label{appendixtab:derived core table}
\begin{tabular}{cccccc}
\hline\hline
n & Core name & FWHM$^{\rm dec}_{\rm 1.3mm}$ & $M_{\rm \tau\gtrsim 1}$ & $T_{\rm dust}$ & $n_{\rm H_2}$ \\
 & W43-MM2\&3\_ALMAIMF$\ast$ & [AU] & [\Msol] & [K] & [$\times 10^{6}$cm$^{-3}$] \\
\hline
  1 & 184736.80-20054.27  & 3070  & $  69.9\pm13.7$ & $65.0\pm10.0$ & 72.85 \\ 
  2 & 184741.71-20028.60  & 1820  & $  44.6\pm8.8 $ & $28.7\pm4.0$ & 223.46 \\ 
  3 & 184739.26-20028.10  & 2760  & $  11.2\pm2.3 $ & $40.0\pm7.0$ & 16.25 \\ 
  5 & 184736.10-20115.98  & 3330  & $  17.8\pm3.3 $ & $27.5\pm4.0$ & 14.58 \\ 
  6 & 184736.03-20120.73  & 1880  & $   8.5\pm1.5 $ & $27.8\pm4.0$ & 38.85 \\ 
  7 & 184736.75-20053.75  & 3500  & $  14.3\pm1.9 $ & $60.0\pm7.0$ & 10.05 \\ 
  9 & 184741.73-20027.42  & 1780  & $  16.0\pm2.9 $ & $28.7\pm4.0$ & 86.32 \\ 
 10 & 184736.28-20050.75  & 1890  & $   2.6\pm0.3 $ & $60.0\pm7.0$ & 11.82 \\ 
 11 & 184740.97-20020.73  & 2260  & $   7.7\pm1.3 $ & $22.6\pm3.0$ & 20.19 \\ 
 12 & 184736.70-20047.55  & 1840  & $  11.2\pm2.1 $ & $27.8\pm4.0$ & 53.79 \\ 
 13 & 184736.15-20047.87  & 1360  & $   8.2\pm1.7 $ & $20.0\pm3.0$ & 98.89 \\ 
 14 & 184735.10-20108.77  & 1810  & $   4.3\pm0.8 $ & $27.8\pm4.0$ & 21.79 \\ 
 15 & 184736.84-20102.61  & 1530  & $   4.7\pm0.9 $ & $27.8\pm4.0$ & 40.31 \\ 
 16 & 184735.69-20032.50  & 1550  & $   3.4\pm0.6 $ & $22.4\pm3.0$ & 27.36 \\ 
 18 & 184740.23-20034.51  & 1640  & $   3.1\pm0.6 $ & $21.6\pm3.0$ & 21.29 \\ 
 20 & 184736.06-20127.82  & 2560  & $   4.9\pm0.8 $ & $23.0\pm3.0$ & 8.79  \\ 
 21 & 184741.39-20036.43  & 3370  & $   0.6\pm0.1 $ & $21.7\pm3.0$ & 0.49  \\ 
 22 & 184736.65-20053.23  & 3180  & $  18.1\pm3.5 $ & $21.6\pm3.0$ & 17.02 \\ 
 24 & 184741.63-20025.37  & 2230  & $   7.5\pm1.3 $ & $29.4\pm4.0$ & 20.39 \\ 
 25 & 184741.83-20029.32  & 4110  & $  19.6\pm3.5 $ & $22.5\pm3.0$ & 8.54  \\ 
 28 & 184736.68-20048.06  & 1860  & $   6.1\pm1.1 $ & $27.8\pm4.0$ & 28.83 \\ 
 30 & 184737.17-20034.45  & 3850  & $   0.7\pm0.1 $ & $21.9\pm3.0$ & 0.36  \\ 
 32 & 184738.30-20041.47  & 1410  & $   1.9\pm0.3 $ & $21.9\pm3.0$ & 20.63 \\ 
 33 & 184736.82-20052.88  & 5890  & $  29.6\pm5.6 $ & $26.9\pm4.0$ & 4.39  \\ 
 34 & 184736.97-20030.47  & 4080  & $   0.7\pm0.1 $ & $22.1\pm3.0$ & 0.29  \\ 
 35 & 184733.73-20100.38  & 2890  & $   4.0\pm0.7 $ & $22.0\pm3.0$ & 5.07  \\ 
 37 & 184739.48-20032.93  & 1370  & $   2.0\pm0.4 $ & $27.8\pm4.0$ & 23.13 \\ 
 38 & 184735.05-20056.75  & 3580  & $   0.4\pm0.1 $ & $21.8\pm3.0$ & 0.28  \\ 
 39 & 184736.14-20046.65  & 1250  & $   2.4\pm0.5 $ & $20.8\pm3.0$ & 38.36 \\ 
 40 & 184734.99-20108.83  & 3180  & $   0.5\pm0.1 $ & $21.8\pm3.0$ & 0.46  \\ 
 41 & 184736.14-20129.16  & 2250  & $   1.9\pm0.3 $ & $29.9\pm4.0$ & 5.10  \\ 
 43 & 184739.22-20027.20  & 5430  & $   7.0\pm1.2 $ & $21.8\pm3.0$ & 1.32  \\
\hline
\end{tabular}
\begin{tablenotes}[para,flushleft]
Notes: FWHM$^{\rm dec}$, deconvolved physical core size at 1.3~mm; $M_{\rm \tau\gtrsim 1}$, optically thick core dust mass measured from the 1.3~mm integrated flux of \cref{appendixtab:core detection table}. Uncertainties include 1.3~mm integrated and peak fluxes and uncertainties (see \cref{appendixtab:core detection table}) and core temperature uncertainty (Col.~5, see Sect.~\ref{sect:mass estimation}); $T_{\rm dust}$, dust temperature measured in \cref{appendixfig:dust temperature map}; $n_{\rm H_2}$, volume density of the core (see Eq.~\ref{eq:density}). The full table is available in electronic form through CDS.
\end{tablenotes}
\end{threeparttable}
\end{table*}}
\end{appendix}

\end{document}